\pdfoutput=1
\documentclass[12pt]{article}

\usepackage{graphicx,psfrag,epsf,color}
\usepackage{amsmath,amssymb,amsfonts}
\usepackage{array}
\usepackage{cite}
\usepackage{multirow}
\usepackage{rotating}

\renewcommand{\arraystretch}{1.25}

\newcounter{MBQ}

\newcommand{\be}{\begin{equation}}
\newcommand{\ee}{\end{equation}}
\newcommand{\bea}{\begin{eqnarray}}
\newcommand{\eea}{\end{eqnarray}}
\newcommand{\bi}{\begin{itemize}}
	\newcommand{\ei}{\end{itemize}}
\newcommand{\ben}{\begin{enumerate}}
	\newcommand{\een}{\end{enumerate}}
\newcommand{\bt}{\begin{tabular}}
	\newcommand{\et}{\end{tabular}}

\newcommand{\lc}{\left[}
\newcommand{\rc}{\right]}
\newcommand{\lp}{\left(}
\newcommand{\rp}{\right)}

\newcommand{\mII}{M_2}
\newcommand{\mIII}{M_3}

\newcommand{\mao}{M_{A}}
\newcommand{\tanb}{\tan\beta}

\newcommand{\FeynHiggs}{\texttt{FeynHiggs}}
\newcommand{\micrOMEGAs}{\texttt{micrOMEGAs}}

\newcommand{\neuI}{\tilde\chi^0_1}
\newcommand{\mlsp}{m_\textrm{LSP}}

\newcommand{\btosgamma}[1]{\mathcal{B}^{#1}\lp\bar B \rightarrow X_s \gamma\rp}
\newcommand{\btosgammabest}[1]{\mathcal{B}^{#1}\lp\bar B \rightarrow X_s \gamma\rp}

\newcommand{\msf}{M_{\rm sf}}
\newcommand{\gev}{{\rm GeV}}
\newcommand{\tev}{{\rm TeV}}

\begin{document}

\allowdisplaybreaks

\begin{titlepage}

\begin{flushright}
{\small
TUM-HEP-1033/15\\
15 January 2016
}
\end{flushright}

\vskip1cm
\begin{center}
{\fontsize{15}{19}\bf\boldmath
Relic density of wino-like dark matter in the MSSM}
\end{center}

\vspace{0.8cm}
\begin{center}
{\sc M.~Beneke$^{a}$, A.~Bharucha$^{b}$, F. Dighera$^{a}$, 
C.~Hellmann$^{a}$, } \\ 
{\sc A.~Hryczuk$^{a,c,d}$, S.~Recksiegel$^{a}$, 
and  P. Ruiz-Femen\'\i a$^{a}$}\\[6mm]
{\it ${}^a$Physik Department T31,\\
James-Franck-Stra\ss e, 
Technische Universit\"at M\"unchen,\\
D--85748 Garching, Germany\\
\vspace{0.3cm}
${}^b$CNRS, Aix Marseille U., U. de Toulon, CPT, UMR 7332,\\
 F-13288, Marseille, France\\
\vspace{0.3cm}
${}^c$National Centre for Nuclear Research,\\
Ho\.za 69, 00-681, Warsaw, Poland\\
\vspace{0.3cm}
${}^d$Department of Physics, University of Oslo, Box 1048,\\ 
NO-0371 Oslo, Norway}
\\[0.3cm]
\end{center}

\vspace{0.7cm}
\begin{abstract}
\vskip0.2cm\noindent
The relic density of TeV-scale wino-like neutralino dark matter in the MSSM 
is subject to potentially large corrections as a result of the Sommerfeld 
effect. A recently developed framework enables us to calculate the 
Sommerfeld-enhanced relic density in general MSSM scenarios, properly 
treating mixed states and multiple co-annihilating channels as well as 
including off-diagonal contributions. Using this framework, including 
on-shell one-loop mass splittings and running couplings and taking into 
account the latest experimental constraints, we perform a thorough study of 
the regions of parameter space surrounding the well known pure-wino scenario: 
namely the effect of sfermion masses being non-decoupled and of allowing 
non-negligible Higgsino or bino components in the lightest neutralino. We  
further perform an investigation into the effect of thermal corrections 
and show that these can safely be neglected. The results reveal a number of 
phenomenologically interesting but so far unexplored regions where the 
Sommerfeld effect is sizeable. We find, in particular, that the relic density 
can agree with experiment for dominantly wino neutralino dark matter 
with masses ranging from 1.7 to beyond 
4 TeV. In light of these results the bounds from Indirect Detection on 
wino-like dark matter should be revisited.
\end{abstract}
\end{titlepage}



\section{Introduction}
\label{sec:introduction}

The so-called ``WIMP miracle'' is the observation that a thermally produced, 
stable, massive particle $\chi$ with electroweak interactions (WIMP) 
naturally accounts for the observed dark matter relic density, if its mass 
is of order of the electroweak or TeV scale. Indeed, on adding a fermionic 
SU(2) triplet to the Standard Model (SM), its tree-level pair 
annihilation into electroweak gauge bosons yields $\Omega_\text{cdm} h^2 =
0.1188$ for $m_\chi \approx 2.2\,$TeV. 
Such models provide attractive dark matter (DM) candidates due to 
their minimal particle content \cite{Cirelli:2005uq}, but further model 
building is required to explain why the mass of the $\chi$ particle should 
be close to the electroweak scale. The minimal supersymmetric 
standard model (MSSM) is a prime example of a model where the DM 
particle mass is tied to the electroweak scale by the desire to temper 
the quantum corrections to the Higgs. The underlying symmetry 
principle then leads to a proliferation of particles and interactions, 
allowing for different successful DM candidates. Within this context
thermal scenarios with DM masses below 1~TeV require additional mechanisms
or accidental degeneracies, e.g.\ resonant annihilation or co-annihilation,  
in order to avoid overproduction in the early Universe. These are also 
becoming somewhat constrained by LHC and dark matter searches, for recent 
analyses see Refs.~\cite{Fowlie:2013oua,Cahill-Rowley:2014boa,Bagnaschi:2015eha,deVries:2015hva}. 
Scenarios with heavier dark matter interpolate to 
minimal models, since the supersymmetric particles form approximate 
electroweak multiplets (except for degeneracies). In particular, when the 
lightest supersymmetric particle (``wino'') is the partner of the electroweak 
gauge bosons, the model is similar to the minimal triplet model, but 
modifications arise due to the mixing with the Higgsino and bino states, 
as well as the interactions with sfermions. It is this ``wino-like'' region 
of the MSSM parameter space, which we focus on in this paper. 

The wino-like region deserves special attention, since the DM relic 
density cannot be calculated reliably from the tree-level annihilation 
cross section. Loop effects from electroweak gauge boson exchange 
are large in non-relativistic scattering before the annihilation of 
TeV-scale dark matter, and lead to the electroweak Sommerfeld 
effect~\cite{Hisano:2004ds, Hisano:2006nn}, which is particularly 
strong in the wino-like region. This has been studied extensively 
in the pure-wino limit~\cite{Hisano:2004ds,Hisano:2006nn,Cirelli:2007xd,Hryczuk:2010zi,Hryczuk:2011vi}, which corresponds 
to the minimal triplet model. To be specific, in the analysis below 
we find that the observed relic density is attained at significantly 
larger mass $m_\chi = 2.88~$TeV when the Sommerfeld effect is accounted 
for, instead of $2.22~$TeV at tree level, when $M_1=3 M_2$, $\mu = 2 M_2$ 
and the common sfermion mass $M_{\rm sf} = 20\,$TeV, which corresponds 
effectively to the pure-wino limit. 
The Sommerfeld effect also displays a resonance at $2.33~$TeV, where the 
relic density is reduced by a factor $3.9$ relative to the computation 
based on the tree-level cross section. 
This highlights the importance of including the Sommerfeld effect 
in full MSSM calculations of the relic density in the wino-like region. 

Away from the pure-wino limit, the lightest neutralino is a mixture 
of wino, Higgsino and bino eigenstates and interacts accordingly, which 
makes the computation of the Sommerfeld effect much more involved. This 
problem was first approached in Refs.~\cite{Hryczuk:2010zi,Hryczuk:2011tq}, 
however a framework that deals systematically with mixed states, multiple 
co-annihilating states and the corresponding off-diagonal reactions was only 
developed in Refs.~\cite{Beneke:2012tg, Hellmann:2013jxa, Beneke:2014gja}, 
which allowed the computation of the relic density including the Sommerfeld 
effect with a relative accuracy similar to state-of-the-art computations 
employing Born cross sections. This was studied in a number of models that 
interpolate from a pure-wino to a pure-Higgsino DM 
particle~\cite{Beneke:2014hja}, but a detailed investigation of the 
MSSM parameter space was left for the future.

We report on this investigation in the present work, focusing on the 
wino-like region of the full MSSM. We note that in this region the 
Sommerfeld effect is not a small correction and should be included in any 
reliable relic density computation and in particular when the 
relic density is correlated with other observational constraints. 
The most important is from indirect dark matter searches. For instance, 
the thermal pure-wino scenario is often said to be excluded (barring some 
astrophysical uncertainties, see~Ref.~\cite{Hryczuk:2014hpa}) 
by the non-observation of a photon line signal from the Galactic Centre
\cite{Cohen:2013ama, Fan:2013faa,Hryczuk:2014hpa}. Other search channels, 
especially the cosmic ray antiprotons and the diffuse gamma rays from dwarf 
spheroidal galaxies, also start to give competitive limits
\cite{Belanger:2012he,Hryczuk:2014hpa,Bhattacherjee:2014dya,Chun:2015mka}.

This conclusion need not hold in the full MSSM, when the mixed nature of 
wino-like dark matter is taken into account. The framework adopted here 
follows Refs.~\cite{Beneke:2012tg, Hellmann:2013jxa, Beneke:2014gja}, 
with several improvements applied relative to Ref.~\cite{Beneke:2014hja}. 
We now include the running of the electroweak couplings from the 
electroweak to the dark matter scale, and use the exact one-loop 
neutralino and chargino on-shell masses to compute the mass splitting, which 
is important in the resonance region. We justify neglecting thermal 
effects due to the fact that the freeze-out happens at temperatures 
close to the electroweak scale. On the practical side, a considerable 
speed-up of the numerical evaluation has been achieved, which now 
allows a systematic investigation of the relevant MSSM parameter space 
in the wino-like region. 

The outline of the paper is as follows: In Section~\ref{sec:setup} we 
define the ranges of the parameters of the phenomenological 
MSSM, and discuss the theoretical and observational constraints we apply to 
select viable models. We further briefly summarize the computation of 
the Sommerfeld correction, the implementation of mass splittings and 
the running of the electroweak coupling. The set-up is rather general, 
but the present version does not include sfermion-neutralino/chargino 
potentials and hence excludes models with sfermion co-annihilation, as 
well as s-channel resonant annihilation, 
in which case the annihilation process 
is not short-distance. Section~\ref{sec:analysis} contains our main 
results. Here we show and discuss, in order, the dependence of the 
relic density and the relative importance of the Sommerfeld effect on 
the sfermion masses $M_{\rm sf}$ (all assumed degenerate for simplicity), 
on the heavy MSSM Higgs bosons, further on the Higgsino 
admixture via the difference $\mu-M_2$ of the Higgsino and wino mass 
parameters of the MSSM, and similarly on the bino admixture. We shall 
see that away from the pure-wino limit, the observed relic density 
is obtained for a wide range of wino-like dark matter particle masses and 
we quantify and explain the parameter dependence. We further 
study the dependence on other MSSM parameters, which generally turns 
out to be minor, except in the vicinity of the Sommerfeld resonance.
We summarize in Section~\ref{sec:summary}.

The investigation of thermal effects is contained in 
Appendix~\ref{app:thermal}. We consider the temperature dependence 
of the electroweak gauge boson masses, which in turn affects the range of 
the electroweak Yukawa potential, and of the neutralino--chargino 
mass difference. The dependence arises from the temperature-dependent 
Higgs vacuum expectation value and the one-loop self-energies. Despite the 
fact that freeze-out may begin in the symmetric phase of the electroweak 
interactions, where the Higgs field has no expectation value and the thermal 
effects are large, we find that the impact on the relic density is 
negligible within other uncertainties. We explain why previous work 
\cite{Cirelli:2007xd, Hryczuk:2010zi} overemphasised the effect. 


\section{MSSM parameters, constraints and 
implementation of the Sommerfeld effect}
\label{sec:setup}

\subsection{MSSM definition and parameter ranges}
\label{sec:params}

We are interested in exploring the parameter space of the CP-conserving, 
minimal flavour violating MSSM defined at the electroweak scale. 
Within this space we focus primarily on the calculation of the relic 
abundance of neutralino dark matter in the close-to-wino region, 
which only depends strongly on a subset of parameters.
Clearly a central role is played by those parameters describing the 
chargino and neutralino sector: the bino mass $M_1$, the wino mass $M_2$ and 
the Higgsino parameter $\mu$.\footnote{Note that we adopt a convention where 
the sign of $M_2$ is positive, but vary that of $M_1$ and $\mu$.} The 
tree level mass matrix for the charginos is given by 
\begin{equation}\label{eq:X}
X=
\left( \begin{array}{cc}
M_2 & \sqrt{2} m_W s_\beta  \\
\sqrt{2} m_W c_\beta  & \mu
\end{array} \right),
\end{equation}
where $s_\beta/c_\beta\equiv\sin\beta/\cos\beta$, $\tanb$ being the ratio of 
the vevs of the two MSSM Higgs doublets, and $m_W$ is the mass of the $W$ 
boson.
The mass matrix for the neutralinos is given by
\begin{equation}\label{eq:Y}
Y =\left( \begin{array}{cccc}
M_1 & 0 & -m_Z c_\beta s_W & m_Z s_\beta s_W \\
0   & M_2 & m_Z c_\beta c_W & -m_Z s_\beta c_W \\
-m_Z c_\beta s_W & m_Z c_\beta c_W & 0 & -\mu \\
m_Z s_\beta s_W & -m_Z s_\beta c_W & -\mu & 0 \end{array} \right).
\end{equation}
where $s_W\equiv\sin\theta_W$, $c_W\equiv\cos\theta_W$ for the Weinberg 
angle $\theta_W$ and $m_Z$ is the mass of the $Z$ boson. On diagonalising 
the hermitian squares of these matrices one obtains the values of the masses 
of the charginos and neutralinos $m_{\tilde{\chi}^+_i}$ and 
$m_{\tilde{\chi}^0_j}$ respectively, numbered $i=1,2,j=1,...,4$ 
in increasing order.

We concentrate on the region where the lightest supersymmetric particle 
(LSP) mass is at the TeV scale, as this 
is where the wino-like neutralino can provide the correct thermal relic 
density and the electroweak Sommerfeld effect is non-negligible. 
Here we assume that either the bino, Higgsino or both are much heavier than 
the wino. We can study the mixing angles of the wino with the bino or the 
Higgsinos and the resulting mass eigenstates by expanding in $m_Z/\mu$ etc. 
The mixing as well as the 
mass difference between the lightest chargino and neutralino can play an 
important role in determining the size of the Sommerfeld enhancement.
In the region where the bino is decoupled, provided that $m_W\ll |\mu|-\mII$, 
the splitting $\delta 
m_{\tilde{\chi}^+_1}\equiv m_{\tilde{\chi}^{+}_1}-m_{\tilde{\chi}^{0}_1}$ 
is given by
\begin{equation}
\label{eq:Higgsino-Wino-Masses1}
\delta m_{\tilde{\chi}^{+}_1}\simeq\frac 
12\frac{ m_W^4 M_2\,(c_\beta^2- s_\beta^2)^2}{(\mu^2-\mII^2)^2}\,.
\end{equation}
If the difference $\delta\mu\equiv |\mu|-\mII$ is too small, 
the splitting is found to be
\begin{equation}
\label{eq:Higgsino-Wino-Masses2}
\delta m_{\tilde{\chi}^+_1}\simeq  \frac{m_Z^2}{8 M_2}
\left(c_W^2\,(1\mp s_{2\beta})\left( 1 
-\frac{\delta\mu}{\sqrt{2}\,(s_\beta\pm c_\beta)\,m_W}\right)
+ 2\,s_W^2\,(1\pm s_{2\beta})\frac{M_2}{ M_1}\right)\,,
\end{equation}
where the upper (lower) sign corresponds to positive (negative) $\mu$.
We have kept the leading sub-leading correction for large $|M_1|$, but 
dropped terms of order $m_Z^2\delta\mu/(M_2 |M_1|)$.

When the Higgsinos are decoupled, if $s_{2\beta}\,m_Z^2\ll 2|\mu|\,
|\delta M_1|$, where $\delta M_1\equiv M_1-\mII$, the mixing between the wino 
and bino depends on
\begin{equation}
\label{eq:binomix}
\theta_{b}=\frac{s_{2\beta}\,s_{2W}\,m_Z^2}{2\mu\,\delta M_1},
\end{equation}
where $s_{2W}\equiv\sin 2\theta_W$. 
Note that for negative $M_1$, $|\delta M_1|$ is not a small quantity, 
the mixing is suppressed and the tree-level splitting between the lightest
chargino and neutralino is negligible; we do not discuss this case further here. Then, depending on whether $s_{2\beta}\,m_Z^2 \ll 2|\mu|\,\delta M_1$
or $s_{2\beta}\,m_Z^2 \gg 2|\mu|\,\delta M_1$,
\begin{eqnarray}
\delta m_{\tilde{\chi}^+_1}&\simeq &\,\theta_b^2 \,\delta M_1
\left(1+\frac{2 M_2}{s_{2\beta}\,\mu}\right)\quad\mbox{or}
\label{eq:Bino-Wino-Masses2a}\\[0.2cm]
\delta m_{\tilde{\chi}^+_1} &\simeq& \begin{cases}
\displaystyle
s_W^2 \frac{m_Z^2}{\mu}\left(s_{2\beta}+\frac{\,M_2}{\mu}\right)
-s_W^2\delta M_1, & 
\text{if } \mu>0\,\,\mbox {or}\,\,\frac{s_{2\beta}|\mu|}{M_2} < 1
\\[0.6cm]
\displaystyle
c_W^2 \frac{m_Z^2}{|\mu|}\left(s_{2\beta}+\frac{\,M_2}{\mu}\right)
-c_W^2\delta M_1, & \text{otherwise}
\end{cases}
\label{eq:Bino-Wino-Masses2b}
\end{eqnarray}
respectively. 
In Eq.~(\ref{eq:Bino-Wino-Masses2a}), there is a clear decrease in 
$\delta m_{\tilde \chi_1^+}$ as $\delta M_1$ increases. Apart from this, one finds that the mass splitting decreases (when $\mu$ is positive) 
as $\tan\beta$ increases, or as $\mu$ increases. Also, under the 
assumptions where Eq.~(\ref{eq:Bino-Wino-Masses2a}) and the first of 
Eq.~(\ref{eq:Bino-Wino-Masses2b}) hold, for the same value of $|\mu|$ 
the mass splitting is always smaller for $\mu<0$ than for $\mu>0$.
The only remaining gaugino is the gluino, the mass of which is 
determined by the parameter $M_3$. The value of this parameter does not 
have much effect on our results, provided it is sufficiently heavy.

The sfermions can play a non-negligible role in the annihilation. The 
sfermion mass matrix is given by 
\begin{equation}
M_{\tilde{f}}=
\left( \begin{array}{cc}
M_{\tilde{f}_L}^2+m_f^2+\widetilde{m}_{Z}^2 (I^f_3-Q_f s_W^2) & 
m_f X^{\ast}_f  \\[.5em]
m_f X_f  & M_{\tilde{f}_R}^2+m_f^2+ \widetilde{m}_{Z}^2\,Q_f s_W^2
\end{array} \right),
\label{sfermion}
\end{equation}
for right- and left-handed sfermion mass parameters 
$M_{\tilde{f}_L}, M_{\tilde{f}_R}$. We make use of the abbreviation 
$\widetilde{m}_{Z}^2\equiv m_Z^2\cos{2\beta}$, and $X_f$ is defined in terms 
of the trilinear coupling $A_f$ via
\begin{equation}
 X_f\equiv A_f-\mu^{\ast} \left\{\cot\beta,\tan\beta\right\},
\end{equation}
where $\cot\beta$ applies for the up-type squarks, $f=u,c,t$, and
$\tan\beta$ applies for the down-type sfermions, $f=d,s,b,e,\mu,\tau$
(we treat the neutrinos as being massless). Note that $m_f$, $Q_f$ and 
$I_3^f$ are the mass, charge and isospin projection of the fermion $f$ 
respectively. We significantly simplify the sfermion sector by adopting a 
common mass parameter $M_{\rm sf}\equiv M_{\tilde{f}_L}=M_{\tilde{f}_R}$.
This simplification is justified in that it does not introduce any 
non-trivial modification of the DM properties, {\it i.e.}~at the TeV scale 
close to the wino limit the dominant contribution to the annihilation cross 
section involves gauge interactions and gauge universality implies equal 
contributions from all flavours. Therefore, the effect of introducing more 
freedom in the sfermion masses can be estimated by a simple rescaling of the 
effect. The sfermion mass parameter is taken to be always larger by at 
least 25\% than $\mII$, in order not to have sfermion co-annihilation 
processes.\footnote{We leave the in-depth analysis of Higgsino or bino LSP 
and the sfermion co-annihilations including our detailed treatment of the 
Sommerfeld effect for future work. For some previous results in these cases 
see e.g. Refs.~\cite{Hryczuk:2010zi,Hryczuk:2011tq,Beneke:2014hja}.} 

The MSSM Higgs sector consists of two doublets $\Phi_1$ and $\Phi_2$, which 
after electroweak symmetry breaking leads to three neutral particles, the 
light neutral and the heavy CP-odd and CP-even Higgs bosons, as well as a heavy charged Higgs boson. Their masses and mixing can be completely described in terms of two parameters, $\tanb$  introduced earlier and the mass $\mao$ of the CP-odd neutral Higgs boson $A^0$, which 
defines the mass scale of the heavy doublet $\Phi_2$. The masses of the remaining particles and the mixing between them can be deduced from these two parameters, on requiring the minimisation of the Higgs potential. Any CP phases that  could arise in the Higgs sector may be rotated away, and the Higgs sector is CP-conserving at tree level.

The ranges of the MSSM parameters are assumed to be as wide as possible 
within the experimentally and theoretically allowed windows and/or until the 
decoupling limit is reached.  The necessary parameters, along with the 
corresponding allowed ranges, are collected in Tab.~\ref{tab:params}. 
Note that the lower bound of $\tan\beta$ is chosen in order to ease satisfying the constraints on the Higgs mass. 
The upper bound is chosen such that the phenomenology is qualitatively unchanged beyond this point. The discussion of the experimental constraints we take into account is given in the following section.
$\mao$ could also potentially influence the relic density. 
It will turn out, however, that this 
dependence is not too strong, except for the case of the mixed wino-Higgsino.
The parameters which are least relevant include the trilinear couplings and the gluino mass parameter $\mIII$.\footnote{Throughout this work we will assume that the gluinos are sufficiently heavy such that co-annihilation with the neutralino can be neglected. For a recent analysis of the relic density including gluino co-annihilation see e.g. Ref.~\cite{Harigaya:2014dwa}.} 
To summarise, the most relevant parameters for our study are the wino, bino 
and Higgsino mass parameters $\mII$, $M_1$ and $\mu$, respectively, and the 
common sfermion mass parameter $M_{\rm sf}$.
Using these inputs, the spectrum is computed using \FeynHiggs\ 2.9.5 
\cite{Heinemeyer:1998yj,Degrassi:2002fi} with a top mass $m_t = 173.2$ GeV. 
After the initial parameter card is generated and the constraints described in Sec.~\ref{sec:constraints} are checked, we include one-loop corrections to the neutralino masses (see Sec.~\ref{sec:1lmass}) and take into account the running of the gauge couplings (see Sec.~\ref{sec:run_couplings}).

\begin{table}[t] \centering
\begin{tabular}{|c|c|}
\hline
Parameter & Range \\
\hline
\hline
$\mII$ & 1 -- 5 TeV \\
$|M_1|$ & $M_2$ -- $3\,M_2$\\
$|\mu|$ & $M_2$ -- $3\,M_2$\\
$M_{\rm sf}$ & 1.25 $\mII$ -- 12 TeV \\
\hline
  $\mao$ & 0.5 -- 10 TeV \\
$\tanb$ & 5 -- 30 \\
\hline
$\lvert A_{f} \rvert$ & 0 -- 8 TeV \\
$\mIII$ & $3\,\mII$ \\
\hline
\end{tabular}
\caption{Ranges of MSSM parameters adopted for the scan, where $f$ 
represents all the SM quarks and leptons.}
\label{tab:params}
\end{table}

\subsection{Constraints}
\label{sec:constraints}

We require a number of theoretical and experimental constraints to be 
satisfied by all the points in the scan. In this section we discuss the 
implementation of these constraints and comment on their relevance on 
limiting the ranges of the MSSM parameters, in particular in the wino-like 
region. All the involved quantities are computed with the use of \FeynHiggs\ 
2.9.5 and \micrOMEGAs~\cite{Belanger:2010gh,Belanger:2013oya}.

\subsubsection{Collider and flavour constraints}

Many current collider and flavour constraints do not limit the region of the 
parameter space where the LSP, in our case the lightest neutralino, is at the 
TeV scale, nevertheless we include all possibly relevant constraints for 
completeness. 

\paragraph{Higgs mass} We require that the light Higgs mass $m_{h^0}$ lies 
within 4\% deviation of the measured central value 
$125.09 \pm 0.21 \pm 0.11$~GeV, from the combination of ATLAS and CMS 
data~\cite{Aad:2015zhl}.  In our numerical analysis we adopt the two-loop 
result for the Higgs mass which we calculate using \texttt{FeynHiggs} 2.9.5. 
Beyond $M_{\rm sf}=6-7$~TeV, the allowed window is slightly ($1-2$~GeV) below 
the estimation of the theoretical uncertainties in the Higgs mass 
determination in the MSSM~\cite{Hahn:2013ria}, given that we do not include 
the resummation of logarithmic corrections arising due to the large hierarchy 
between the top and the stop masses. The approximate formula for the Higgs 
mass at one-loop level reads \cite{Carena:1995bx,Haber:1996fp}:
\begin{equation}
 m^2_{h^0}\simeq m_Z^2\cos^2 2\beta+\frac{3}{4\pi^2}\frac{m_t^4}{v^2}
 \left[\log\frac{M_{\rm sf}^2}{m_t^2} +\frac{X_t^2}{M_{\rm sf}^2} 
 \left(1-\frac{X_t^2}{12\,M_{\rm sf}^2}\right)\right].
\end{equation}
From this expression we deduce that the main implication of the Higgs mass constraint is to impose that either the stop masses are a few TeV, or the stop mixing is large. The first condition is often satisfied in the scenarios we consider, and when not the mixing can easily be chosen such that this constraint is satisfied. Note that as $X_t$ does not play a significant role in the relic density computation in the wino-like region, and that for $\tan\beta\gtrsim 5$, $\cos 2\beta$ is close to $-1$, the Higgs mass constraint does not have much impact on our results. For $M_{\rm sf}>6-7$ TeV the effect of the neglected corrections to the Higgs mass could therefore be compensated by a change in $X_t$, leaving the relic density unaltered. 
 
\paragraph{$\boldsymbol \rho$ parameter} We require that the value of 
$\Delta_\rho$ computed in the MSSM~\cite{Heinemeyer:2004gx} does not exceed 
two standard deviations from the SM expectation~\cite{Agashe:2014kda}: 
\be
\rho_0 = 1.0004 \pm 0.00024, \quad \rm{therefore} \quad  
\Delta_\rho < 0.00048.
\ee
Since the SUSY contribution can only be large when the mass splitting in the 
sfermion $SU(2)$ doublets is large, and in the scenario we consider all the 
sfermion doublets are nearly degenerate, it does not have a significant 
effect on our parameter space.

\paragraph{$\boldsymbol {b\rightarrow s \gamma}$} In general MSSM scenarios this branching ratio provides a strong constraint, as the contribution from broken SUSY is generically large,
while the SM prediction is compatible with measurement. The experimental~\cite{Stone:2012yr} and SM theory~\cite{Misiak:2006ab} 
 values, with the corresponding uncertainties, we use are 
 \bea
 \btosgamma{\textrm{exp}}\ & =\ & \lp 3.37 \pm 0.23 \rp\times 10^{-4}, \nonumber \\
 \btosgamma{\textrm{SM}}\ & =\ & \lp 3.15 \pm 0.23 \rp\times 10^{-4}.\nonumber
 \eea
 The SUSY contribution $\Delta \btosgamma{}$ is computed with \FeynHiggs\ 
 and the implemented criterion reads
 \be
 \lc \lp \btosgammabest{\textrm{SM}} + \Delta \btosgamma{} \rp - \btosgammabest{\textrm{exp}} \rc^2 < \lp 3\sigma^\textrm{exp} \rp^2 + \lp \sigma^\textrm{SM} \rp^2.
 \ee
 
There are three classes of diagrams which contribute to $b\to s\gamma$ in the 
MSSM: these are diagrams involving either charged Higgs bosons, charginos or 
gluinos. The first always interfere constructively with the SM contribution, 
and decouple as the Higgs mass increases beyond the TeV scale. The chargino 
contribution can take either sign, depending on the sign of $\mu$ and $A_t$, 
but also decouples with increasing $|\mu|$ and $M_2$. At the scales that are 
relevant to this study, {\it i.e.}~above 1 TeV, in general the MSSM 
contribution lies within the uncertainties.
 
\paragraph{$\boldsymbol{B_s\to\mu^+\mu^-}$}
The correction to $B_s \to\mu^+\mu^-$ from SUSY should also lie within the 
errors from the experimental measurement and the SM calculation. To this end, 
we check whether the result of the calculation in the 
MSSM~\cite{Heinemeyer:2004gx} is consistent with the combined CMS and LHCb 
result, $(2.9\pm 0.7)\times 10^{-9}$~\cite{CMSandLHCbCollaborations:2013pla}. 
The 3 sigma error on the experimental result is added to the uncertainty on the theoretical result in quadrature, where the updated SM prediction is $(3.56\pm 0.30)\times 10^{-9}$, using latest values on the $B^0_s$ lifetime and relative $B^0_s$ decay width difference~\cite{Buras:2012ru,CMSandLHCbCollaborations:2013pla}.
We note that as we consider the wino-like region with masses of the LSP of $\mathcal{O}(\mathrm{TeV})$, and masses of the heavy Higgs bosons also of $\mathcal{O}(\mathrm{TeV})$, this constraint does not have much influence on our parameter space. Another related constraint is of course the branching ratio of $B\to\tau\nu$, measured precisely at the B-factories~\cite{Lees:2012ju,Adachi:2012mm}. However, we do not consider this constraint as the parameter space of interest in our analysis, in particular the large masses of the charged Higgs bosons and values of $\tan\beta$, do not result in MSSM contributions beyond the combined experimental and theoretical uncertainty~\cite{Altmannshofer:2012ks}.

\paragraph{$\boldsymbol{g_\mu-2}$}  The experimental and SM theory values adopted for the muon anomalous magnetic moment, $a_\mu = \frac{g_\mu - 2}{2}$, are given by~\cite{Agashe:2014kda},
\bea
a_\mu^\textrm{exp}\ & =\ & \lp 1165920.91 \pm 0.63 \rp \times 10^{-9}, 
\nonumber \\
a_\mu^\textrm{SM}\ & =\ & \lp 1165918.03 \pm 0.48 \rp \times 10^{-9}, 
\nonumber
\eea
and we require that $\Delta a_\mu$, the MSSM contribution, satisfies
\be
-\sigma^\textrm{exp-SM} < \Delta a_\mu < \lp a_\mu^\textrm{exp} - 
a_\mu^\textrm{SM} \rp + \sigma^\textrm{exp-SM},
\ee
where for the error on the difference between experimental and SM values 
we take
\be
\sigma^\textrm{exp-SM} = \sqrt{(3\sigma^\textrm{exp})^2+
(\sigma^\textrm{SM})^2}.
\ee
This means that we do not insist that the MSSM contribution explains the 
deviation between the experimental and SM theory values. Note that as the 
SUSY contribution is proportional to $\tanb$ and inversely proportional to 
the square of the masses of the sparticles, it is typically strongly 
suppressed in the region of interest where $M_{\rm sf}$ lies at the TeV scale. 

\subsubsection{Theoretical constraints}

\paragraph{Higgs potential} 
Theoretical consistency demands that the scalar potential is free from charge 
and/or colour breaking minima (CCB). For the tree-level scalar potential in 
the MSSM, the corresponding criteria read~\cite{Frere:1983ag,Claudson:1983et}
\bea
&&A^2_t < 3\lp \cos^2\!\beta\ \mao^2 + \frac{m_Z^2}{2}\cos 2\beta + 
2 M_{\rm sf}^2\rp,  \nonumber\\[0.1cm]
&&A^2_{b (\tau)} < 3\lp \sin^2\!\beta\ \mao^2 - \frac{m_Z^2}{2}\cos 2\beta + 
2 M_{\rm sf}^2 \rp,
\eea
One can always choose the trilinear couplings low enough such that the CCB 
constraint is satisfied without altering the nature of the neutralino. 
 
\paragraph{s-channel resonances} Our calculation relies on the factorisation of the annihilation cross section into the short-range tree-level annihilation and the long-range potential interaction. However, this factorization does not hold in the case that the final light particles are produced through an s-channel propagator which is resonant, as such a contribution cannot be attributed to the short-distance part of the annihilation. Therefore, we need to exclude regions of parameter space where this may occur. In the MSSM this means that we need to avoid s-channel resonances through the Higgs bosons, and to be conservative we assume that the masses of the heavy Higgses lie outside the interval 
 \be \label{eq: noresonance_condition}
 m_{H^0,A^0,H^+} \notin \lc 1.7\ m_{\neuI} ,\ 2.3\ m_{\neuI}\rc.
 \ee
It follows that in this work we are not in a position to study the $H$- and $A$-funnel regions \cite{Drees:1992am,Nath:1992ty}.

\subsubsection{Cosmological and Direct DM Detection constraints}

In choosing suitable points to calculate the Sommerfeld effect on the relic density, we insist that certain basic constraints are fulfilled. First we require that the lightest neutralino $\neuI$ is the LSP. We further insist on compatibility with Direct Detection bounds. The details of how these conditions are imposed is described in this subsection. We choose not to include any limits coming from Indirect Detection experiments or measurements of the CMB, as although these may be relevant they are subject to large systematic uncertainties and their discussion goes beyond the scope of this work; we plan to address such constraints in the future. 
 
\paragraph{Direct Detection} We require that the DM-nucleon spin-independent cross 
section $\sigma^\textrm{SI}$ is less than twice the LUX limit \cite{luxSI}. 
The theoretical prediction of this cross section within the MSSM is obtained 
using \micrOMEGAs . 
The spin-independent cross section is sensitive to the Higgs exchange between the LSP and the quarks of the nucleon. The interaction with the Higgs relies on the LSP containing both gaugino and Higgsino components, and therefore this constraint is most relevant for the scenarios we study where $|\mu| \sim M_2$. 
Note that the limits of the spin-dependent cross section coming from Direct Detection experiments and neutrino signals from the Sun are always much weaker than those coming from spin-independent results for the scenarios we are interested in here.

\subsection{One-loop mass splittings}
\label{sec:1lmass}

The differences in mass between the LSP and the heavier neutralinos and 
charginos can have an effect on the relic density. The most 
relevant case is the small mass difference between the lightest chargino 
and neutralino state, $\tilde \chi_1^+$ and $\tilde \chi_1^0$, 
respectively. In order to be consistent with the accuracy of the rest of the 
calculation we calculate these masses at one-loop. In doing so we adopt an 
on-shell renormalisation scheme,  which is described here in brief. For 
further details we refer the reader to Refs.~\cite{Fritzsche:2002bi,Fowler:2009ay,Fowler:2010eba,Bharucha:2012nx,Bharucha:2012re}.

The mass matrix in the chargino sector is renormalised via $X \to X+\delta X$,
where $\delta X$ is defined by
\begin{equation}\label{eq:delX}
\delta X=
\left( \begin{array}{cc}
\delta M_2 & \sqrt{2} \delta (m_W s_\beta)  \\
\sqrt{2} \delta (m_W c_\beta)  & \delta \mu
\end{array} \right),
\end{equation}
containing the renormalisation constants (RCs) for the wino parameter $M_2$ and Higgsino parameter $\mu$, {\it i.e.}~$\delta M_2$ and $\delta \mu$. In addition, the matrix $\delta X$ contains the RCs of $c_\beta$ and $s_\beta$, {\it i.e.}~$\delta c_\beta$ and $\delta s_\beta$ (which can be expressed in terms of $\delta \tan\beta$), and of the $W$ boson mass $m_W$, $\delta m_W$. Definitions of and expressions for $\delta \tan\beta$ and $\delta m_W$ can be found in Ref.~\cite{Bharucha:2012re}.
The neutralino mass matrix, $Y$, is renormalised in a similar manner via $Y \to Y+\delta Y$,
where $\delta Y$ is defined  in analogy to $\delta X$ in Eq.~\eqref{eq:delX}
and further contains RC of the bino parameter $M_1$, $\delta M_1$.
In the on-shell scheme, we must fix the RCs $\delta M_1$, $\delta M_2$
and $\delta \mu$ (as in e.g.~Ref.~\cite{Fowler:2009ay}) by requiring that three out of the total six
physical masses of the charginos and neutralinos satisfy on-shell conditions, {\it i.e.}~that the tree-level masses, $m_{\tilde{\chi}_i}$, coincide with the
one-loop renormalised masses, $M_{\tilde{\chi}_i}=m_{\tilde{\chi}_i}+\Delta m_{\tilde{\chi}_i}$,
\begin{align}
\label{eqn:deltami}\Delta
m_{\tilde{\chi}_i}&\equiv-\frac{m_{\tilde{\chi_i}}}{2}\mathrm{Re}\left[\hat{\Sigma}
^L_{ii}(m_{\tilde{\chi}_i}^2)+\hat{\Sigma}^{R}_{ii}(m_{\tilde{\chi}_i}^2)\right]-\frac
{1}{2}\mathrm{Re}\left[\hat{\Sigma}^{SL}_{ii}(m_{\tilde{\chi}_i}^2)+\hat{\Sigma}^{SR}
_{ii}(m_{\tilde{\chi}_i}^2)\right]\quad=\,0.
\end{align}

Note that we define the coefficients 
$\hat\Sigma^{L/R}_{ij}(p^2)$ and $\hat\Sigma^{SL/SR}_{ij}(p^2)$ of the self energy via
\begin{equation}
 \hat\Sigma_{ij}(p^2)=\displaystyle{\not}p\, P_L 
\hat\Sigma^L_{ij}(p^2)+\displaystyle{\not}p\, P_R  
\hat\Sigma^R_{ij}(p^2)
+P_L  \hat\Sigma^{SL}_{ij}(p^2)+ P_R \hat\Sigma^{SR}_{ij}(p^2).
\label{eqn:Lorentzse}
\end{equation}
The left- and right-handed vector and scalar coefficients,
$\displaystyle\hat{\Sigma}^{L/R}_{ij}(p^2)$ and
$\displaystyle\hat{\Sigma}^{SL/SR}_{ij}(p^2)$ of the renormalised self-energy 
are defined analogously. Expressions for the renormalised self-energies can 
be found in e.g.~Ref.~\cite{Bharucha:2012re}.
The mass shifts for the remaining three chargino and neutralino masses
are therefore given by $\Delta m_{\tilde{\chi}^\pm_i}$ and 
$\Delta m_{\tilde{\chi}^0_j}$ in Eq.~\eqref{eqn:deltami}. For the calculation 
of these mass shifts we used the program 
\texttt{FeynArts}~\cite{Hahn:2000kx,Hahn:2001rv}, together with the packages
\texttt{FormCalc}~\cite{Hahn:1998yk} and 
\texttt{LoopTools}~\cite{Hahn:1998yk}, 
using the model files presented in Ref.~\cite{Fritzsche:2013fta}.

\renewcommand*{\arraystretch}{1.4}
\begin{table}
\begin{center}
\begin{tabular}{|c|c|}
\hline
Scenario & Particles on shell\\
\hline
\hline
 \,$M_2<|M_1|<|\mu|$ \,&\, $\tilde\chi_1^+$, $\tilde\chi_2^0$, $\tilde\chi_3^0$ \,\\
 \,$M_2<|\mu|<|M_1|$ \,&\, $\tilde\chi_1^+$, $\tilde\chi_2^0$, $\tilde\chi_4^0$â \,\\
\hline
\end{tabular}
\caption{Choice of particles whose masses are required to be on shell for the various scenarios corresponding to the possible orderings of $M_1$, $M_2$ and $\mu$ that we consider.\label{tab:massren}}
\end{center}
\end{table}

The choice of which masses should be chosen on shell is non-trivial, as 
certain choices can lead to unphysical divergences when e.g.~$|M_1|=M_2$ or 
$|\mu|=M_2$, and we follow the prescription discussed in 
Refs.~\cite{Bharucha:2012nx,Chatterjee:2011wc} as follows to avoid this 
situation as far as possible. We therefore employ the NNC scheme, that is, 
two neutralinos and one chargino are chosen 
on-shell, of which the chargino should be wino-like, and the neutralinos should be bino and Higgsino-like. Note however that there is an ambiguity here given that there are two Higgsino-like 
neutralinos. In this work we are particularly interested in the region where 
the neutralino has a large wino component, {\it i.e.}~$M_2<|M_1|,|\mu|$, 
and may in addition contain a sizeable bino or Higgsino component. We therefore find that in order to obtain results free from scheme-dependent 
divergences, the choice of particles whose masses are required to be on shell should be made as in Tab.~\ref{tab:massren}.
This corresponds to the Higgsino closer in mass to the wino being on shell.
Note that when all three parameters are very close ($<0.1\%$ splittings) the situation may arise that the ordering of the neutralinos changes, and one should exercise caution in these regions. 
This has been accounted for in the code.


\subsection{Running couplings}
\label{sec:run_couplings}

Due to the multi-scale nature of the considered problem, the running
of the coupling constants has to be treated consistently. 
In different parts of the calculation the couplings should be taken 
at a different energy scale $Q$, in particular $Q = m_Z$ for the potential 
interactions, $Q = m_\textrm{LSP}$ for the mass splittings in the 
neutralino/chargino sector and $Q = 2\, m_\textrm{LSP}$ for the short-range 
annihilations.
 
We perform the running in the unbroken $SU(2)_L \times U(1)_Y$ theory, 
since most of the running occurs above the electroweak scale. The starting 
values of the $SU(2)_L$ and $U(1)$ couplings at $Q = m_Z$ are taken as 
$\alpha_2(m_Z) = 0.034723$ and $\alpha_1(m_Z)=0.009986$, respectively. Since 
the short-range annihilation is evaluated at tree-level, 
we run the couplings to $Q = 2\, m_{\rm LSP}$ with the one-loop
renormalisation group equation. In the computation of the one-loop 
mass splittings discussed above, the couplings are evaluated at 
$Q=m_{\rm LSP}$.
The energy range from $m_Z$ to $2\mlsp$ that we are interested in 
can be divided into five regions\footnote{
We neglect the sfermions' contribution to the beta functions, when the 
sfermion mass lies between $\mlsp$ and $2\mlsp$.
The error introduced in this way is small: for a 2.5 TeV LSP
the total running of $\alpha_2$ from $m_Z$ up to $2\mlsp$ is around 5-6$\%$,
and the maximum contribution from sfermions (when they are all decoupled at
their smallest allowed mass 1.25 $\mlsp$) is only $0.4\%$.}
where the beta functions are constant, delimited by the scales  
$\mao,\ |M_1|,\ M_2,\ |\mu|,$ at which we decouple respectively the 
heavy Higgs doublet $\Phi_2$, the bino, the wino, 
and the Higgsinos.
 
At the required level of accuracy there are no threshold effects to be 
considered and the leading order beta function $\beta_{0,i}$ at a scale 
$Q$ is given by
\be
 \beta_{0,i} = 
 \frac{11}{3}\textrm{Tr}\lc T^{(i)\,2}_A \rc 
 - \frac{2}{3}\sum_f \textrm{Tr}\lc T^{(i)\, 2}_f \rc
 - \frac{1}{3}\sum_s \textrm{Tr}\lc T^{(i)\, 2}_s \rc,
 \ee
 where $T^{(i)}_R$ are the generators of the group $i$ in
 the representation $R$ and the three terms correspond respectively 
 to gauge bosons (always in the adjoint representation $A$), fermions,
 and scalars. 
 The sums extend only to particles with mass smaller than $Q$, and the
 contributions are listed in Tab.~\ref{tab: beta_functions}.

\vskip0.2cm 
\renewcommand{\arraystretch}{1.7}
 \begin{table}[h!] \centering
 \begin{tabular}{|c|c|c|c|}
 \hline
\multirow{2}{*}{ Particles }  & $\textrm{Tr}\lc T^{(i)\, 2}_A \rc$ & 
$\sum_f \textrm{Tr}\lc T^{(i)\, 2}_f \rc$ & 
$\sum_s \textrm{Tr}\lc T^{(i)\, 2}_s \rc$ \\
& $U(1)_Y\quad SU(2)_L$	& $U(1)_Y\quad SU(2)_L$ & $U(1)_Y\quad SU(2)_L$ \\  
                     
\hline
\hline
 SM & 0 \hspace{0.8cm} 2 & \!\!\!\! 10 \hspace{0.71cm} 6 & $\frac{1}{2}$ \hspace{0.7cm} $\frac{1}{2}$ \\
 \hline
 $\Phi_2$ & 0 \hspace{0.8cm} 0 &  0 \hspace{0.8cm} 0 & $\frac{1}{2}$ \hspace{0.7cm} $\frac{1}{2}$ \\
 \hline
 $\tilde B$ & 0 \hspace{0.8cm} 0 &  0 \hspace{0.8cm} 0 & 0 \hspace{0.8cm} 0 \\
 \hline
 $\tilde W$ & 0 \hspace{0.8cm} 0 &  0 \hspace{0.8cm} 2 & 0 \hspace{0.8cm} 0 \\
 \hline
 $\tilde H$ & 0 \hspace{0.8cm} 0 &  1 \hspace{0.8cm} 1 & 0 \hspace{0.8cm} 0 \\
 \hline
 \end{tabular}
 \caption{Contributions to the $U(1)_Y$ and $SU(2)_L$ beta functions.}
 \label{tab: beta_functions}
 \end{table}
\renewcommand{\arraystretch}{1.25}


\subsection{Annihilation matrix implementation}

The rate at which neutralinos and charginos annihilate into the (light) 
standard model particles in the early Universe is a necessary input for the 
calculation of the present-day amount of dark matter. For a given 
two-particle state $\tilde\chi_i\tilde\chi_j\equiv[\tilde\chi \tilde\chi]_a$ formed out of two 
neutralino or chargino species, the annihilation 
rate including long-distance Sommerfeld 
corrections can be parametrised as~\cite{Beneke:2014gja}
\begin{eqnarray}
\sigma^{[\tilde\chi \tilde\chi]_a \to \,{\rm light}} \, v_\text{rel} 
& =& \, S_a [\hat f_h(^{1}S_0)] 
     \; \hat  f_{aa}(^{1}S_0)
 + \, S_{a}[\hat f_h(^{3}S_1)] 
     \; 3 \,\hat  f_{aa}(^{3}S_1)
\nonumber\\
&&  \hspace*{-2cm} + \, \frac{\vec{p}_{a}^{\,2}}{M_{a}^2} \, 
     \Big( \, S_{a} [\hat g_\kappa(^{1}S_0)]  \; \hat  g_{aa}(^{1}S_0)
          +  S_{a}[\hat g_\kappa(^{3}S_1)] \; 3 \, \hat  g_{aa}(^{3}S_1)
\nonumber\\
&&  \hspace*{-0.5cm}
         + \,S_{a} \Big[\frac{\hat f(^{1}P_1)}{M^2}\Big] \; \hat  f_{aa}(^{1}P_1)
         + S_{a} \Big[\frac{\hat f({}^3P_{\cal J})}{M^2} \Big] \; \hat f_{aa}(^{3}P_{\cal J})
     \Big)
\ ,\qquad
\label{eq:SFenhancedsigma}
\end{eqnarray}
up to higher orders in $\vec{p}_{a}^{\,\,2}=2\mu_{ij}(\sqrt{s}-M_{a})+\dots$, 
the relative momentum of the annihilating 
particles in their centre-of-mass frame,
with $M_{a}$, $\mu_{a}$ the total and reduced mass, respectively, 
of the two-particle state. 
The quantities $\hat f_{ab}(^{2S+1}L_J),\,\hat g_{ab}(^{2S+1}L_J),\,\dots$ 
are the  absorptive part of the Wilson coefficients of
local four-fermion operators which reproduce the short-distance annihilation 
of the chargino and neutralino pairs into SM and light Higgs final states
in the non-relativistic EFT 
framework~\cite{Beneke:2012tg,Hellmann:2013jxa,Beneke:2014gja}. 
They were  determined by matching the tree-level MSSM
amplitudes for the process $[\tilde\chi\tilde\chi]_a \to X_AX_B \to [\tilde\chi\tilde\chi]_b$
with SM and Higgs intermediate states $X_AX_B$ in 
Refs.~\cite{Beneke:2012tg,Hellmann:2013jxa}.\footnote{We have dropped an upper index
``$\chi\chi\to \chi\chi$'' used in Refs.~\cite{Beneke:2012tg,Hellmann:2013jxa} for 
the notation of  the Wilson coefficients.} 
The definition of the various
Wilson coefficients appearing in Eq.~(\ref{eq:SFenhancedsigma}) can be found
in Ref.~\cite{Beneke:2014gja}. The Sommerfeld factors $S_a[\dots]$ 
in Eq.~(\ref{eq:SFenhancedsigma}) account for the long-distance interactions of the 
two-particle states prior to the short-distance annihilation. Details on the computation
of these factors are given below. 
The tree-level annihilation rate with no long-distance corrections is 
readily recovered  by setting all the Sommerfeld factors 
in Eq.~(\ref{eq:SFenhancedsigma}) to one. The 
tree-level annihilation cross section thus obtained depends only on the
diagonal entry of the Wilson coefficients corresponding to channel $[\tilde\chi\tilde\chi]_a$, 
{\it i.e.} $\hat f_{aa}(^{2S+1}L_J),\,\hat g_{aa}(^{2S+1}L_J),\,\dots$. 
As shown in Eq.~(\ref{eq:SFdef}) below, the computation of the Sommerfeld factors 
also requires knowledge of the off-diagonal terms,
$\hat f_{ab}(^{2S+1}L_J),\,\hat g_{ab}(^{2S+1}L_J),\,\dots$, with $a\ne b$,
since the interference of loop diagrams where the two-particle states that undergo
short-distance annihilation are different are accounted for in the
Sommerfeld-corrected cross section.

A word on the notation for labelling the two-particle states is relevant here.
The two-particle states $\tilde\chi_{i}\tilde\chi_{j}$ formed out of charginos and 
neutralinos are denoted by a single label $a=1,\dots N_{|Q|}$, where $N_{|Q|}$ 
is the total number of states (channels) for each electric-charge sector, 
$|Q|=0,1,2$, corresponding to neutral ($\tilde\chi^0\tilde\chi^0,\,\tilde\chi^+\tilde\chi^-$), 
single-charged ($\tilde\chi^0\tilde\chi^\pm$) and double-charged ($\tilde\chi^\pm\tilde\chi^\mp$) 
sectors. If all four neutralinos and the two charginos are considered,
in the charge-0 sector the single label runs over the 14 different states 
\begin{equation}
\tilde\chi^0_1 \tilde\chi^0_1,\,\tilde\chi^0_1 \tilde\chi^0_2,\,\tilde\chi^0_1 \tilde\chi^0_3,\dots,\,\tilde\chi^0_3 \tilde\chi^0_4,\,\tilde\chi^0_4 \tilde\chi^0_4,\,
\tilde\chi^+_1\tilde\chi^-_1,\,\tilde\chi^+_1\tilde\chi^-_2,\,\tilde\chi^+_2\tilde\chi^-_1,\,\tilde\chi^+_2\tilde\chi^+_2 \;, 
\label{eq:neutralstates}
\end{equation}
whereas in the charge $\pm 1$ sectors we have 8 channels each,
\begin{equation}
\tilde\chi^0_1 \tilde\chi^\pm_1,\,\tilde\chi^0_1 \tilde\chi^\pm_2,\dots,\,\tilde\chi^0_4\tilde\chi^\pm_1,\,\tilde\chi^0_4\tilde\chi^\pm_2 \;, 
\label{eq:singlychargedstates}
\end{equation}
and just three each in the charge $\pm 2$ sectors,
\begin{equation}
 \tilde\chi^\pm_1\tilde\chi^\pm_1,\,\tilde\chi^\pm_1\tilde\chi^\pm_2,\,\tilde\chi^\pm_2\tilde\chi^\pm_2 \;. 
\label{eq:doublychargedstates}
\end{equation}
The coefficients $\hat f_{ab}(^{2S+1}L_J)$ for each partial wave can then be considered as the entries of a matrix whose
dimension is equal to the number of channels in each sector. Since the coefficients $\hat f_{ab}(^{2S+1}L_J)$ have the property 
$\hat f_{ba}(^{2S+1}L_J)=[ \hat f_{ab}(^{2S+1}L_J) ]^*$, such annihilation matrices turn out to be hermitian.
The computation of each of the  annihilation matrices 
appearing in the annihilation cross section
formula~(\ref{eq:SFenhancedsigma}), requires 
the evaluation of $105$, $2\times 36$ and $2\times 6$ distinct entries for neutral, single- and double-charged sectors, respectively.
Ten of such matrices are needed for a complete calculation of the Sommerfeld-corrected annihilation
cross section including ${\cal O}(v^2)$ corrections (see Ref.~\cite{Beneke:2014gja} for details on this),
making up a total number of 1890 independent entries.
In the CP-conserving case, the annihilation cross sections of the charged-conjugated sectors, $\tilde\chi^0\tilde\chi^+$ and $\tilde\chi^0\tilde\chi^-$, 
$\tilde\chi^+\tilde\chi^+$ and $\tilde\chi^-\tilde\chi^-$, become equal, and the number of independent annihilation matrix entries is reduced to 1470.

A code to obtain the analytic results for the entries of the annihilation 
matrices at ${\cal O}(\alpha_2^2)$ in the MSSM has been developed following 
the conventions and recipes 
of Refs.~\cite{Beneke:2012tg,Hellmann:2013jxa,HellmannPhDthesis}.
The expressions account for the sum of all possible $X_AX_B$ exclusive states 
with $X_{A/B}$ being a SM particle (including the light Higgs) or heavy 
MSSM Higgs (the mass of the state $X_AX_B$ must however be smaller than 
$2m_{\rm LSP}$).\footnote{Strictly speaking, one should allow for 
$m_{X_AX_B}<M_I$ when we are dealing with the co-annihilation cross section 
of the external 2-particle state $I$, but this would require having a 
different set of annihilation matrices for each co-annihilation channel, 
which is impractical.} For the neutral, single- 
and double-charged sectors, the number of exclusive final states is 31, 16 
and 3, respectively, including the possible heavy Higgs final states; 
a complete list can be found in Appendix~A of Ref.~\cite{Beneke:2012tg}. 
Despite coming from the product of tree-level amplitudes,
the analytic expressions for the Wilson coefficients are very large, which is traced back to the fact that there
are several diagrams with different topologies and/or virtual intermediate particles
contributing to a given exclusive state, and because of the non-relativistic expansion 
performed. Recall as well that we keep the general dependence on all MSSM parameters in the coefficients. The numerical evaluation of all matrix entries for a given MSSM parameter set 
is done using pre-compiled functions within {\sc Mathematica}, 
taking on average approximately 300 sec of CPU time. If only the 
annihilation matrices necessary for the leading-order cross section,
$\hat f_{ab}(^{1}S_0)$ and $\hat f_{ab}(^{3}S_1)$, are evaluated, the cost 
in CPU time reduces to less than 40~sec per model.

We should mention here a modification of a part of the analytic expressions
for the Wilson coefficients given in Refs.~\cite{Beneke:2012tg,Hellmann:2013jxa} that we have 
implemented in the present code. The Wilson coefficients obtained
in Refs.~\cite{Beneke:2012tg,Hellmann:2013jxa} describe $[\tilde\chi \tilde\chi]_a \to [\tilde\chi \tilde\chi]_b$
annihilation amplitudes expanded in powers of  $\sqrt{s}-M$, where $M\equiv (M_a+M_b)/2$ is the
average of the masses of the two-particle states taking place in the short-distance part of the 
annihilation process (for diagonal 
reactions, $a=b$, this is just an expansion around the $[\tilde\chi\tilde\chi]_a$ threshold).
When the annihilation proceeds through s-channel boson exchange, such an 
expansion implies for the boson propagator (with generic mass $m_\phi$) that
\begin{equation}
\frac{1}{s-m_\phi^2} = \frac{1}{M^2-m_\phi^2} 
\Big( 1 - \frac{2M \, (\sqrt{s}-M) }{M^2-m_\phi^2} + \dots \Big) \,, 
\label{eq:Higgsprop}
\end{equation}
up to linear terms in $\sqrt{s} - M$. The first term on the right-hand side 
of Eq.~(\ref{eq:Higgsprop}) contributes to the leading-order Wilson 
coefficients, whereas the second goes to the $S$-wave $v^2$-suppressed ones.
For heavy Higgs exchange, the following problem may arise: Once radiative 
corrections are included, any (virtual) states $a,b$ can 
participate in the short-distance part, such that we can find a situation 
where $M$ gets very close to the Higgs mass $m_\phi\approx \mao$, producing 
arbitrarily large contributions in the right-hand side of 
Eq.~(\ref{eq:Higgsprop}). Those resonance contributions are spurious, since the 
annihilating cross section of the external state 
$[\tilde\chi\tilde\chi]_I$ in the non-relativistic regime should be expanded for energies close to the mass of that state,
{\it i.e.}~around $\sqrt{s}=M_I$, which produces terms
from s-channel contributions proportional to $1/(M_I^2-m_\phi^2)$ instead of those in Eq.~(\ref{eq:Higgsprop}).
For the relevant co-annihilation channels, the latter terms cannot become resonant in our analysis because
we have explicitly excluded Higgs masses inside the range $[1.7\,m_{\tilde\chi_1^0},2.3\,m_{\tilde\chi_1^0}]$, see Eq.~(\ref{eq: noresonance_condition}). 
Therefore, the problem of spurious resonances is absent if we have a 
set of annihilation matrices for each co-annihilation channel $I$
where the s-channel propagators have been expanded around $\sqrt{s}=M_I$. In practice, that solution is 
unfeasible, since the number of co-annihilations channels in a mixed scenario can be rather large and evaluating several annihilation matrices would increase the required CPU time beyond reasonable limits. 
We can adopt, however, another solution that avoids the occurrence of spurious resonances
in s-channel propagators that only requires minimal changes in the Wilson coefficients 
obtained in Refs.~\cite{Beneke:2012tg,Hellmann:2013jxa}. It amounts to modifying the expanded s-channel
propagators from the Wilson coefficients such that they correspond to their expansion around 
$\sqrt{s}=2 m_{\tilde\chi_1^0}$, regardless of which is the external co-annihilating state. We note
that since  the relevant channels that are included in the long-distance radiative corrections
are very close in mass (see next section), the differences between the annihilation amplitudes expanded around 
$2m_{\tilde\chi_1^0}$ or around any of the other masses of the
co-annihilating states are in any case negligible, and the suggested prescription is a very good approximation. 
The necessary modifications can be immediately read off by rewriting the 
right-hand side of Eq.~(\ref{eq:Higgsprop}) using $M=2 m_{\tilde\chi_1^0} + (M-2 m_{\tilde\chi_1^0})$:
\begin{eqnarray}
&&\frac{1}{M^2-m_\phi^2} \bigg( 1 - \frac{2M \, (\sqrt{s}-M) }{M^2-m_\phi^2} \bigg) 
\nonumber\\[1mm]
&&=\frac{1}{4 m_{\tilde\chi_1^0}^2-m_\phi^2} \bigg( 1 - \frac{4m_{\tilde\chi_1^0} \, (M - 2m_{\tilde\chi_1^0}) }{4m_{\tilde\chi_1^0}^2-m_\phi^2} - \frac{4 m_{\tilde\chi_1^0}^2 \, (\sqrt{s}-M) }{4 m_{\tilde\chi_1^0}^2-m_\phi^2}  \bigg) 
\, , 
\label{eq:Higgspropmod}
\end{eqnarray}
where we have dropped terms of second order in the small quantities 
$(M - 2m_{\tilde\chi_1^0})$ and $(\sqrt{s}-M)$. We notice that the dependence 
on $M$ cancels out in the second line of Eq.~(\ref{eq:Higgspropmod}), and the
resulting expression matches the expansion of the Higgs propagator 
$1/(s-m_\phi^2)$ around $\sqrt{s}=2 m_{\tilde\chi_1^0}$. The replacements 
that have to be performed in the Wilson coefficients 
of Refs.~\cite{Beneke:2012tg,Hellmann:2013jxa} thus read:
\begin{eqnarray}
\text{LO Wilson coeffs.:} &&  \frac{1}{M^2-m_\phi^2} \; \longrightarrow \;
\frac{1}{4 m_{\tilde\chi_1^0}^2-m_\phi^2} 
\bigg( 1 - \frac{4m_{\tilde\chi_1^0} \, (M - 2m_{\tilde\chi_1^0}) }{4m_{\tilde\chi_1^0}^2-m_\phi^2} \bigg) \, , 
\nonumber
\\[1mm]
\text{$v^2$ Wilson coeffs.:} &&  \frac{1}{M^2-m_\phi^2} \;  \longrightarrow \;
\frac{1 }{4 m_{\tilde\chi_1^0}^2-m_\phi^2} \; , 
\nonumber
\\[1mm]
 &&  \frac{2M}{(M^2-m_\phi^2)^2} \;  \longrightarrow \;
\frac{4 m_{\tilde\chi_1^0} }{(4 m_{\tilde\chi_1^0}^2-m_\phi^2)^2}  \; . 
\label{eq:replaceHiggsprop}
\end{eqnarray}
The factor of $2M$ in front of the square of a scalar propagator gets 
replaced by $4m_{\tilde\chi_1^0}$ in the $S$-wave $v^2$-suppressed Wilson 
coefficients  to get exactly the form in the right-hand side of 
Eq.~(\ref{eq:Higgspropmod}), though the difference between both expressions is 
formally of higher order. In self-energy contributions, the replacement 
(\ref{eq:replaceHiggsprop}) in the scalar propagator of LO Wilson 
coefficients produces an ${\cal O}(v^4)$ term from the product of the right 
and left s-channel propagators in the diagram, which is consistently 
dropped in our code in order to keep the expansion of Wilson coefficients 
to ${\cal O}(v^2)$ everywhere.


\subsection{Sommerfeld-corrected cross section}

The annihilation cross sections for the processes 
$[\chi\chi]_a = \chi_i\chi_j \to X$, Eq.~(\ref{eq:SFenhancedsigma}), are 
computed by multiplying every term in the 
partial wave expansion of the Born cross section 
by its specific Sommerfeld factor
\begin{equation}
S_a[\hat f(^{2S+1}L_J)] = 
\frac{ 
\left[\psi^{(L,S)}_{ca}\right]^* \hat f_{bc}(^{2S+1}L_J)\,\psi^{(L,S)}_{ba}}
{\hat f_{aa}(^{2S+1}L_J)|_{\rm LO}}\ .
\label{eq:SFdef}
\end{equation}
When the Sommerfeld factors are neglected, Eq.~(\ref{eq:SFenhancedsigma}) 
reproduces the Born annihilation cross section including 
${\cal O}(v^2)$ terms. The Sommerfeld factors are computed by 
solving the Schr\"odinger equation for a system of coupled two-particle 
states with the leading-order Yukawa and Coulomb potentials generated 
by the exchange of electroweak gauge bosons, Higgs bosons\footnote{In 
practice, we include the Higgs-exchange potential only when 
the Higgs mass is less than $m_{\rm LSP}/2$.} 
and the photon. For further details including notation, we refer 
to Ref.~\cite{Beneke:2014gja}.

The calculation can be done separately in the sectors of two-particle 
states with different electric charge $0,\pm 1,\pm 2$. Since we 
restrict ourselves to the CP-conserving MSSM, the annihilation cross sections 
for the negatively charged two-particle states are identical to the 
corresponding positively charged ones, and do not have to be 
calculated explicitly.

In every 
charge-sector, the Sommerfeld factors are computed for all two-particle 
states with mass less than $1.2\times 2 m_{\rm LSP}$ 
unless the number of such states is larger than four, in which 
case the four lightest two-particle states are selected. For the 
other, heavier two-particle states, we employ the Born cross sections. 
Furthermore, in the computation of the Sommerfeld factor we 
include the light states (at most four) exactly in the 
solution of the Schr\"odinger equation and the others approximately 
in the last loop near the annihilation vertex as described 
in~\cite{Beneke:2014gja}. The mass cut at  $1.2\times 2 m_{\rm LSP}$  
is motivated by the fact that heavier states are either strongly 
Boltzmann-suppressed and irrelevant for freeze-out or they are 
sufficiently off-shell within the ladder diagrams to not 
contribute substantially to the Sommerfeld effect of the lighter 
states. The restriction to at most four light states is motivated 
by  CPU considerations, since the time needed for the matrix 
Schr\"odinger equation solution increases rapidly with the number 
of stated treated exactly. The restriction is certainly sufficient 
for models close to the pure-wino case, when the degenerate states 
are $\tilde\chi_1^0\tilde\chi_1^0$, $\tilde\chi_1^+\tilde\chi_1^-$ in the neutral sector, 
and $\tilde\chi_1^0\tilde\chi_1^+$, $\tilde\chi_1^+\tilde\chi_1^+$ in the charge-1 and 
and charge-2 sectors, respectively. When the LSP acquires a substantial 
Higgsino or bino component, the number of degenerate states 
increases and may exceed four in the neutral and charge-1 sector. 
An example of a strongly mixed wino-Higgsino LSP model has been 
analysed in Ref.~\cite{Beneke:2014hja}, which demonstrated that in this 
case the effect of the additional states is accurately reproduced 
by the approximate treatment in the last loop before the annihilation. 
In the analysis of strongly mixed wino-Higgsino LSP models with a 
nearly decoupled bino discussed below, all possible 10 neutral states 
fall below the 
mass cut $1.2\times 2 m_{\rm LSP}$ in much of the interesting region. 
We checked on a subset of 1575 analysed model points that the relic 
density is always accurately reproduced by the approximate treatment. 
The largest difference we find is 4\%, but it is below 1\% in 96\% of these 
points, and most of the times closer to the permille level. In any case, 
this is not a restriction, since the code can always be run with the 
full set of states treated exactly, at the expense of an increase in 
CPU time of about a factor of ten.

The Sommerfeld factors are computed from the asymptotic behaviour 
at $r\to \infty$ of radial solutions of the 
Schr\"odinger equation with boundary conditions near the origin. 
In practice, evolution of the differential equation system to 
large $r$ is costly, and a finite value of $r_\infty$ must be chosen. 
We determine this value by requiring that the Sommerfeld factor 
changes by less than $0.3\%$, when $r_\infty$ is doubled. This 
accuracy is often difficult to achieve for very small velocities $v$, 
defined by $E = m_{\rm LSP} v^2 = \sqrt{s}-2 m_{\rm LSP}$ or 
near values, where new two-particle channels with mass above 
$2 m_{\rm LSP}$ open, especially for $\tilde\chi^+\tilde\chi^-$ states which 
experience the long-range Coulomb interaction. Hence we 
fix $x_\infty = r_\infty/(m_{\rm LSP} v)$ 
to 20 (50), when $v<0.03$ (within 0.0002 of a threshold). This 
can lead to local inaccuracies of several percent. However, we find 
that the deviation from the exact result is oscillatory, and mostly 
averages out in the thermal average. Once again, this treatment is 
not necessitated by a limitation of the code but a convenience, since 
one can set always $x_\infty$ to larger values if needed.

We generate tables of annihilation cross sections 
$(\sigma v_{\rm rel})_a$ of two-particle states with on average 
around 50 velocity points  chosen adaptively from $10^{-4} \ldots 1$ 
with more sampling 
points near thresholds and the characteristic velocities near 
the freeze-out temperature. We interpolate these functions 
and compute the thermally averaged effective cross section 
$\langle \sigma_{\rm eff} v\rangle$, summed over all co-annihilating 
two-particle states for around 60 suitably chosen values 
of $x=m_{\rm LSP}/T$ between 1 and $10^8$. This table is interpolated 
and the interpolating function is employed in the Boltzmann equation
\begin{align}
\frac{dY}{dx} = -\sqrt{\frac{\pi}{45G}}\,\frac{g_*^{1/2}\,m_1}{x^2} 
\, \langle \sigma_{\rm eff} v \rangle ( Y^2 - Y_{\rm eq}^2 )
\,,
\label{eq:boltzeqY2}
\end{align}
for $Y=n/s$. Here $G$ is the gravitational constant, and 
the parameter $g_*^{1/2}$ is defined in the standard way as
\begin{align}
g_*^{1/2} = \frac{h_{\rm eff}}{g_{\rm eff}^{1/2}} 
\bigg( 1+ \frac{T}{3h_{\rm eff}} \, \frac{dh_{\rm eff}}{dT} \bigg)
\,
\label{eq:gstar}
\end{align}in terms of the effective degrees of freedom 
$g_{\rm eff}$ and $h_{\rm eff}$ of the energy and entropy densities:
\begin{align}
\rho = g_{\rm eff}(T) \frac{\pi^2}{30}\,T^4 
\quad,\quad 
s= h_{\rm eff}(T) \frac{2\pi^2}{45}\,T^3 
\,.
\label{eq:rhos}
\end{align}
For $g_{*}^{1/2}(T)$ and $h_{\rm eff}(T)$ 
we use the values derived in Ref.~\cite{Gondolo:1990dk}, 
which can be found conveniently
tabulated as a function of temperature among the
package files of the automated programs {\sf DarkSUSY}~\cite{Gondolo:2004sc} 
and {\tt micrOMEGAs}~\cite{Belanger:2010gh,Belanger:2013oya}. 
Other numerical values needed for the computation of the relic density are 
$T_0=2.7255$~K and $\rho_{\rm crit}=1.05368\times 10^{-5} h^2$~GeV~cm${}^{-3}$,
both taken from Ref.~\cite{Beringer:1900zz}. 

Given the annihilation matrices, the calculation of all Sommerfeld factors, 
cross section tables, thermal averages and, finally, the evolution of 
the Boltzmann equation through freeze-out takes about 400 sec of CPU time, 
leading to a total computation time (including the evaulation of the 
annihilation matrices) of somewhat above 10 min per MSSM parameter point.


\section{Analysis}
\label{sec:analysis}

The departure from the pure-wino limit can be obtained by lowering the 
sfermion masses and/or introducing non-negligible Higgsino or bino fractions 
of the lightest neutralino. Therefore we organise the analysis and results in 
three parts: effect of the sfermion masses (Sec.~\ref{sec:Res_sfer}), 
Higgsino admixture (Sec.~\ref{sec:Res_H}) and bino admixture 
(Sec.~\ref{sec:Res_B}). The residual dependence on remaining parameters is 
discussed in Sec.~\ref{sec:Res_other}.


\subsection{Impact of sfermions}
\label{sec:Res_sfer}

\begin{figure}[p]
  \centering
 \includegraphics[width=.49\textwidth]{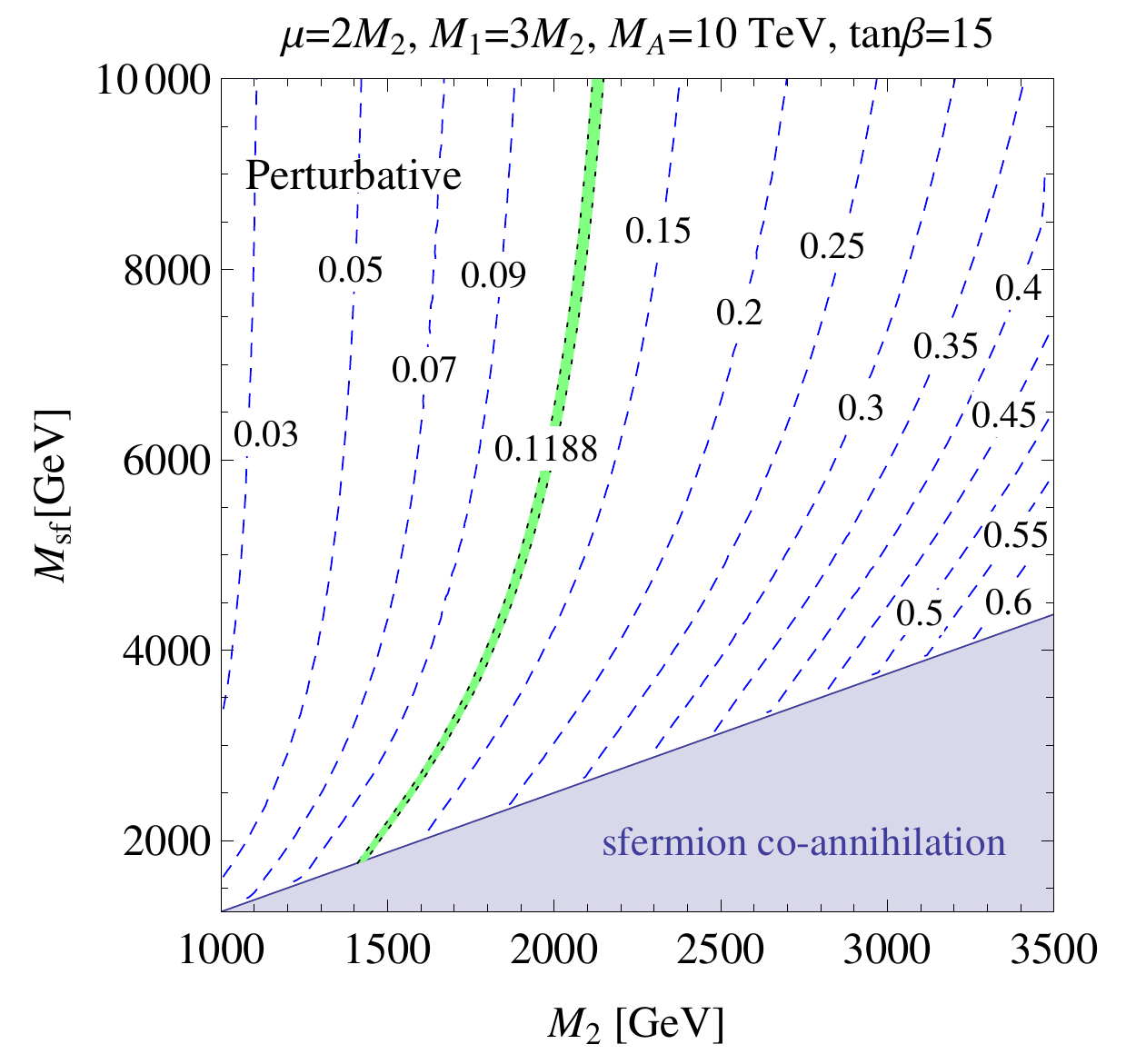} \includegraphics[width=.49\textwidth]{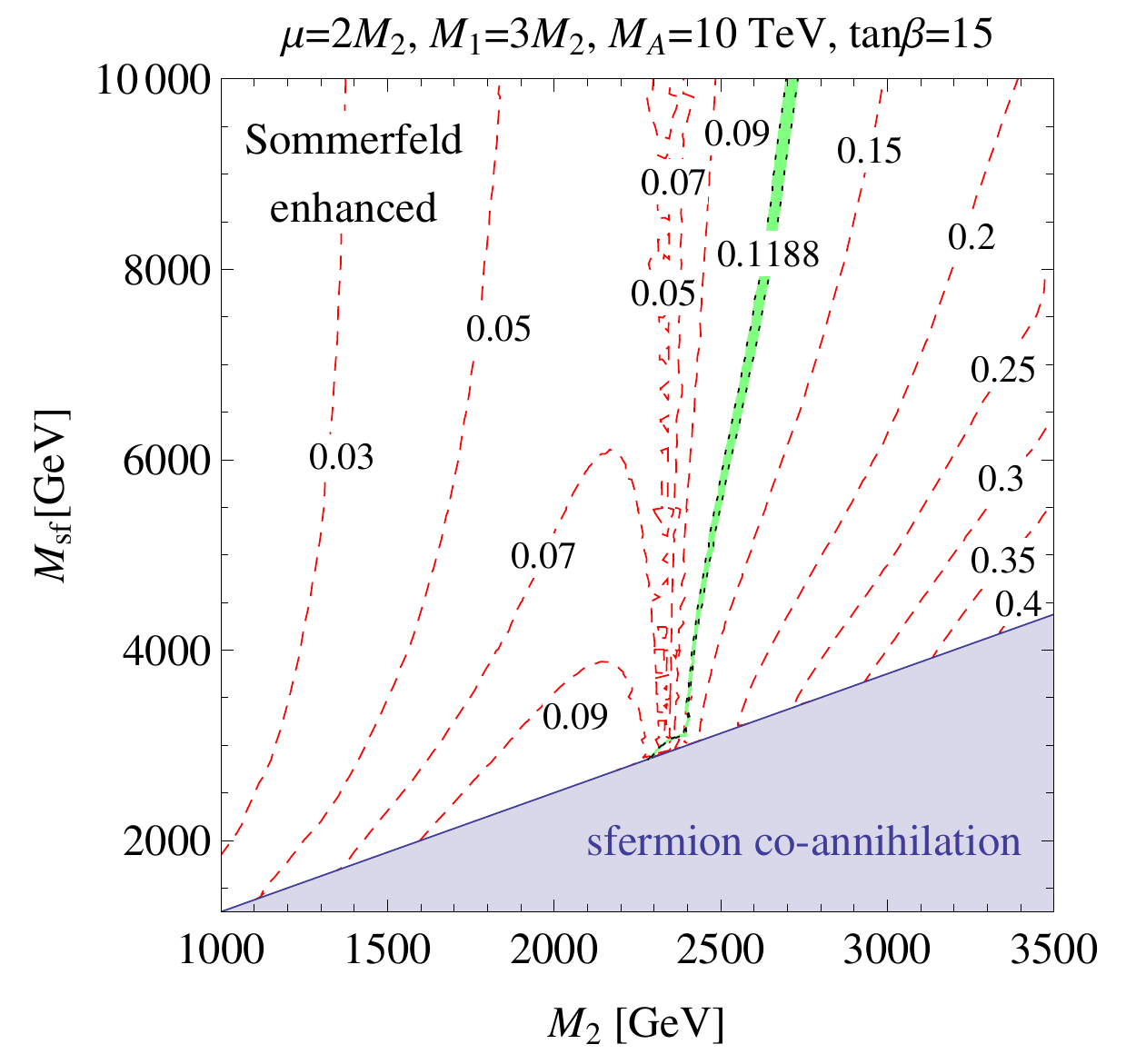}
 \caption{The contours of constant relic density: perturbative (left) and Sommerfeld enhanced (right). The (green) bands show the region within $2\sigma$ of the observed dark matter abundance. The grey area indicates the region in parameter space where the co-annihilations with sfermions are potentially relevant and which is not studied in this work. Other parameters are as indicated, with $A_i =8$~TeV and $X_t$ is fixed by the measured Higgs mass.\label{fig:m2-msf}}
\end{figure}

\begin{figure}[p]
  \centering
 \includegraphics[width=.6\textwidth]{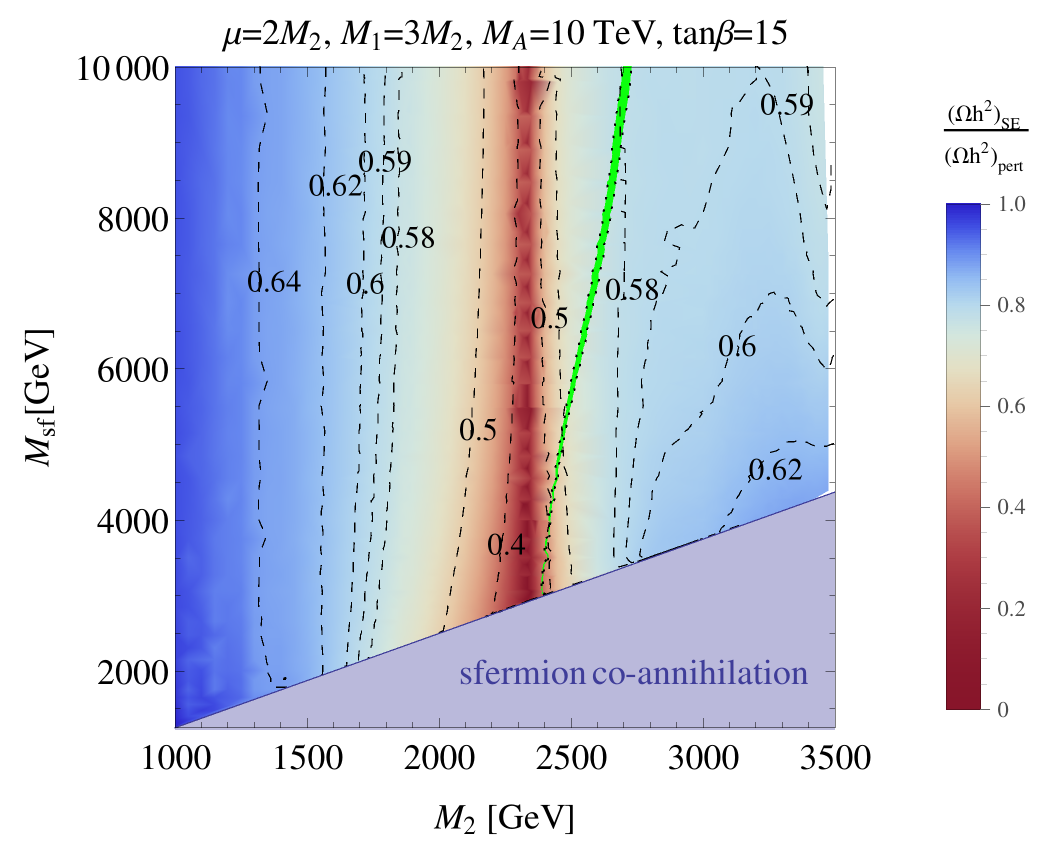}
 \caption{The ratio of the relic density including Sommerfeld enhancement to the perturbative result is shown via a density map as well as black dashed contours. The (green) band indicates the region within $2\sigma$ of the observed dark matter abundance. Other parameters are as indicated, with $A_i=8$~TeV and $X_t$ is fixed by the measured Higgs mass.
\label{fig:m2-msf_SE}
}
\end{figure}

The role played by the sfermions in the production of the thermal neutralino relic density is threefold: \textit{i)} they appear in the t- and u-channel annihilation into SM fermions, \textit{ii)} they introduce corrections to the neutralino and chargino masses indirectly, via loop effects, and \textit{iii)} if light enough, they can contribute to the effective annihilation cross section through additional co-annihilation channels. The last of these is beyond the scope of this work and we leave a detailed analysis of general sfermion co-annihilation regions with the inclusion of the Sommerfeld enhancement for future work. Therefore in our results for the perturbative and Sommerfeld corrected relic density, shown in Fig.~\ref{fig:m2-msf} and the ratio of these results shown in Fig.~\ref{fig:m2-msf_SE}, 
we require that all the sfermions are at least 25\% heavier than the LSP.\footnote{Note that the horizontal axis of most of our plots is chosen to be $M_2$, which in general lies within a few $\gev$ of $m_{\rm LSP}$.} From points \textit{i)} and \textit{ii)}, the indirect effect of changing the spectrum is sub-dominant, even for the regions of parameter space where the Sommerfeld effect exhibits a resonance and where the resulting cross section is extremely sensitive to the mass difference between $\tilde{\chi}_1^\pm$ and $\tilde{\chi}_1^0$. The reason is that the main contribution to this quantity comes from loops involving gauge bosons and the ones with sfermions are suppressed by their large masses.


The direct impact is on the other hand quite important. Decoupled sfermions mean that the only contributions to the effective co-annihilation cross section from the processes with SM fermion final states arise due to the s-channel annihilation through gauge or Higgs bosons. When the sfermions become lighter the t- and u-channel processes start to be non-negligible. This is especially relevant for the co-annihilation channels, while the direct LSP annihilation to SM fermions is helicity or p-wave suppressed. These t- and u-channel diagrams involving sfermions interfere destructively with the s-channel gauge boson exchange effectively \textit{lowering} the co-annihilation cross section \cite{Baer:2005zc}.

At the TeV scale the degeneracy between the charginos and neutralinos is more pronounced resulting in co-annihilation channels not being Boltzmann suppressed. Therefore the effective annihilation cross section is strongly affected by the co-annihilation channels e.g.~from the processes $\tilde{\chi}^+_1 \tilde{\chi}^-_1\to \bar f f$, $\tilde{\chi}^0_1 \tilde{\chi}^\pm_1\to \bar f' f$. Due to interference between the sfermion t-channel and $W$ boson s-channel diagrams, the lower the sfermion masses, the smaller the contribution from these processes, leading to lower total annihilation cross section and higher thermal relic density. In other words, the contours of constant relic density move towards lower $\mlsp$ values as the sfermion masses decrease. This is indeed what is observed in the left panel of Fig.~\ref{fig:m2-msf}, where the contours of constant perturbative relic density are plotted in the $M_2$--$M_{\rm{sf}}$ plane for the case of a wino-like LSP ($\mu=2 M_2$, $M_1=3 M_2$). In particular note that, by varying the sfermion masses one can obtain the perturbative thermal relic density in agreement with the observed abundance over a large range ($\sim 800$ GeV) of LSP masses. It is also worth pointing out that the fact that the contours become denser as $M_2$ increases is a simple result of the approximate quadratic dependence of the relic density on the wino mass.

The situation becomes more involved at the non-perturbative level, as shown in the right panel of Fig.~\ref{fig:m2-msf}. The main features that were previously discussed for the perturbative case are still present, but two important modifications arise. First, the contours are seen to be shifted towards larger values of $M_2$. This is simply the effect of the Sommerfeld enhancement on the annihilation cross section, such that one requires a wino mass of around 2.9~TeV rather than 2.2 TeV in order to obtain the correct thermal relic density in the decoupled sfermion case. The size of the shift however depends on the masses of the sfermions, in particular the lowest wino mass giving the correct relic density $\Omega h^2\rvert_\textrm{exp} = 0.1188\pm 0.0010$ \cite{plancknew} 
without sfermion co-annihilations is around 2.3 TeV. This is related to the second effect, namely the resonance in the Sommerfeld enhancement, which is also responsible for lowering the constant relic density contours in the sfermion mass at $\mlsp$ of around 2.3-2.4 TeV. The presence of the resonance is most clearly seen in Fig.~\ref{fig:m2-msf_SE} where the impact of the Sommerfeld effect on the relic density is shown. It can be seen that, as expected, the Sommerfeld effect gets stronger for larger values of $M_2$ until the resonance region is reached, and that in the resonance region the relic density can be suppressed by nearly an order of magnitude. What is worth stressing is that the Sommerfeld effect is also approximately independent of the value of the sfermion masses. This can be easily understood by noting that the largest impact of the Sommerfeld effect comes from its contribution on the $\tilde{\chi}^0_1\tilde{\chi}^0_1$ annihilation, which does not depend in any significant way on the nature of the sfermions. We also note that the Sommerfeld effect changes the relic density by almost a factor of two in the region where the observed relic density is attained (green bands in the figures), and by an even larger factor for smaller $M_{\rm sf}$, when the 
observed relic density is produced near the Sommerfeld resonance.

The generic behaviour of the results for the relic density as a function of sfermion masses shown and discussed above holds when one departs from wino-like neutralino as well, but with the details depending on the precise neutralino composition and the spectrum of the sfermions. The latter comes from the fact, that while the coupling of the sfermions with the wino is purely gauge, the one with the Higgsino is Yukawa-type, and therefore discriminates the three generations, as well as squarks from sleptons. The analysis of such scenarios, with Higgsino- and bino-like neutralinos, will be provided in the future.


\subsection{Higgsino admixture}
\label{sec:Res_H}

The Higgsino-wino mixing predominantly depends inversely on the difference 
between $\mu$ and $M_2$, as discussed in Sec.~\ref{sec:params}.
Increasing the Higgsino component of the predominantly wino-like LSP has 
several effects on the relic density: 
\textit{i)} it modifies the LSP annihilation cross section due to different 
couplings of the wino and Higgsino components, \textit{ii)} it changes the 
relevant number and weights of the 
co-annihilation channels and finally \textit{iii)} it significantly alters the Sommerfeld effect.
The first two effects are very well known, we therefore concentrate on the 
non-perturbative effects. We choose to parametrise the Higgsino admixture via 
the difference between the input parameters $\mu$ and $M_2$. For definiteness, 
we restrict ourselves to positive $\mu$ in the following analysis. 
In the $\mlsp$ range considered, values of $\mu-M_2\gtrsim 500$~GeV 
lead to nearly decoupled Higgsinos, and the LSP is practically purely wino-like, while 
values of around $300\text{--}500$~GeV correspond to a Higgsino fraction of around a few~\%, growing up to 50\% for $\mu=M_2$.

\begin{figure}[p]
  \centering
 \includegraphics[width=.49\textwidth]{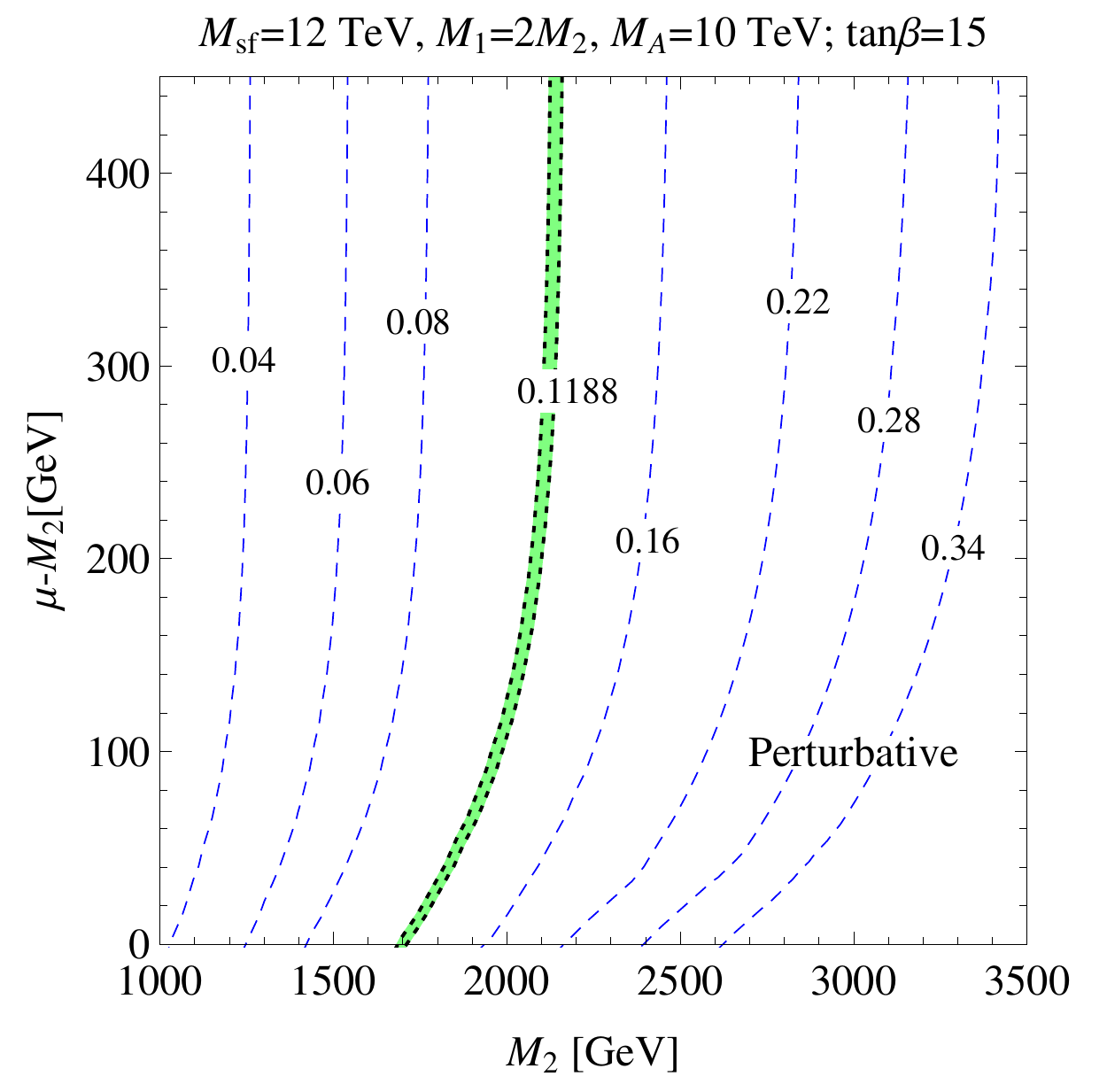}
  \includegraphics[width=.49\textwidth]{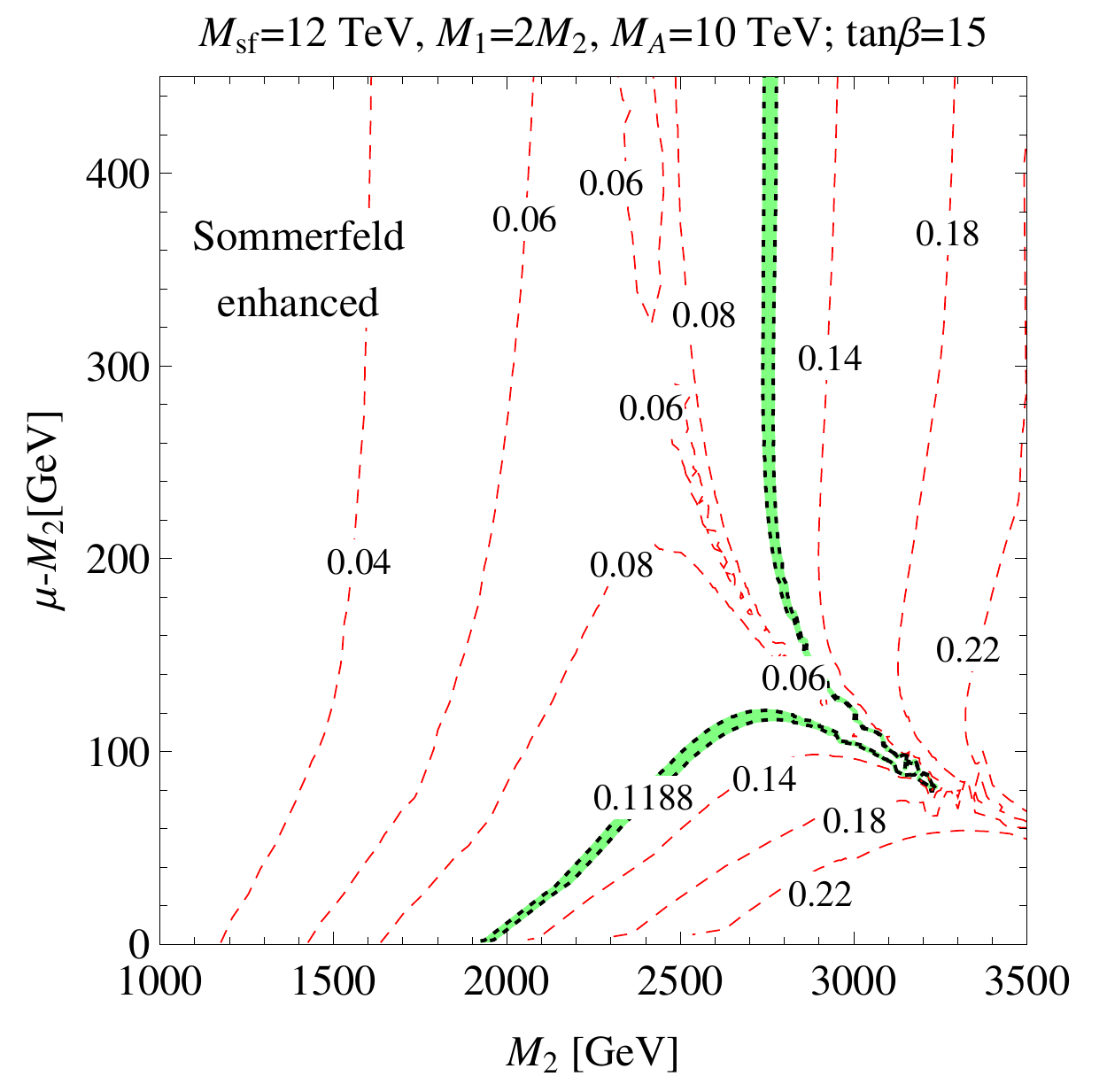}
 \caption{Contours of constant relic density are shown for the case of the perturbative (left) and Sommerfeld enhanced (right) calculation.
 The (green) bands indicate the region within $2\sigma$ of the observed dark matter abundance. Other parameters are as indicated, with $A_i=8$~TeV and $X_t$ is fixed by the measured Higgs mass. \label{fig:m2-mu_RD}}
\end{figure}

\begin{figure}[p]
  \centering
 \includegraphics[width=.6\textwidth]{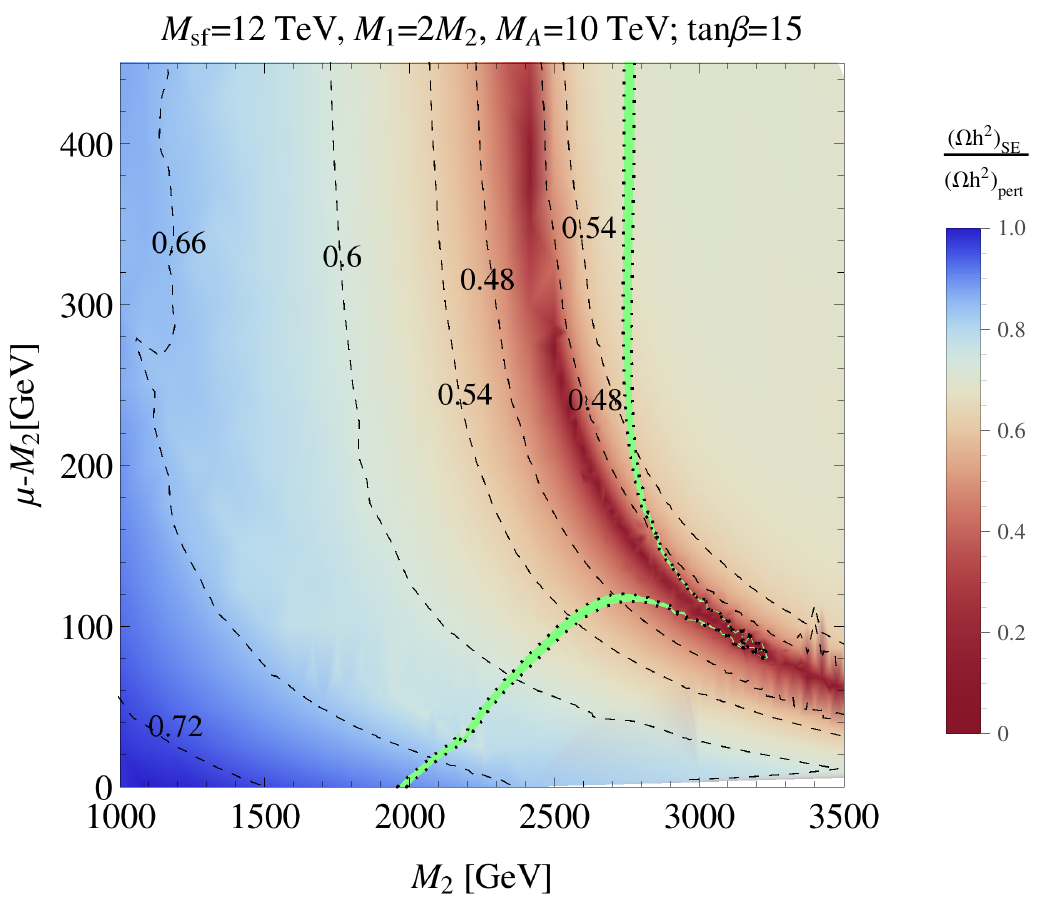}
 \caption{The impact of the Sommerfeld enhancement of the relic density shown as a density map as well as via the black dashed contours. The (green) band indicates the region within $2\sigma$ of the observed dark matter abundance. Other parameters are as indicated, with $A_i=8$~TeV and $X_t$ is fixed by the measured Higgs mass.\label{fig:m2-mu_SE}}
\end{figure}

\begin{figure}
 \centering
 \includegraphics[width=.5\textwidth]{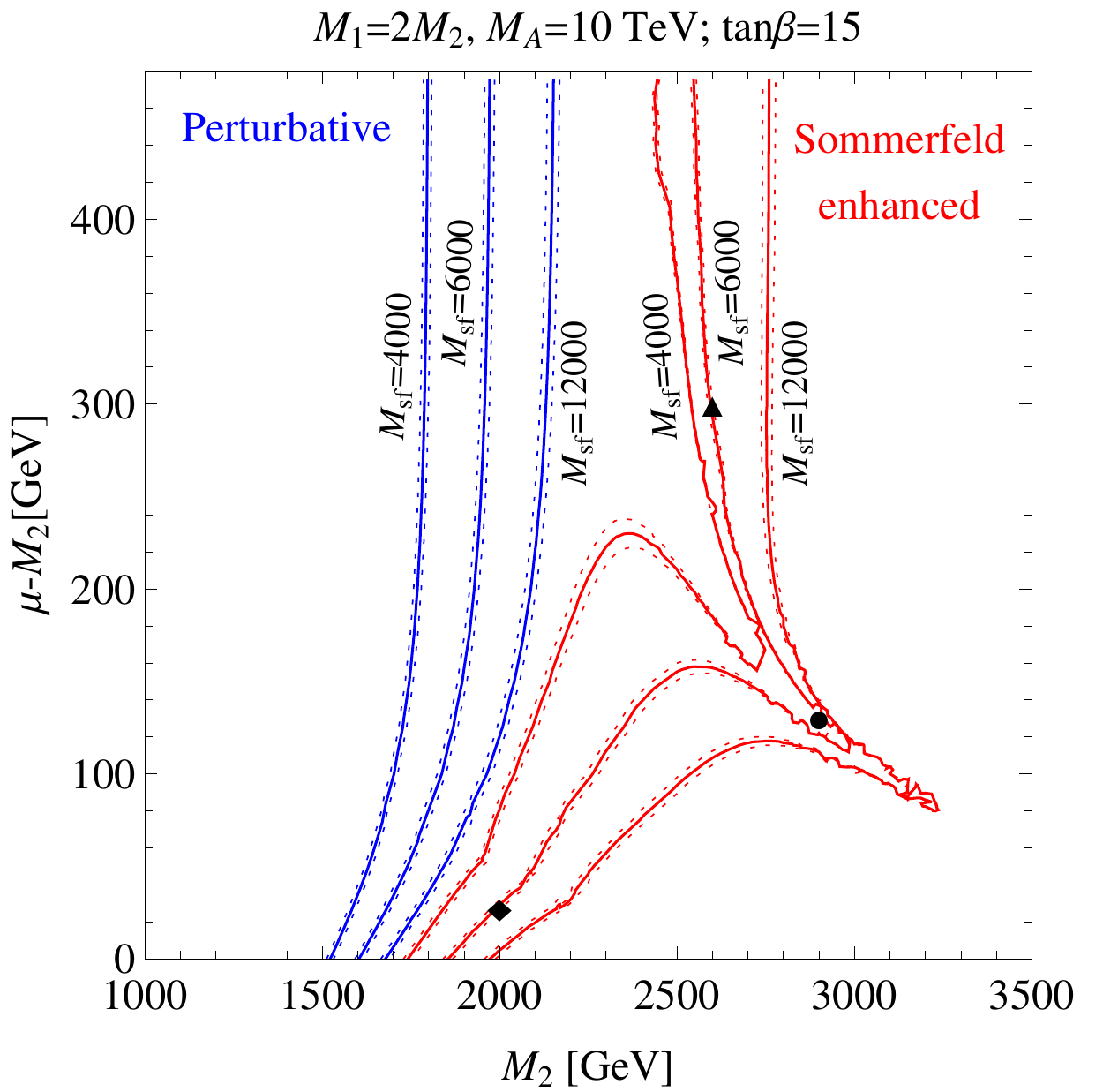}
 \caption{Contours providing the correct relic density are shown for the case of the perturbative (blue) and Sommerfeld enhanced (red) calculation for three different values of the common sfermion mass parameter. Other parameters are as indicated, with $A_i=8$~TeV and $X_t$ is fixed by the measured Higgs mass to a different value depending on $\msf$. The black markers denote the three points studied in Sec.~\ref{sec:Res_other}. \label{fig:m2-mu}}
\end{figure}

The results of the analysis are displayed on Figs.~\ref{fig:m2-mu_RD} to \ref{fig:m2-mu}. In Fig.~\ref{fig:m2-mu_RD} 
the contours of constant relic density are shown in the $M_2$ vs.~$(\mu-M_2)$ plane for the perturbative (left plot, blue lines) and Sommerfeld enhanced (right plot, red lines) cases. In the upper region of the plot the contour lines flatten as we recover the pure wino scenario, while in the lower 
region they tend to lower values of $\mlsp$ because the large Higgsino fraction suppresses the annihilation cross section.
Equivalently, for a fixed LSP mass, increasing the Higgsino admixture increases the relic density. At the perturbative level 
this is mainly a consequence of the lower value of the coupling to gauge bosons, while in the case of the 
Sommerfeld effect it also is a result of the larger mass splitting between the LSP and the lightest chargino.
In Fig.~\ref{fig:m2-mu_SE} we show the ratio of the above plots in order to display the impact of the Sommerfeld enhancement over the  $M_2$ vs. $(\mu-M_2)$ plane.


In Fig.~\ref{fig:m2-mu} we show those contours giving the correct thermal relic density, for three different values of the sfermion mass parameter
$\msf$, and show both the perturbative (left three lines, blue) and Sommerfeld-corrected 
(right three lines, red) results in one plot. The $\msf=12~$TeV lines correspond to the (green) 
bands in the previous Figs.~\ref{fig:m2-mu_RD} and~\ref{fig:m2-mu_SE}.  We change here from displaying contours of constant relic density to the correct relic density in order to highlight the effect of the sfermion mass parameter. Note that, in agreement with what was discussed in the previous section,
the lower the sfermion masses, the larger the relic density and hence the lower the LSP mass at correct relic density -- both for the perturbative and non-perturbative results. 

Comparing the perturbative result to the full one in Fig.~\ref{fig:m2-mu}, 
one observes the following:
\begin{itemize}
\item[i)] the contours shift to higher masses, indicating the decrease of 
the relic density with respect to the perturbative result; in particular the 
mass of the lightest neutralino (when $\mu>M_2$) giving correct 
thermal relic density is around 1.7~TeV,
\item[ii)] for LSP masses in the range 2--3~TeV the contours take a non-trivial form, which can be understood in terms of the resonance 
in the Sommerfeld enhancement. This resonance leads to much more efficient annihilation, strongly suppressing the relic density; 
it appears at different positions in $\mlsp$ depending on the neutralino composition. In particular, on increasing the Higgsino fraction, the 
resonance occurs for heavier LSPs, which is mainly due to the increasing of the mass splitting between the lightest chargino and neutralino, the decreasing coupling, and the fact that the resonance depends on the splitting through
(mass splitting)/$(m_{\rm LSP} \alpha_2^2)$.
\end{itemize}

The position of the peak of the resonance was clearly visible in Fig.~\ref{fig:m2-mu_SE}.
It is therefore evident that in Fig.~\ref{fig:m2-mu} the contours of the correct relic density cluster around this peak for higher values of $M_2$ and
lower values of $\mu-M_2$. This is easily understood when one recalls that in this region the neutralino at a perturbative level has a thermal
abundance larger than that observed by a factor of a few. The proximity to the resonance enhances the cross section, reducing the relic density to
agree with the measured value.
In particular, it follows that the largest value of $M_2$ giving the correct thermal relic density is close to 3.3 TeV, approximately 20\% higher than
that for the pure-wino scenario.

Note also that, in contrast to the pure-wino scenario with decoupled sfermions, a region of parameter space exists where the thermal relic density 
is obtained in very close vicinity to the resonance, leading to strong bounds on such scenarios coming from dark matter 
indirect searches. Previously only limiting wino and Higgsino cases have been studied from this perspective with the inclusion of  
the Sommerfeld effect \cite{Hisano:2003ec,Hisano:2005ec,Cirelli:2007xd}, and even slightly mixed scenarios remain unexplored.\footnote{The only related 
works available in the literature \cite{Roszkowski:2014iqa,Catalan:2015cna,Bramante:2015una} are considering the Sommerfeld effect in an approximate 
way and/or without inclusion of recent developments \cite{Beneke:2012tg,Hellmann:2013jxa,Beneke:2014gja}.}
It also follows that some regions of the pMSSM parameter space can be effectively constrained by non-observation 
of any dark matter signal in cosmic or $\gamma$-rays. The precise analysis of such phenomenologically interesting regions 
will be presented in upcoming work~\cite{IDpaper}.


\subsection{Effect of the heavy Higgs bosons}
\label{sec:MA}

In the MSSM, the only particles beyond the SM having positive R-parity are 
the additional Higgs bosons. These can therefore act as an s-channel mediator 
and, if light enough, as end-products of the (co-)annihilation. As the effect 
of these Higgs bosons is greatest when the Higgsino mass parameter 
is close to $M_2$, 
we discuss this first before moving on to the wino-bino mixed case.
The additional Higgs bosons can affect the relic density in the following 
two ways:
\begin{itemize}
\item by contributing to the (co-)annihilation rate via 
s-channel diagrams, particularly if $\mao$ lies in the vicinity of 
$2\,m_{\rm LSP}$, thereby typically reducing the relic density
\item by providing additional final states with one heavy Higgs plus one 
gauge or light Higgs boson, or with two heavy Higgs bosons, if the 
combined mass of the final state lies below $2 m_{\rm LSP}$, which  
leads to a reduction in the relic density. 
\end{itemize}
The former is only relevant when the coupling of the annihilating particles 
to the Higgs bosons is non-negligible. This requires one of the two 
annihilating neutralinos or charginos to contain a considerable gaugino 
component and the other a considerable Higgsino component. For 
$\tilde{\chi}_0$ annihilation this implies that the LSP is mixed. We remind 
the reader that we do not consider the resonant annihilation region 
when $\mao$ is inside the interval 1.7--2.3$\,m_{\rm LSP}$ as explained in 
Sec.~\ref{sec:constraints}. As for the latter point, the heavy Higgs and 
gauge boson final state is obtained via a s-channel gauge boson, or a 
t-channel neutralino or chargino. This is again more relevant when the LSP 
contains some Higgsino admixture, as this also allows the coupling of 
neutralinos to $Z$ bosons. However, in contrast to the case of the heavy 
Higgs boson in the s-channel, this contribution does not vanish when the 
Higgsino decouples, as a coannihilating chargino and neutralino can 
annihilate into a heavy Higgs and gauge boson via a s-channel $W$ boson 
even in the pure-wino limit.

\begin{figure}[p]
\centering
\includegraphics[width=.49\textwidth]{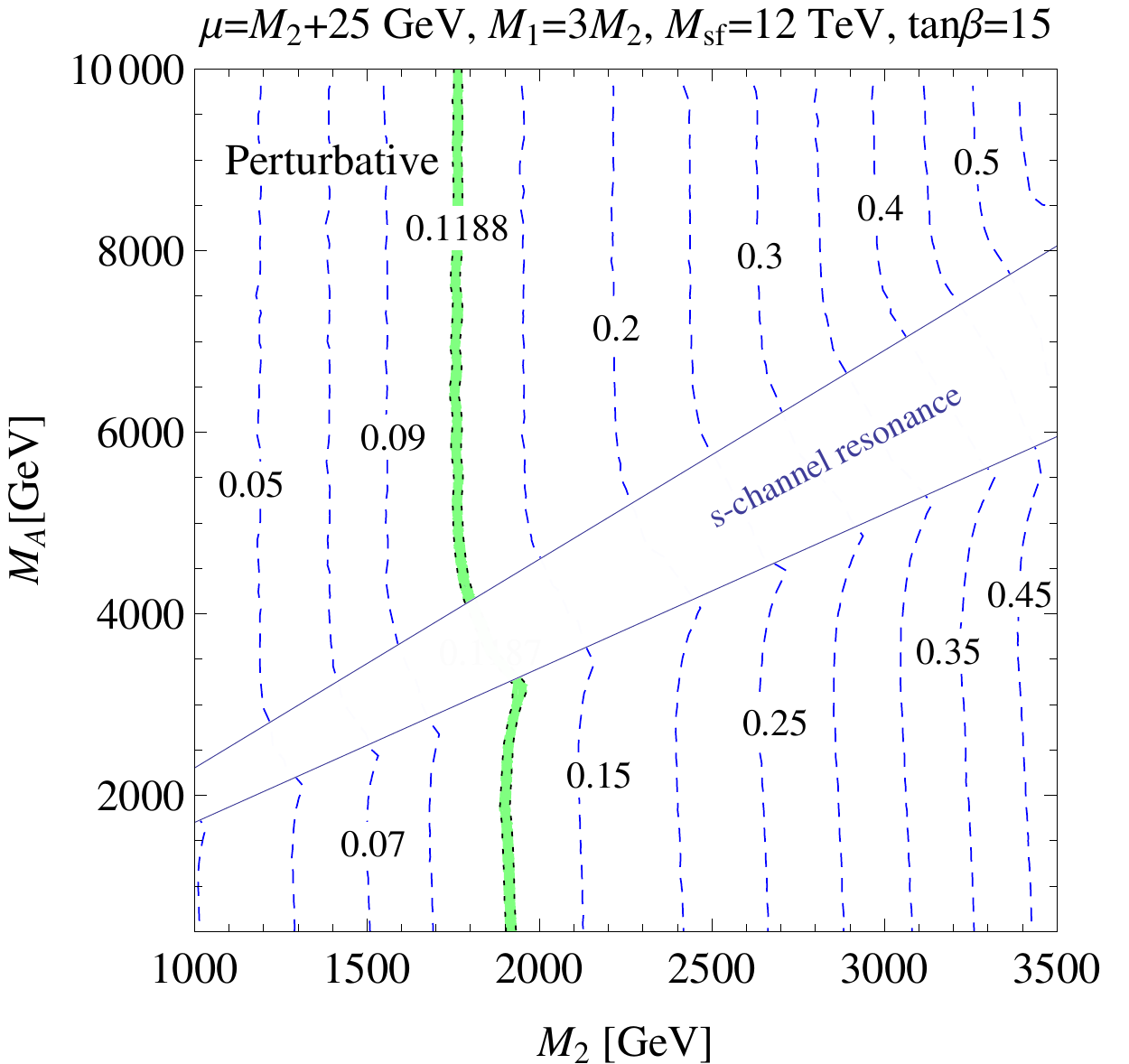}
\includegraphics[width=.49\textwidth]{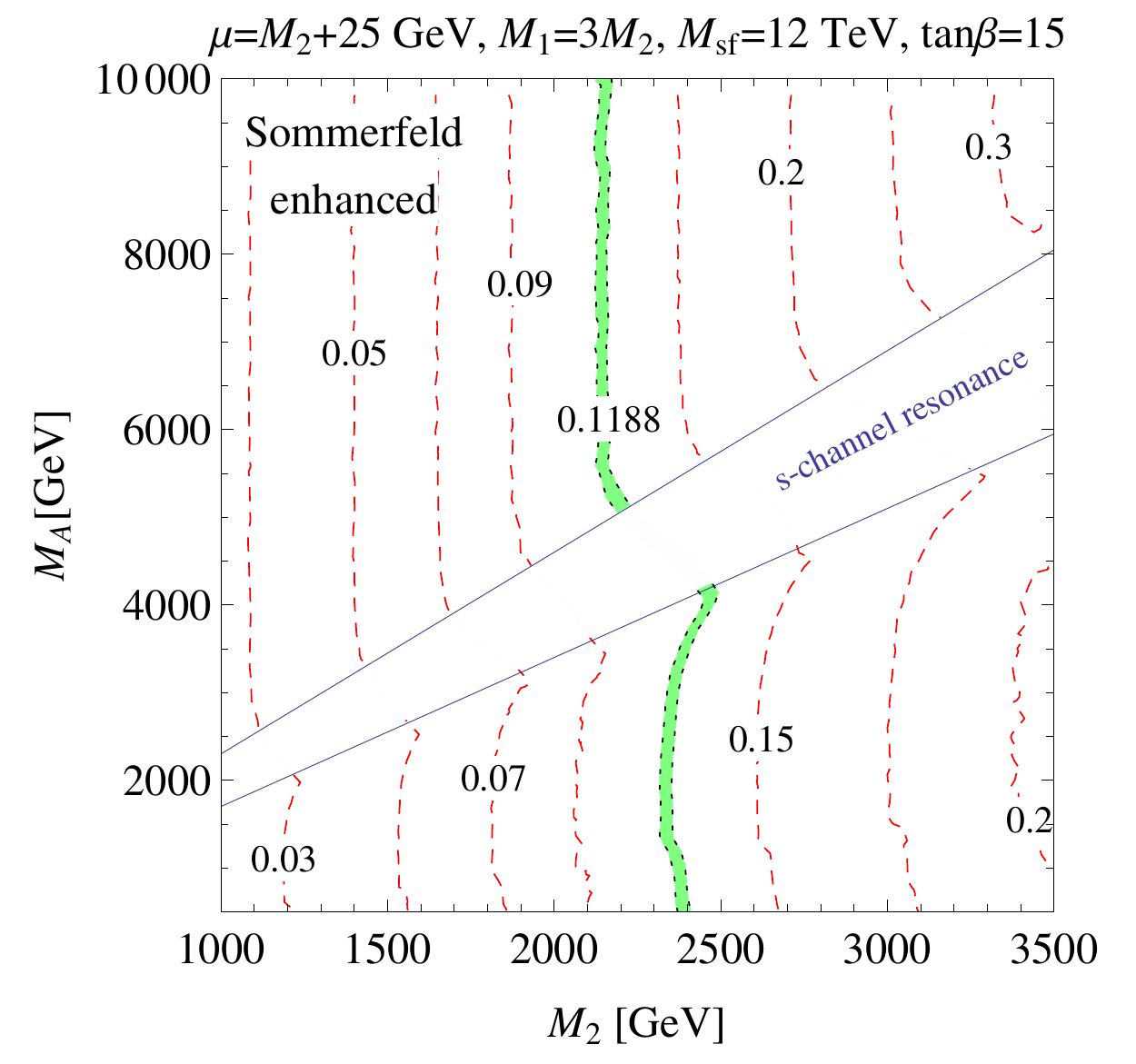}
\caption{Contours of constant relic density are shown for the case of the 
perturbative (left) and Sommerfeld enhanced (right) calculation. The (green) 
bands indicate the region within $2\sigma$ of the observed dark matter 
abundance. Other parameters are as indicated, with $A_i=8$~TeV and $X_t$ is 
fixed by the measured Higgs mass. 
\label{fig:m2-ma_RD}}
\end{figure}

\begin{figure}[p]
\centering
\includegraphics[width=.5\textwidth]{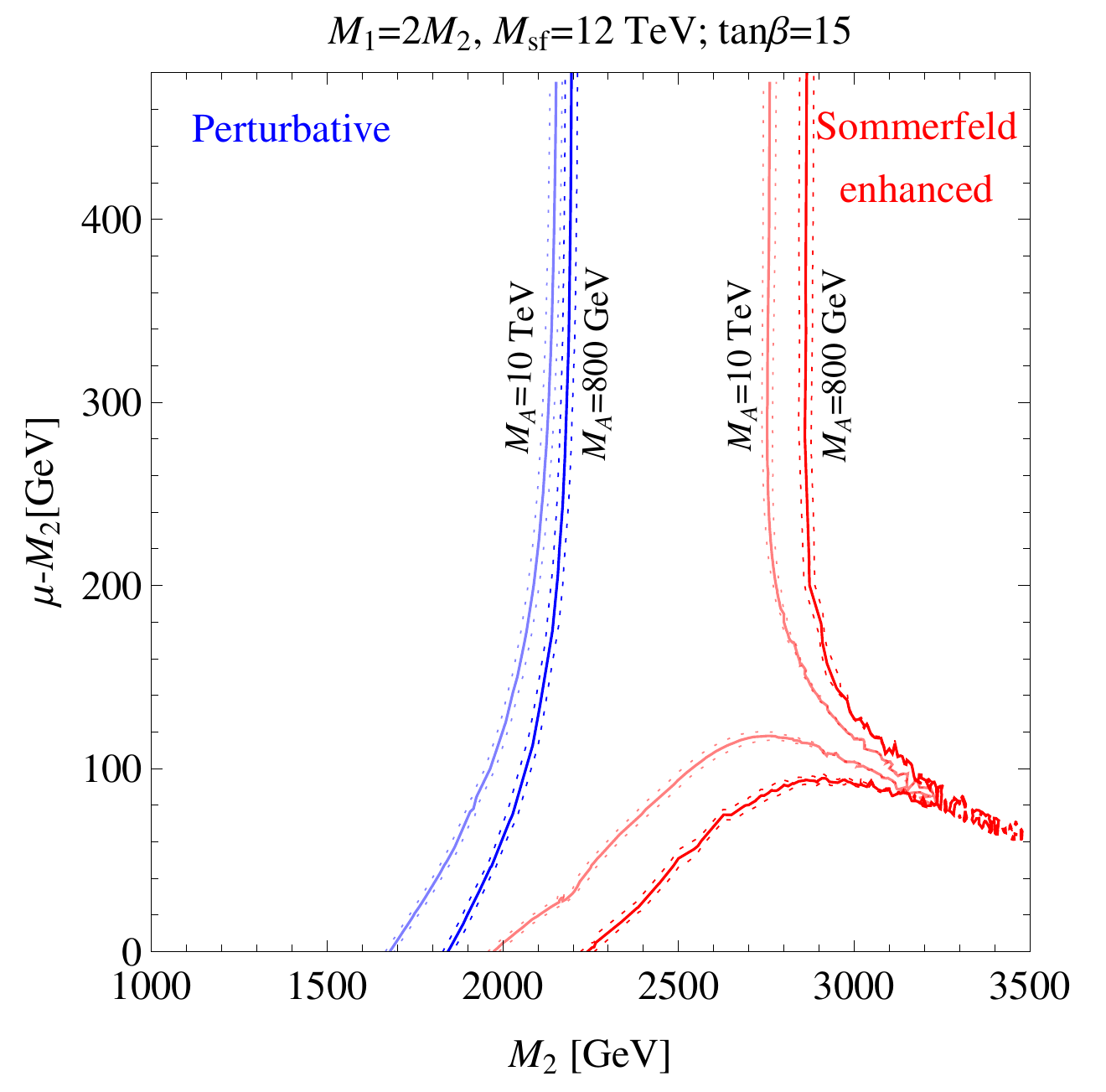}
\caption{Contours providing the correct relic density are shown for the case 
of the perturbative (blue) and Sommerfeld enhanced (red) calculation for 
three different values of the heavy Higgs mass $M_A$. Other parameters are as 
indicated, with $A_i=8$~TeV and $X_t$ is fixed by the measured Higgs mass.
\label{fig:m2-mu_MA}}
\end{figure}

We explore these issues in Fig.~\ref{fig:m2-ma_RD} where we show contours of constant relic density in the $\mII$ vs. $\mao$ plane both at the perturbative level (left) and on taking account of the Sommerfeld effect (right). The region corresponding to the measured relic density is shown by the green band. As the s-channel resonance cannot be accurately calculated in our framework we do not provide results near $\mao=2\,\mII$. It is seen that for the perturbative case above the excluded region the lines are approximately vertical, just bending slightly towards higher values of $\mII$ on approaching this region. Below the excluded area we find that there is a slight shift to the right as the heavy Higgs bosons are accessible in the final state. The difference in $\mII$ giving the correct relic density is approximately $150\,\gev$ when $\mao$ changes from $10\,\tev$ to $500\,\gev$. For the Sommerfeld-enhanced case the result is qualitatively similar, however the difference in $\mII$ giving the correct relic density is around $250\,\gev$ for the same change of $\mao$. 

In Fig.~\ref{fig:m2-mu_MA} we further investigate the effect of the heavy Higgs bosons on the contours showing the correct relic density in the $\mII$ vs. $\mu-\mII$ plane. The blue lines show the perturbative result while the red lines include the Sommerfeld enhancement. We see that on decreasing $\mao$ from $10\,\tev$ to $800\,\gev$, the shift in the value of $\mII$ giving the correct density is indeed dependent on the proximity of $\mu$ to $\mII$, increasing from 50 to $150\,\gev$ in the perturbative case and 100 to $250\,\gev$ 
in the Sommerfeld-enhanced case. As mentioned earlier, an increased Higgsino admixture allows a stronger coupling to the Higgs  and $Z$ bosons in the s-channel (where $Z$ bosons can give rise to heavy Higgs bosons in the final state), increasing the effect of the heavy Higgs boson. Nevertheless when the Higgsino is decoupled a dependence on $\mao$ persists; for the $\mao=800\,\gev$ contours coannihilation via a $W$ boson to a final state containing a heavy Higgs boson and a gauge boson is allowed but not for the $\mao=10\,\tev$ contours. 


\subsection{Bino admixture}
\label{sec:Res_B}

The bino only mixes with the wino via the off-diagonal terms in the Higgsino block of the neutralino mass matrix. 
It follows that the mixing is weak, depending of course on the Higgsino parameter $\mu$, and is further sensitive to $\tan\beta$ and the sign of $M_1$ and $\mu$ as seen in Eq.~\eqref{eq:binomix}. In order that the wino-like neutralino contains a substantial bino component, either $\mu$ should be of the same order as $M_1$ and $M_2$ or the $M_1$ and $M_2$ parameters should be highly degenerate. For example, when $\delta M_1=M_2-M_1=10$~GeV, $\mu=2\,M_2$ and $\tanb=15$ the mixing is about 1\%, decreasing to 0.1\% when $\delta M_1=100$~GeV.
Such situations may arise and are worth studying as the resulting features are of phenomenological interest. In this section we focus on the 
second case $M_1\sim M_2 \ll \mu$, since the first ($M_1\sim M_2\sim \mu$) falls into the category of a mixed wino-Higgsino state and shares the gross features with the case of a decoupled bino analysed 
in the previous section. We further assume $M_1>0$, since $M_1<0$ entails 
an essentially decoupled bino. 

When $M_1$ is close to $M_2$, the perturbative relic density is affected both by the modification of the LSP annihilation 
cross section due to the change in composition and by the co-annihilation 
with the bino-like NLSP, with mass close to $M_1$. On top of that the Sommerfeld effect 
is modified, analogously to the Higgsino-wino mixed scenario, by the weakened coupling and larger mass splitting between $\tilde{\chi}^0_1$ and $\tilde{\chi}^\pm_1$ as given in Eqs.~\eqref{eq:Bino-Wino-Masses2a}, \eqref{eq:Bino-Wino-Masses2b}. Qualitatively the behaviour observed
on increasing the bino component is largely the same as in the Higgsino case, with
an important quantitative difference: a larger sensitivity of the results to the remaining parameters.  

\begin{figure}[p]
  \centering
 \includegraphics[width=.49\textwidth]{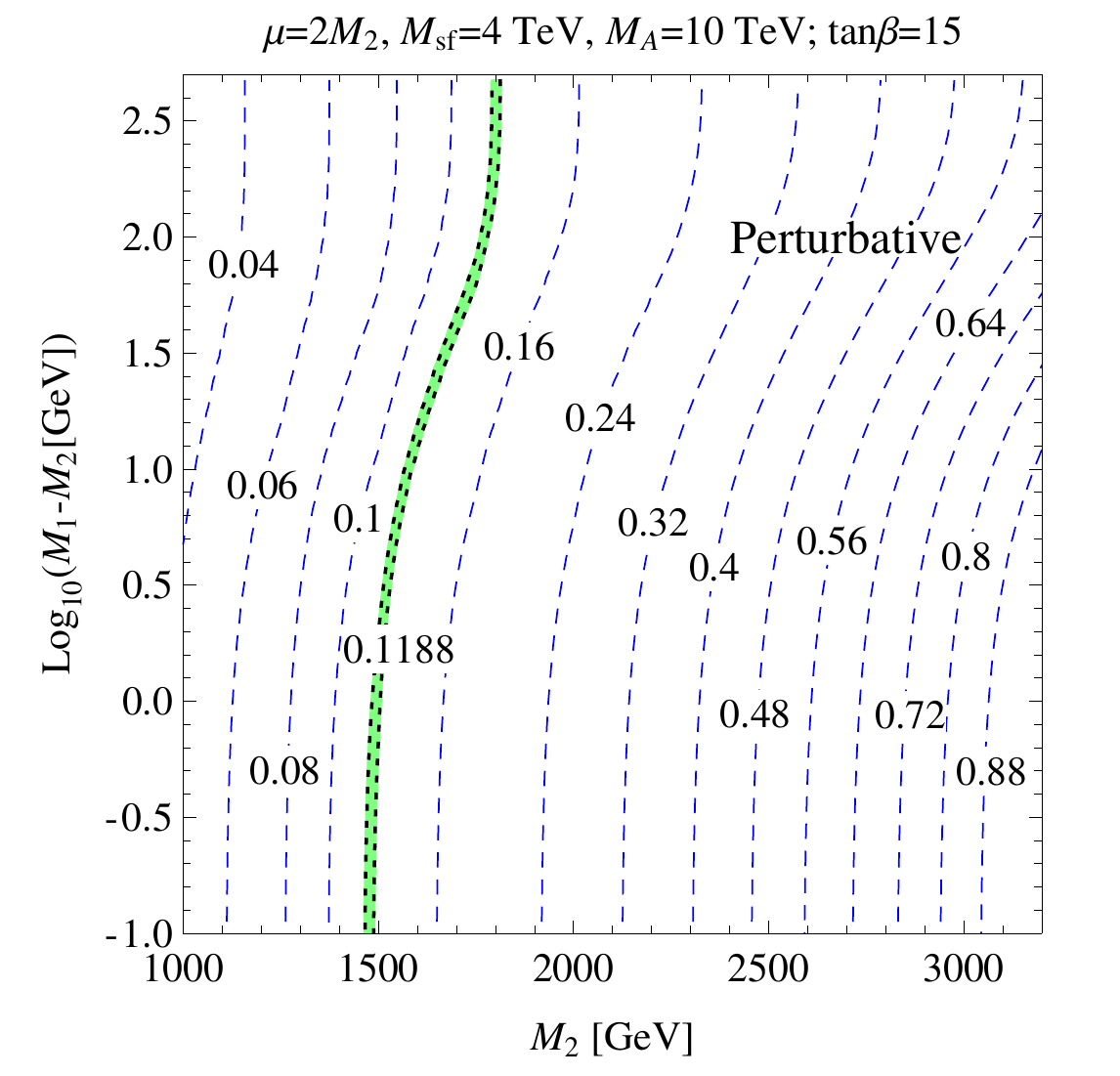}
  \includegraphics[width=.49\textwidth]{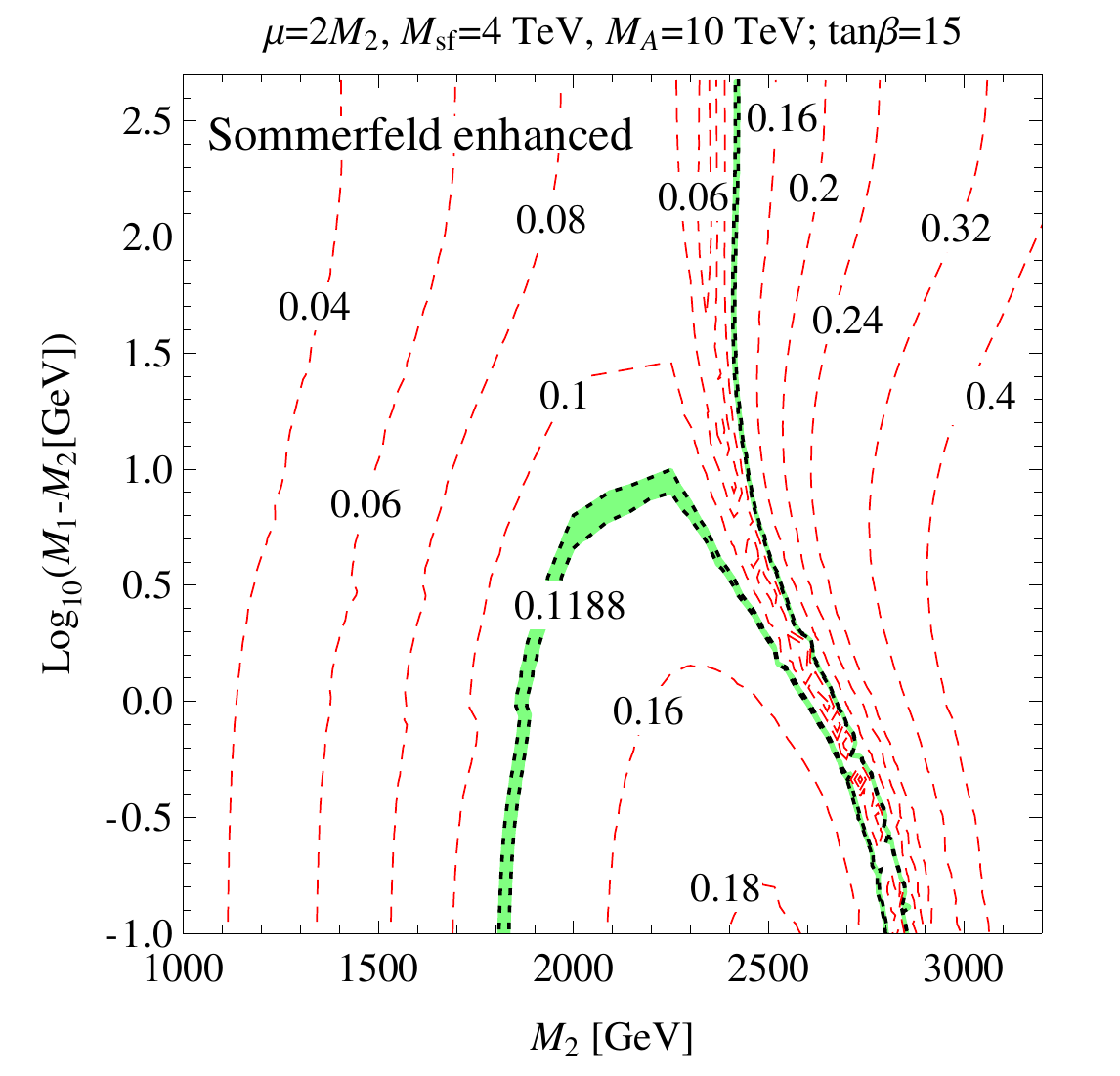}
 \caption{Contours of constant relic density are shown for the case of the perturbative (left) and Sommerfeld enhanced (right) calculation.
 The (green) bands indicate the region within $2\sigma$ of the observed dark matter abundance. Other parameters are as indicated, with $A_i=8$~TeV and $X_t$ is fixed by the measured Higgs mass.  \label{fig:m2-m1_msf4TeV}}
\end{figure}
\begin{figure}[p]
  \centering
 \includegraphics[width=.6\textwidth]{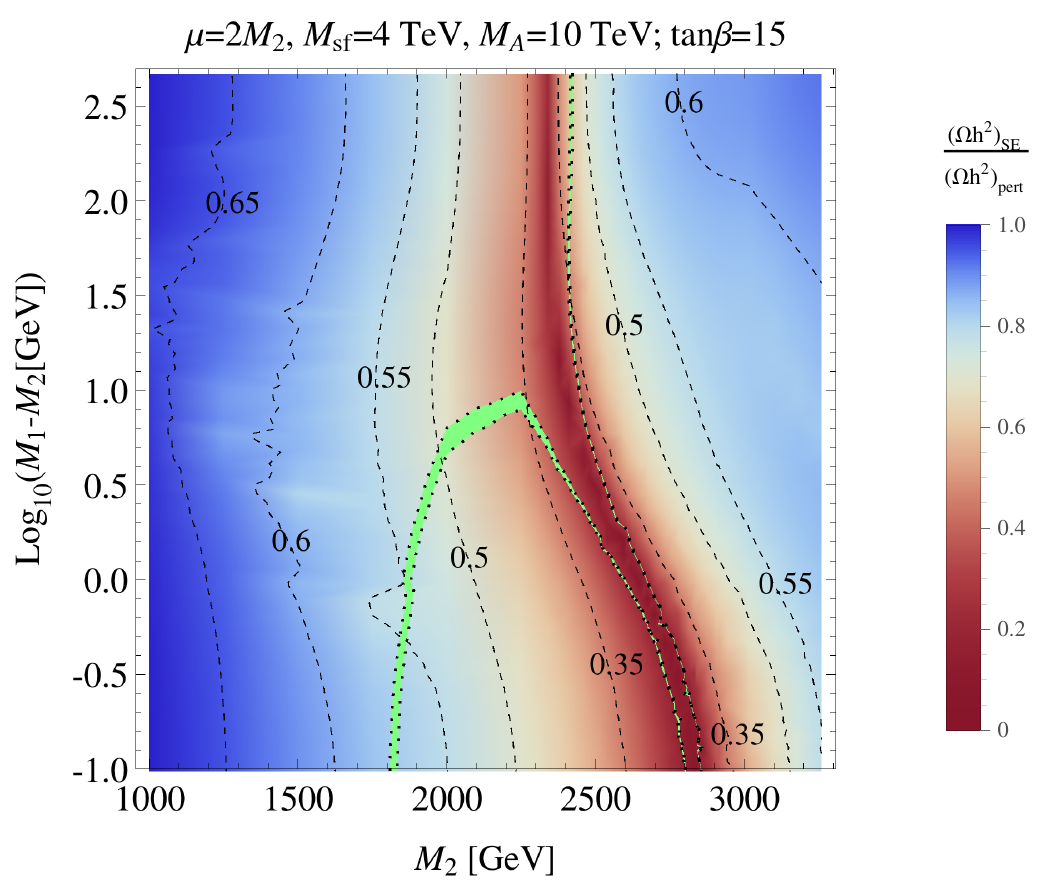}
 \caption{The impact of the Sommerfeld enhancement of the relic density shown as a density map as well as via the black dashed contours. The (green) band indicates the region within $2\sigma$ of the observed dark matter abundance. Other parameters are as indicated, with $A_i=8$~TeV and $X_t$ is fixed by the measured Higgs mass.\label{fig:m2-m1_SE}}
\end{figure}
 
Contours of constant relic density in the $M_2$ vs.~$(M_1-M_2)$ plane for $M_1>0$, and both the perturbative and Sommerfeld enhanced case
are displayed in Fig.~\ref{fig:m2-m1_msf4TeV}. Note that the logarithmic scale for the vertical axis, chosen due 
to the weak mixing between the bino and the wino, changes the appearance of the resonance
with respect to the Higgsino case. 
As one increases the bino component the mass of the LSP resulting in the correct relic density is approximately 1500 GeV rather than $1800$ GeV for the perturbative case. This changes rather dramatically when the Sommerfeld enhancement is taken into account, notably for strong mixing, i.e. $M_1-M_2\lesssim 10$ GeV, there are three values of $M_2$ which give the correct relic density. One can interpret the larger two of these values as a result of the resonance in the Sommerfeld enhancement.

In Fig.~\ref{fig:m2-m1_SE} we study the ratio between the relic densities shown in Fig.~\ref{fig:m2-m1_msf4TeV}, in terms of a density plot with contour lines overlaid. The correct relic density for the full calculation including the Sommerfeld enhancement is highlighted by the (green) band. We observe the maximal effect of the Sommerfeld enhancement is in fact in the region where the relic density agrees with observation, in particular when the difference between $M_1$ and $M_2$ is below approximately 10 GeV, the Sommerfeld enhanced relic density in agreement with that observed is three times smaller than the perturbative result at the same parameter values. Note that over the entire region covered by the plot the effect of the Sommerfeld enhancement is greater than 30\%.

\begin{figure}
  \centering
 \includegraphics[width=.5\textwidth]{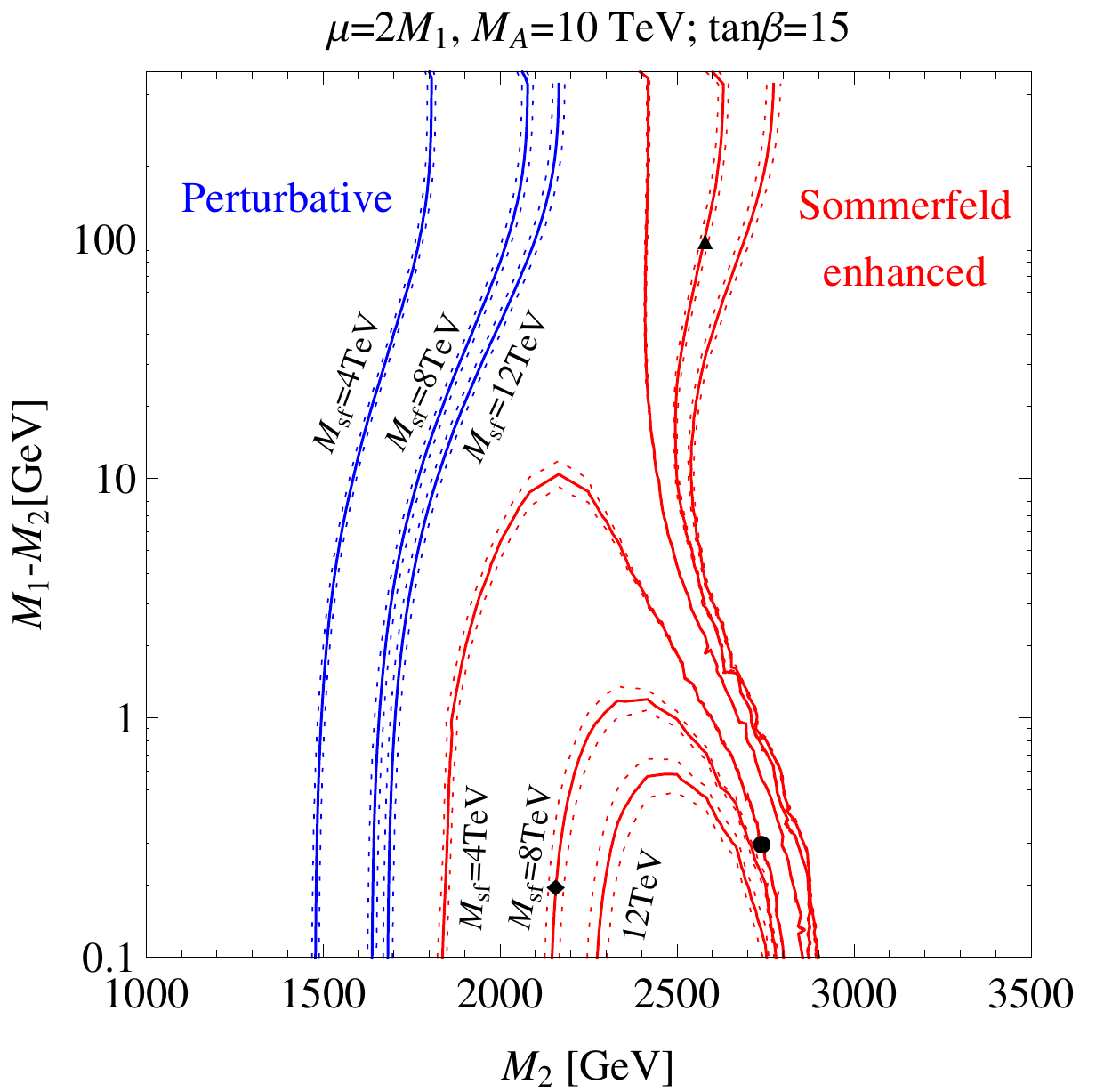}
 \caption{Contours of correct relic density: perturbative (blue) and Sommerfeld enhanced (red) are shown for 3 different values of the sfermion mass parameter. Other parameters are as indicated, with $A_i=8$~TeV and $X_t$ is fixed by the measured Higgs mass. The black markers denote the three points studied in Sec.~\ref{sec:Res_other}.\label{fig:m2-m1}}
\end{figure}

\begin{figure}
  \centering
 \includegraphics[width=.5\textwidth]{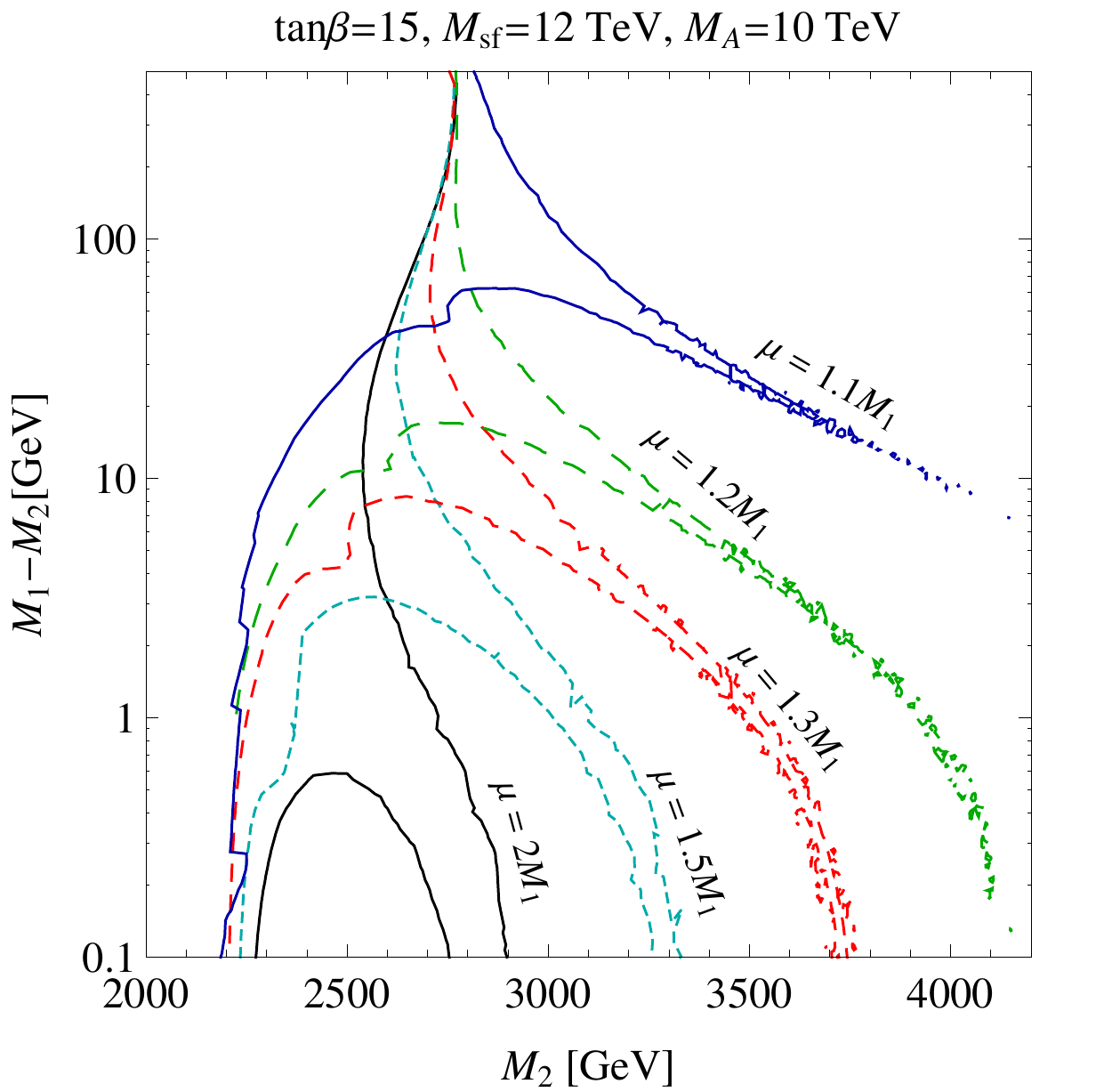}
 \caption{Contours of correct relic density for different values of $\mu$. Other parameters are as indicated, with $A_i=8$~TeV and $X_t$ is fixed by the measured Higgs mass.  \label{fig:m2-m1_mu}}
\end{figure}

In Fig.~\ref{fig:m2-m1} contours with the correct relic abundance for three 
choices of the sfermion mass parameter are shown. The results again resemble the Higgsino admixture case, up to differences already commented on. Note that in the region around the resonance, the effect of the sfermion masses is less pronounced than elsewhere.
One observes that the lowest mixed wino-bino neutralino mass giving the observed 
relic density is around 1.8 TeV for $\msf=4$ TeV, marginally higher than the wino-Higgsino case. The highest value is 2.9 TeV (for $\msf=12$ TeV) compared to 3.3 TeV in Fig.~\ref{fig:m2-mu}. However, as can be seen in  Fig.~\ref{fig:m2-m1_mu}, the highest value of $M_2$ resulting in the correct relic density is strongly dependent on the value of the $\mu$ parameter, as this mediates the mixing. This dependence is demonstrated via contours for five different choices of $\mu$. The contour for $\mu=1.1 M_1$ bears a closer resemblance to the wino-Higgsino case, as suggested earlier. Note that as $\mu$ decreases, the lightest chargino-neutralino mass splitting for a given point in the plane 
increases,\footnote{See Fig.~\ref{fig:masssplit} in the Appendix.} resulting in the resonance moving to higher values of $\mlsp$. 
Due to the presence of the resonance, it appears that by making an appropriate choice in $\mu$  and $M_1$ the entire region could be covered, at least for values of $M_2$ from 2100 to 4200 GeV if not even higher. All these points would be on or around the resonance, having implications for Indirect Detection. Moreover, interestingly adding a bino component to the LSP can extend the possible neutralino masses giving observed dark matter abundance up to and even beyond 4.1 TeV.

\subsection{Residual dependence on other parameters}
\label{sec:Res_other}

In the previous Secs.~\ref{sec:Res_sfer} to \ref{sec:Res_B} certain parameters were fixed in order to obtain a clearer understanding of the dependence of the results on the central parameters $M_1$, $M_2$, $\mu$, $\msf$, as well as 
on $\mao$. 
However, it is important to confirm whether these are indeed the most relevant, and to investigate the effect of the other parameters, e.g. $\tanb$, which was so far neglected.

\begin{figure}[t]
  \centering
   \includegraphics[width=.5\textwidth]{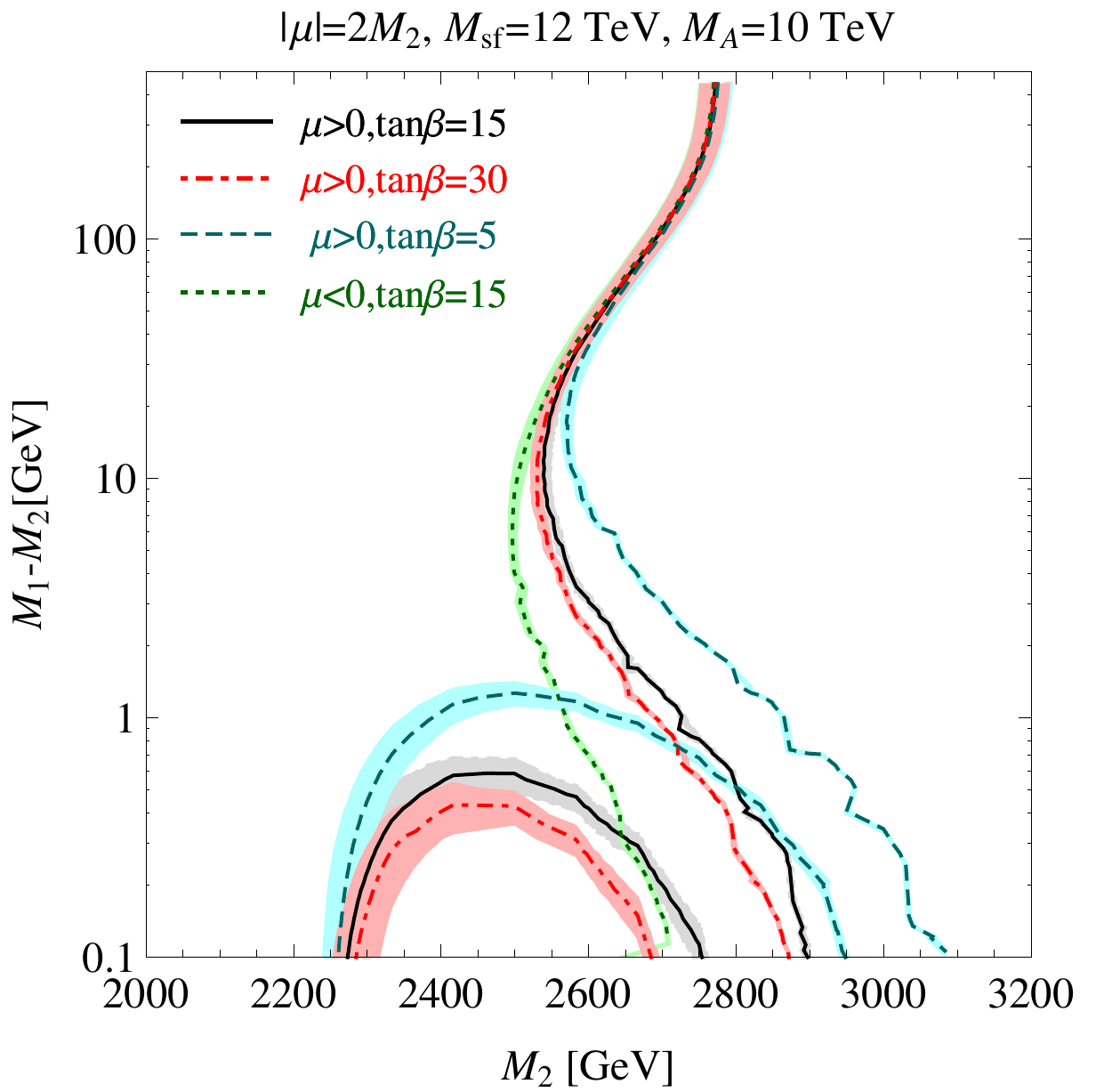}
 \caption{The contours of correct relic density for different choices of $\tanb$ and the sign of $\mu$. Other parameters are as indicated, with $A_i=8$~TeV and $X_t$ is fixed by the measured Higgs mass.\label{fig:m2-m1_other}
}
\end{figure}

One case in which additional parameters may play a significant role is when the lightest neutralino is a wino-bino mixture as in Sec.~\ref{sec:Res_B}.
This is because, as seen in Eqs.~\eqref{eq:binomix} to 
(\ref{eq:Bino-Wino-Masses2b}), the mixing of the wino with the bino, and the splitting between the lightest neutralino and
chargino is sensitive to $|\mu|$, the sign of $\mu$ and $\tanb$. The dependence on $\mu$ was already examined in Sec.~\ref{sec:Res_B}, and the results
can be found in Fig.~\ref{fig:m2-m1_mu}.
Here we further consider the effect of the sign of $\mu$ and the choice of $\tanb$. Our results can be found in Fig.~\ref{fig:m2-m1_other}. The
benchmark choice for the results presented in previous subsections was $\tanb=15$ and $\mu>0$; in addition here we consider $\mu>0$ with $\tanb=5,30$
and $\mu<0$ with $\tanb=15$. Large deviations from the benchmark scenario are seen in the resonant region. This can be understood by examining the
expressions for the mass splitting between the lightest neutralino and chargino in Eqs.~(\ref{eq:Bino-Wino-Masses2a}) and (\ref{eq:Bino-Wino-Masses2b}). 
The splitting is seen to
increase when $\tanb$ decreases, and also when $\mu>0$ compared to $\mu<0$, resulting in the position of the resonance moving towards higher values of
$\mlsp$, i.e.~the correct relic density is observed for higher $M_2$.

\begin{figure}[p]
  \centering
 \includegraphics[width=.4\textwidth]{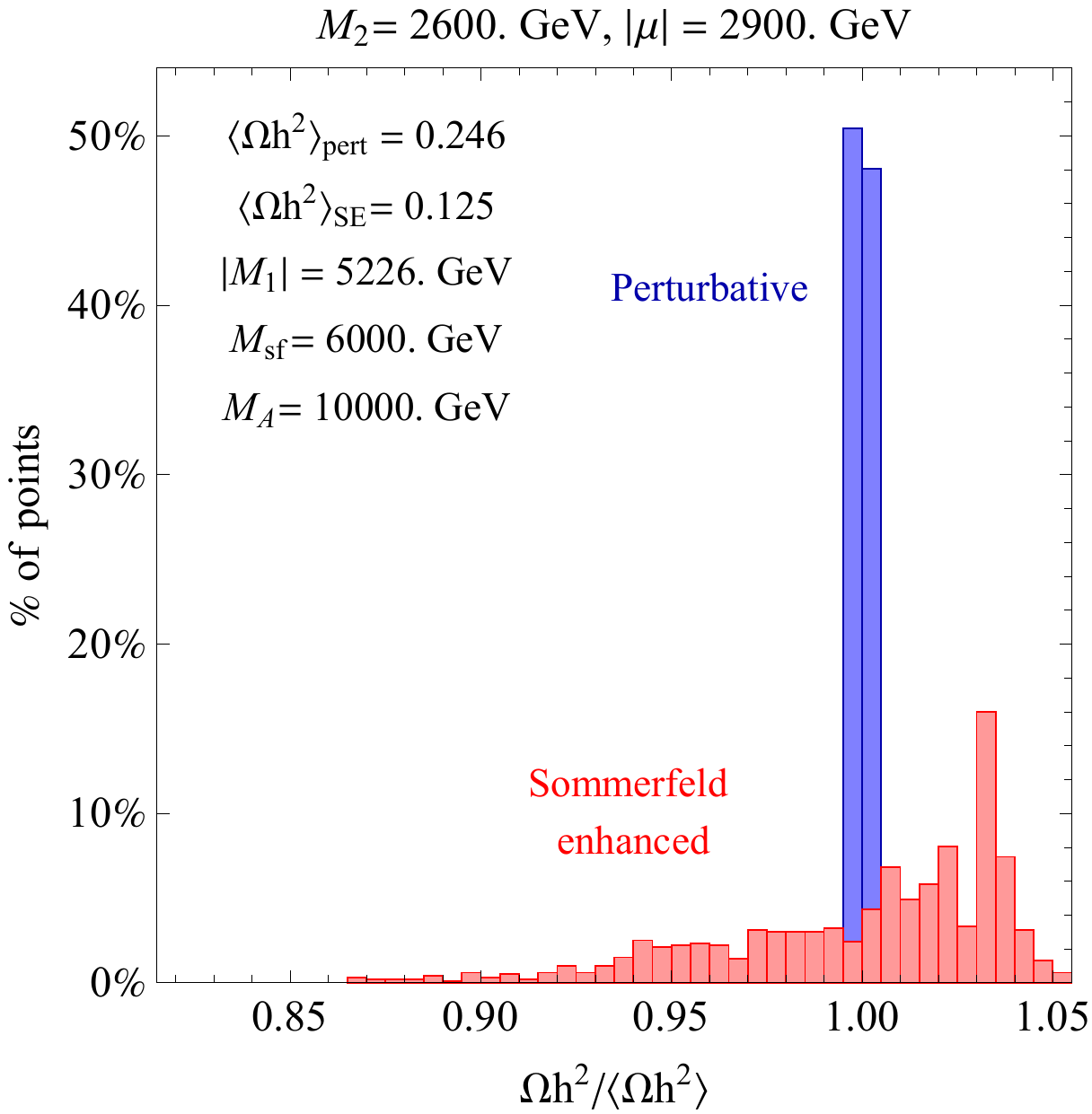}\hspace{.5cm}\includegraphics[width=.4\textwidth]{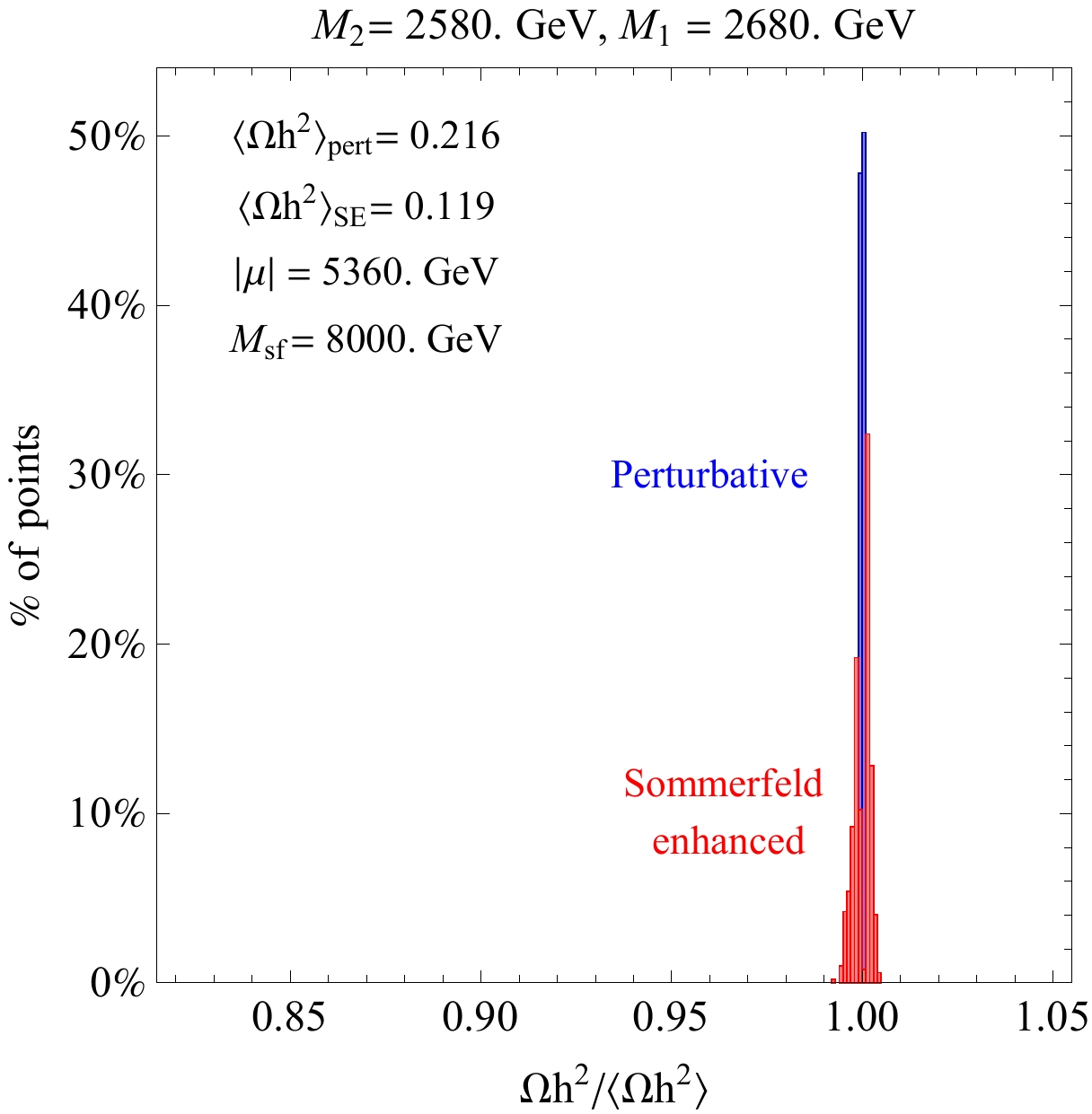}\\
   \includegraphics[width=.4\textwidth]{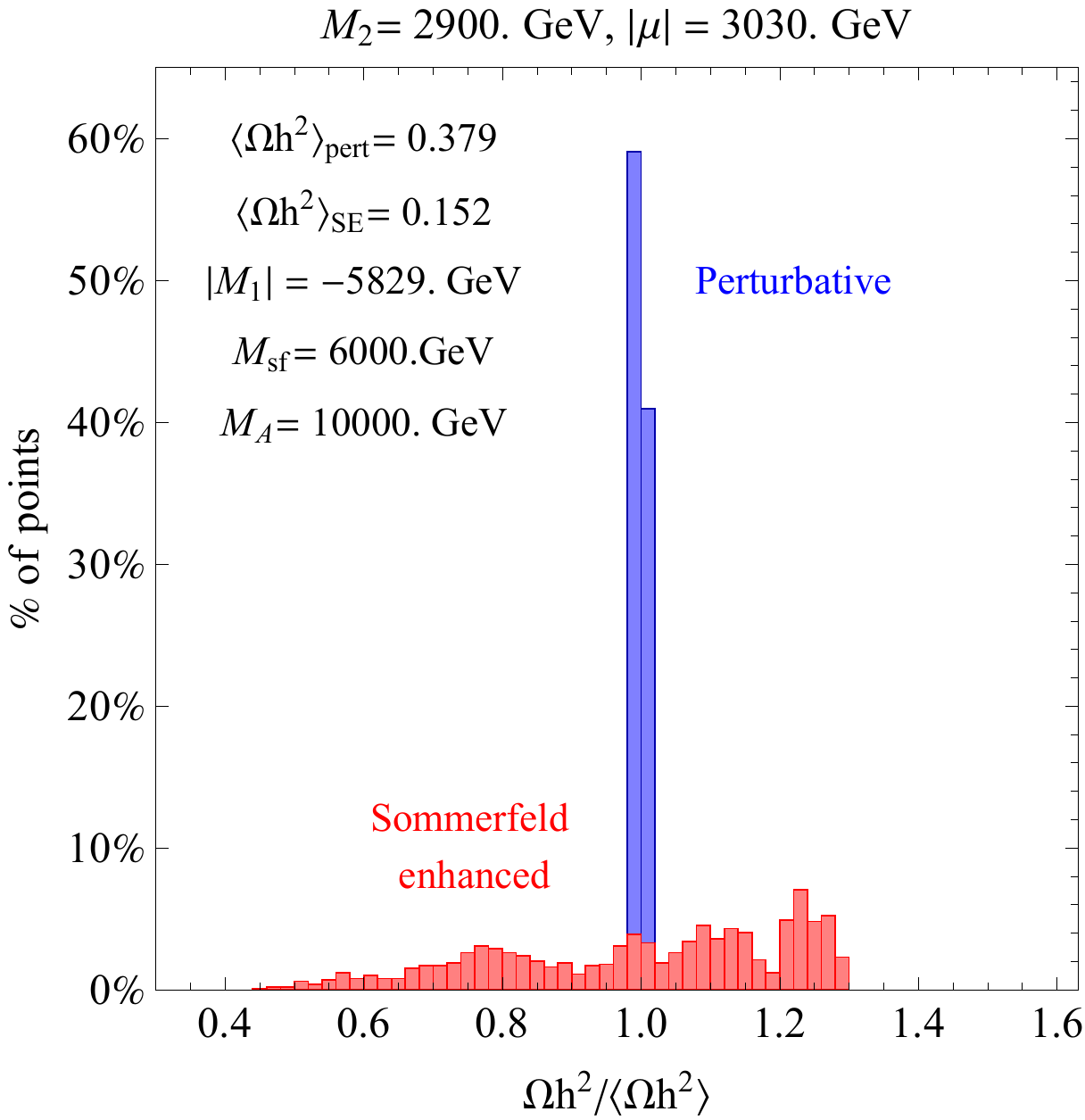}\hspace{.5cm}\includegraphics[width=.4\textwidth]{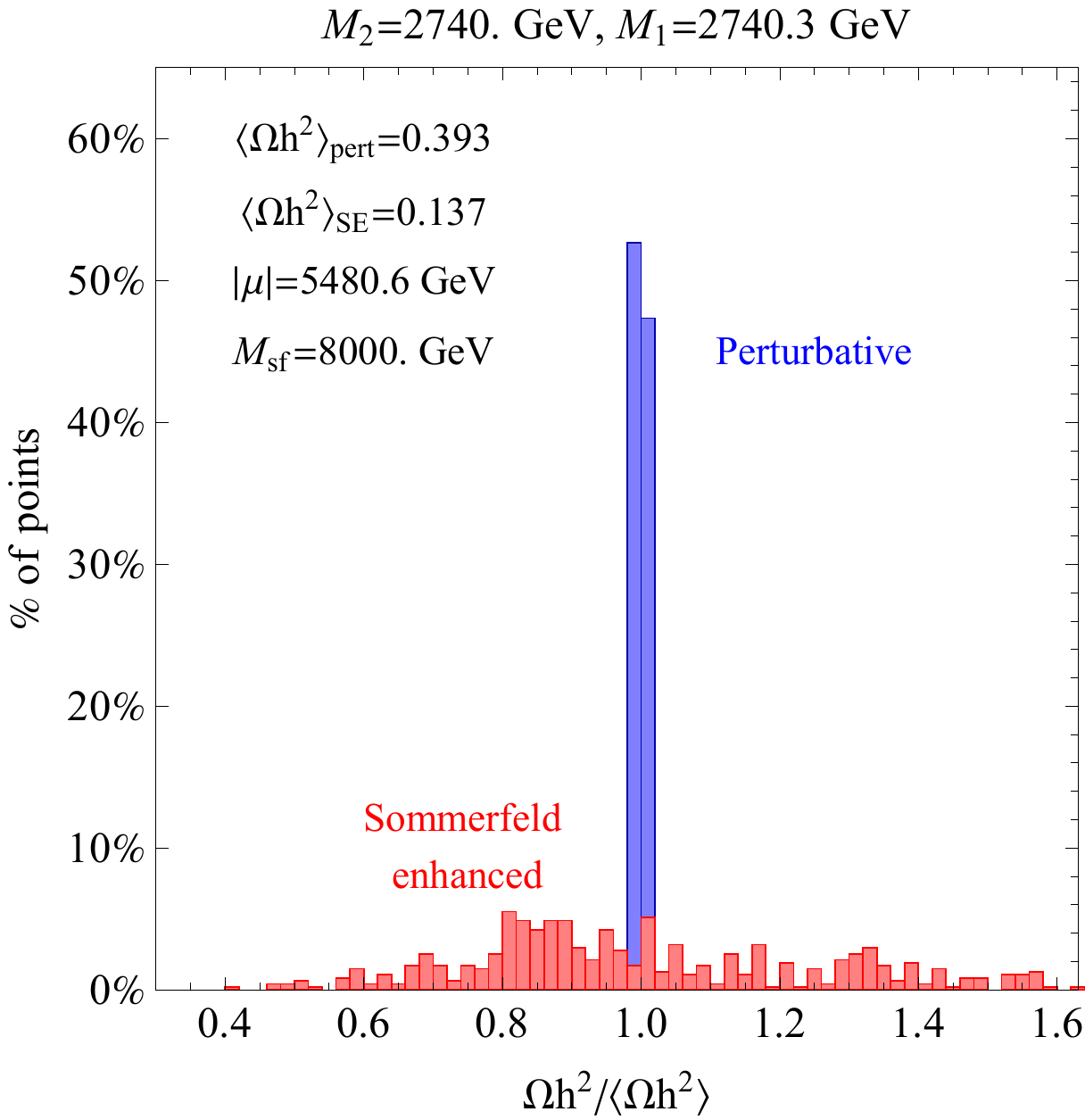}\\
  \includegraphics[width=.4\textwidth]{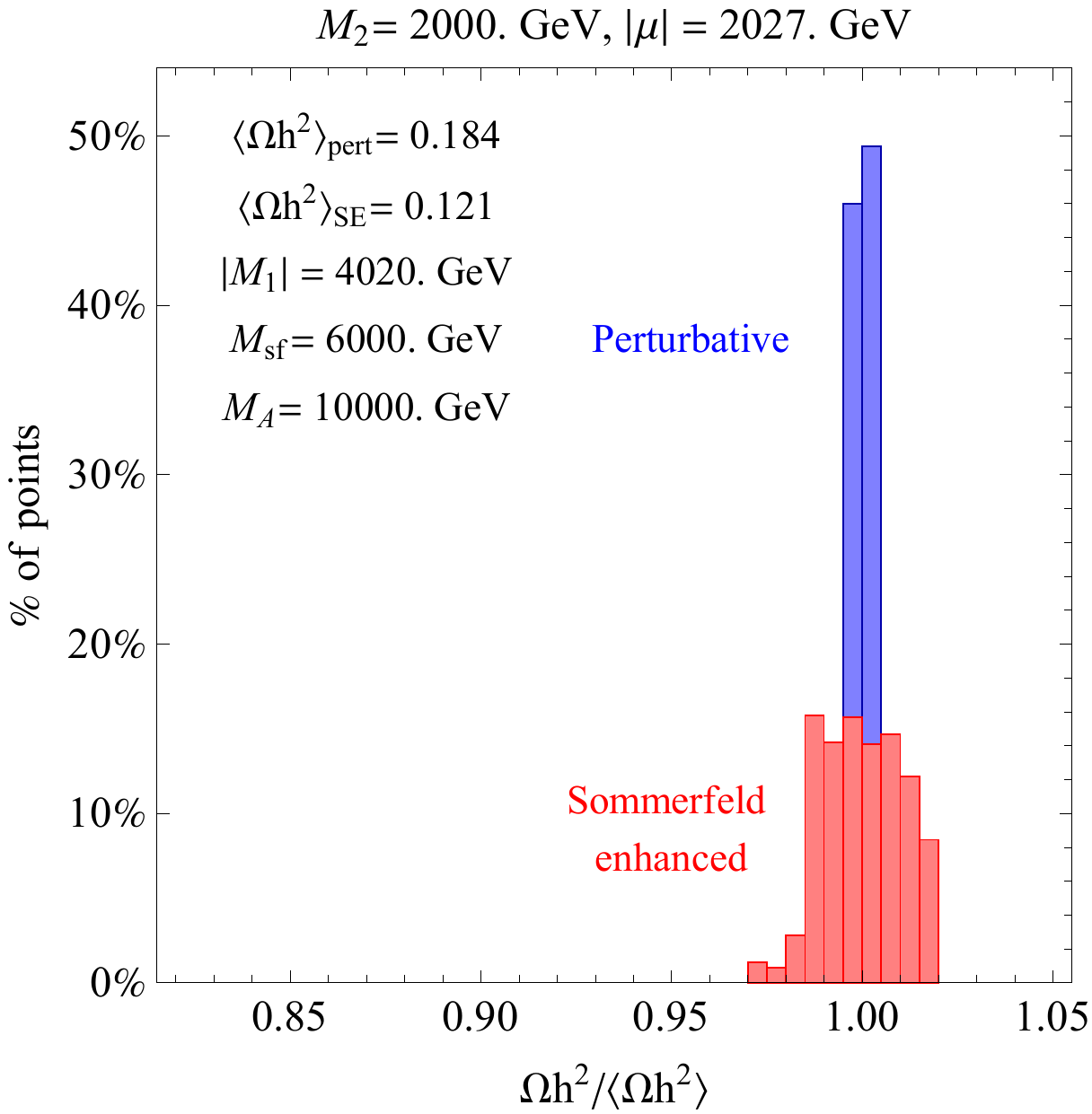}\hspace{.5cm}\includegraphics[width=.4\textwidth]{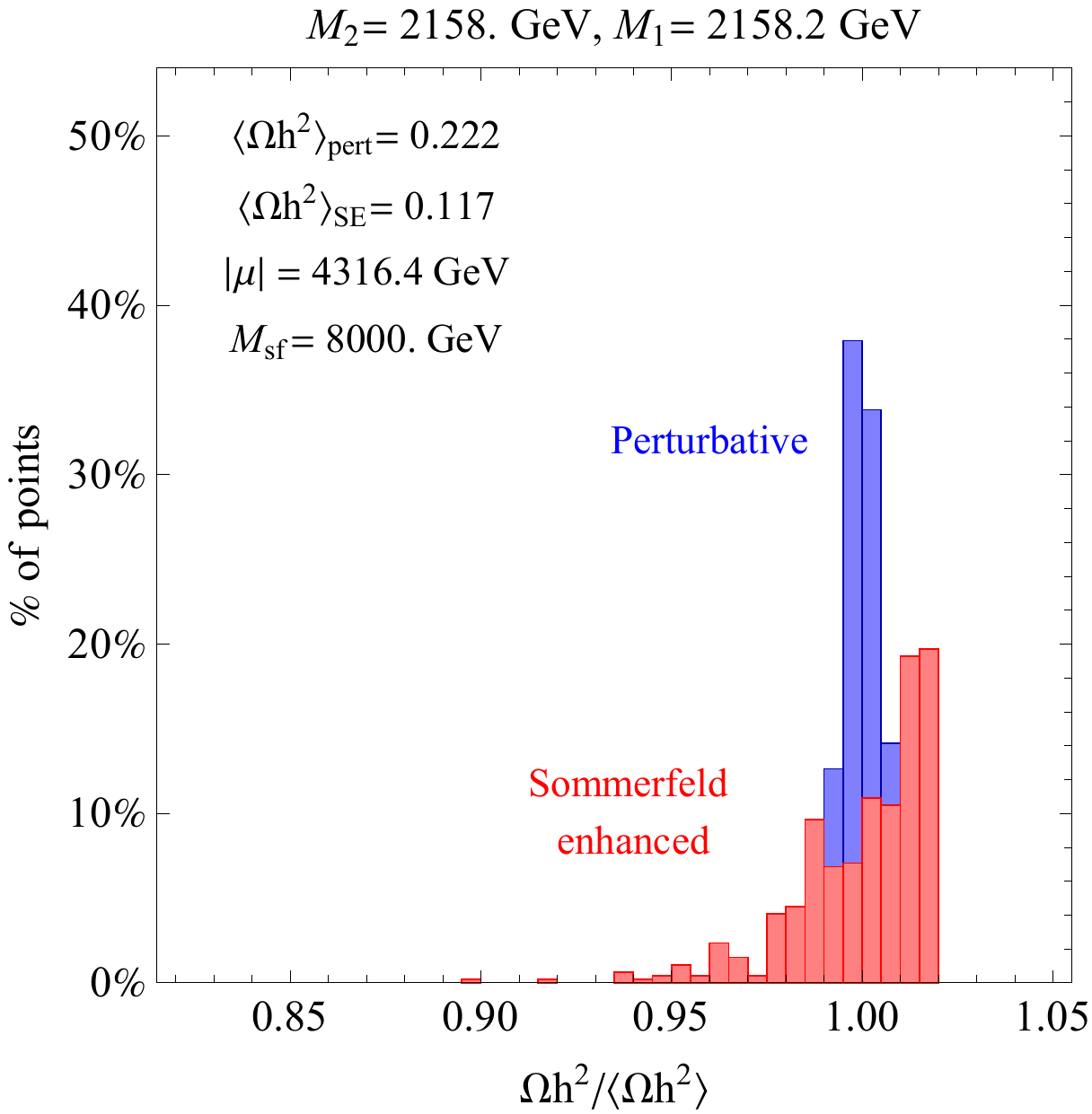}
 \caption{Histograms showing the impact of the remaining parameters on the relic density for wino-like LSPs with a varying Higgsino (left) and bino (right) admixtures. The relic density is normalised to the mean for each case respectively.\label{fig:His_Higgsino}}
\end{figure}

We further examine the sensitivity to the remaining parameters for both the 
cases of mixed wino-Higgsino and wino-bino LSPs in 
Fig.~\ref{fig:His_Higgsino}. To this end we show histograms of the 
percentage of points in bins of $\Omega h^2/\langle\Omega h^2\rangle$ where $\langle\Omega
h^2\rangle$ is the mean value of the relic density, both for the perturbative and the full calculation. We choose six wino-like points, three of which
contain bino admixtures and three Higgsino admixtures of varying degree. 
The points in the $(M_2,\mu-M_2)$ and $(M_2,M_1-M_2)$ planes corresponding to 
the left and right hand plots from top to bottom are marked (up to signs 
in $\mu$, $M_1$) in Figs.~\ref{fig:m2-mu} and \ref{fig:m2-m1} by the 
triangle, circle and diamond, respectively, i.e. the Higgsino or bino  
component increases from top to
bottom panels. For each of these points we fix the values of the central parameters as indicated in Fig.~\ref{fig:His_Higgsino} and compute the relic density for 1000 different realisations of the remaining parameters $\mao$ (for the 
case of wino-bino mixing), 
$\tan\beta$ and $A_f$ within the ranges given in Tab.~\ref{tab:params}, with $M_3$ 
fixed to $3\,M_2$, assuming a uniform distribution before the constraints 
are imposed.

The perturbative results are shown by the dark-grey/blue histograms. We see that in all cases the distribution is strongly peaked near the central value with the variation of order of at most a few per cent, and that the distribution widens with the departure from the pure-wino limit. The situation changes when considering the full result (light-grey/red histograms), as all the distributions become broader and asymmetric. The position of the resonance in the Sommerfeld effect is greatly sensitive to values of the neutralino mass, couplings and the chargino. Therefore, slight changes in these values caused by different choice of remaining less relevant MSSM parameters, especially close to the resonance, can lead to observable differences in the relic density. Indeed, the broadening of the full result with respect to the perturbative one is strongest in the middle panel (due to the vicinity of the circle benchmark point to the resonance) and in the upper left-hand plot being not far from the resonance as well. The asymmetry in the distributions is caused by the fact that deviations around central parameters may go towards or away from the resonance, leading to larger or smaller Sommerfeld effects respectively. The bottom line of this analysis is that away from the resonance the residual MSSM parameters have a very mild impact, justifying our choice of central parameters, while in the vicinity of the resonance regions the variation is very significant. 

\begin{figure}
\centering
\includegraphics[width=.6\textwidth]{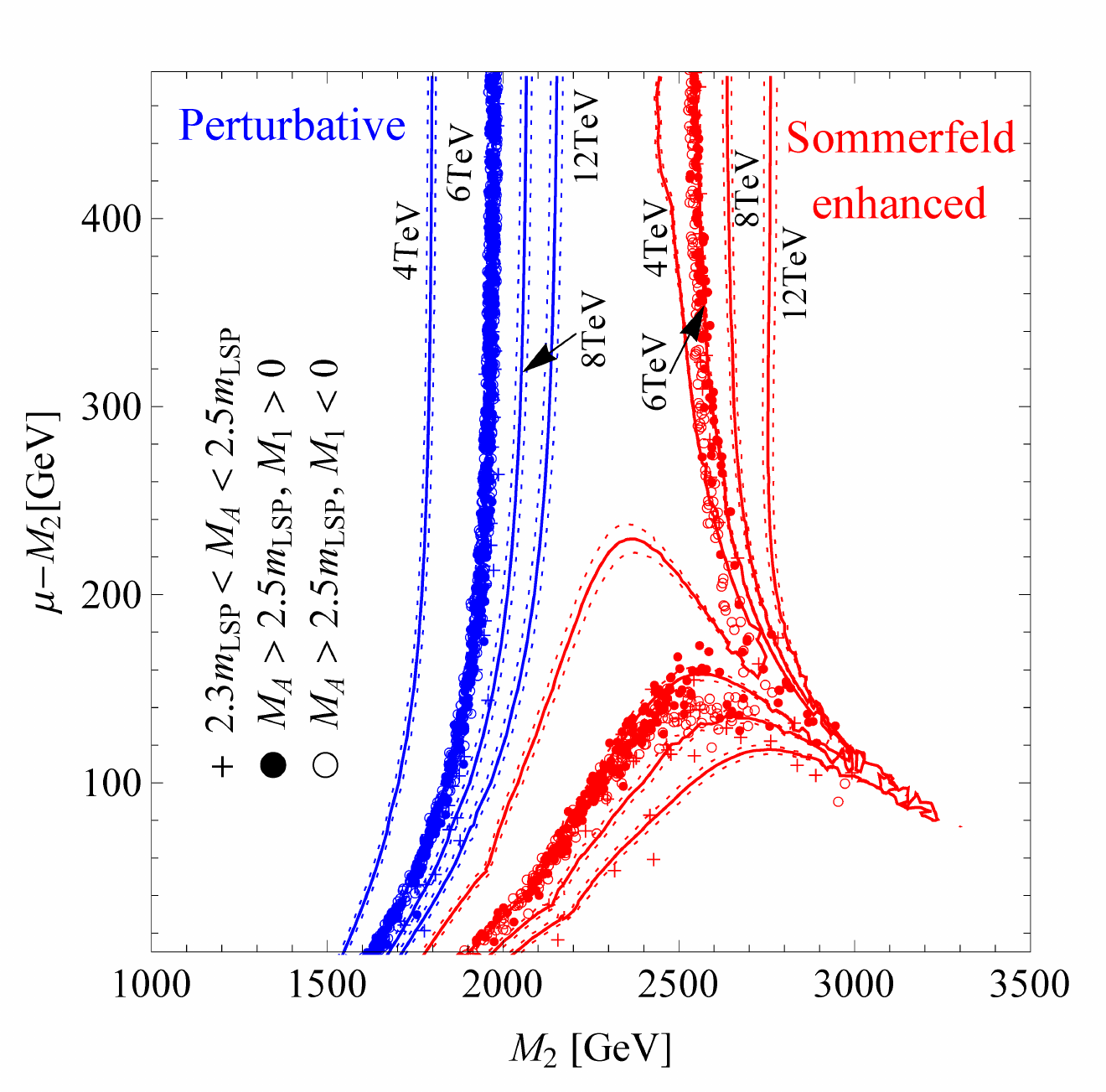}
\\\vspace{.4cm}
\centering\includegraphics[width=.6\textwidth]
{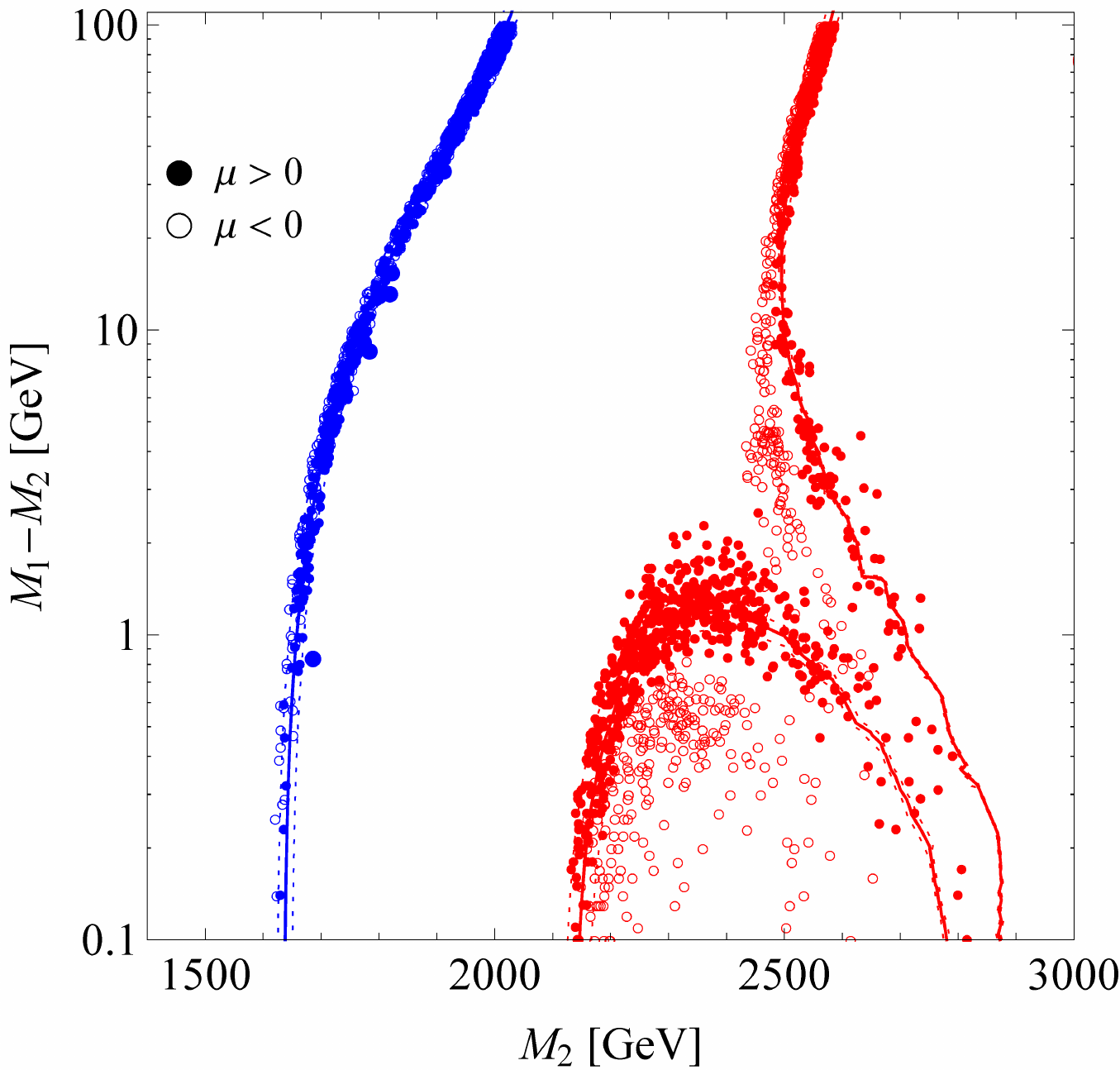}
\caption{Plots showing points satisfying the relic density constraint obtained 
on varying the parameters $\tan\beta$, $\mao$ and $A_f$ for wino-like LSPs 
with varying Higgsino (upper) and bino (lower) admixtures. The points are 
overlaid on contours for fixed values of these parameters and $\msf$ as 
indicated.\label{fig:Add-Pars}}
\end{figure}

To study the dependence on the residual parameters even further, we have generated a
large number of points (50000 and 90000 for the Higgsino and bino case,
respectively), where we considered the wino mass in the range $\, M_2 \in \{1,3.5\}~\tev$ and (different from Tab.~\ref{tab:params} and the analyses in the 
previous sections) fixed the gluino mass parameter via $M_3=2M_2$. 
The sfermion masses were fixed to the values given below, 
but we varied all other
parameters in the following ranges:
\begin{equation}
\frac{X_t}{\msf} \in \{0.5,3\},\, A_{f}\in \{0,8\}~\tev,\, 
\mao \in \{1,10\}~\tev,\, \tan \beta \in\{5,30\},
\end{equation}
where $f$ in $A_f$ includes all fermions except the top. 
In addition, for the Higgsino case we chose:
\begin{equation}
\msf=6~\tev,\, |M_1|=2.01 M_2,\, M_2 \in \{1,3.5\}~\tev,\, 
\mu \in \{M_2,M_2+0.5~\tev\},
\end{equation}
and for the bino case:
\begin{equation}
\msf=8~\tev,\, M_1 \in \{M_2, M_2+0.1~\tev\},\, |\mu| = 2 M_1.
\end{equation}
From the generated points, we selected those where either the perturbative
or the Sommerfeld enhanced relic density was found to be between 0.1168 and 0.1208,
i.e.\ within two sigma of the central value. In Fig.~\ref{fig:Add-Pars} we overlay these points on
the relevant plots shown earlier, Figs.~\ref{fig:m2-mu} and \ref{fig:m2-m1} for the Higgsino (upper plot) and
bino (lower plot) case, respectively.

We observe that those points for which the perturbative relic density lies 
within 2$\sigma$ of the central value are located very close to the 
respective sfermion mass contours, the spread of the points being comparable to twice the width of the 1$\sigma$ contours. We conclude
that the dependence on the residual parameters is very mild in this case.
As could be expected, when including the Sommerfeld enhancement, the
residual parameters can have a larger effect, especially close to
the resonance. In order to investigate this effect further, for the 
wino-Higgsino case we have divided these points according to whether 
$2.3\mlsp<\mao<2.5\mlsp$, $\mao>2.5\mlsp$ and $M_1>0$, or $\mao>2.5\mlsp$ 
and $M_1<0$. These are indicated in Fig.~\ref{fig:Add-Pars} by the
cross, filled circle and open circle respectively. The former division is made in order to isolate those points in proximity to the
heavy Higgs funnel region and the latter due to the effect of the sign on $M_1$ on the lightest neutralino--chargino mass splitting. We do not plot the 
points with $\mao < 1.7\mlsp$ in this case. 
For the wino-bino case we separated the points according to whether $\mu>0$ or $\mu<0$, indicated in Fig.~\ref{fig:Add-Pars} by the filled circle and
open circle respectively, as the sign of $\mu$ also plays a role in the size 
of the lightest neutralino--chargino mass splitting.

The effect of the residual parameters is sub-dominant with respect to e.g.\ that of the sfermion masses, but both the sign of $\mu$ and $M_1$ are
seen to play a role in the resonance region for the wino-bino and wino-Higgsino cases, respectively. This can be understood in terms of the expressions for
the mass difference between the lightest chargino and neutralino in Eqs.~(\ref{eq:Higgsino-Wino-Masses1}), (\ref{eq:Higgsino-Wino-Masses2}) and
(\ref{eq:Bino-Wino-Masses2a}), (\ref{eq:Bino-Wino-Masses2b}) to which the resonance is sensitive. As the splitting increases the position of the resonance moves towards higher values of $\mlsp$.
Whether the heavy Higgs is below, above, or, in particular, close to the excluded window also has a noticeable effect for the case of wino-Higgsino mixing, and this
extends beyond the resonance region and holds for the perturbative case as well. This is because for states with larger mixing the coupling to the
heavy Higgs is enhanced, and therefore when $\mao$ decreases the s-channel 
annihilation cross section increases, and one has to go to higher values of $M_2$ to obtain the
correct relic density. This is not relevant for the wino-bino case, where 
the dependence on the value of $\mao$ is negligible.

To summarise, we find that the assumption that our results of the previous sections were more or less independent of certain parameters was largely
justified. Only for the wino-Higgsino case there is some dependence on the value of $\mao$, and in the resonance region the sensitivity to
these parameters is somewhat enhanced, particularly to the values of $\tanb$ and the sign of $\mu$ for the case of wino--bino mixing and $M_1$ for the
case of wino--Higgsino mixing.


\section{Summary}
\label{sec:summary}

We have studied the Sommerfeld effect on the relic density of neutralino dark matter beyond the pure-wino limit. This involved a scan of parameter space for three scenarios where the lightest neutralino contained a large wino component: one with non-decoupled sfermions, and the remaining with either a Higgsino or bino admixture. We aimed to determine how in these scenarios (a) the mass of the LSP where the relic density constraint is satisfied and (b) the size of the Sommerfeld enhancement is altered in comparison to the pure-wino case.

The calculation of the Sommerfeld enhancement for the scenarios in question required a consistent treatment of mixed neutralinos including multiple co-annihilation channels and off-diagonal contributions as well as of $\mathcal{O}(v^2)$ contributions. As the Sommerfeld effect, in particular the position of the resonance, depends strongly on the mass splittings between the neutralinos and charginos, we used a dedicated on-shell renormalisation scheme scheme for one-loop masses. As the relic density depends strongly on the precise value of the gauge coupling, we adopted running couplings. Further we have argued that the size of thermal corrections is sufficiently below the uncertainty of our calculation, i.e.~the percent level, that they can be neglected. Finally, all points in MSSM parameter space considered were checked for consistency with current experimental measurements. Our calculation was carried out by a code which will be made available to the public: this will be presented in more detail in a separate publication.

Due to t-channel interference, we found that when the sfermions are non-decoupled they reduce the annihilation cross section such that the relic density constraint is satisfied at lower values of $m_{\rm LSP}$, from 2.9 TeV for decoupled sfermions down to 2.4 TeV. For the mixed neutralino scenarios we found a much larger dependence than expected on those parameters affecting the mass splitting between the lightest chargino and neutralino. This was particularly evident in the mixed bino-wino region, where the position of the resonance was seen to be sensitive to $\mu$, $\tan \beta$ and the sign of $\mu$. As the splitting increases we observed that the resonance lies at higher values of $m_{\rm LSP}$. This led to a large range of neutralino masses from 1.8 to beyond 4 TeV satisfying the relic density constraint. For the Higgsino-wino mixed region the position of the resonance depends primarily on $\mu-M_2$, but whether the mass of the heavy Higgs boson lies below or above $2\,m_{\rm LSP}$ also plays a significant role. Here we found values of $m_{\rm LSP}$ ranging from 1.7 to 3.3 TeV. 

This is the first time that the sensitivity of the Sommerfeld enhancement to MSSM parameters for mixed neutralinos has been studied systematically and to such accuracy, and the large range of possible $m_{\rm LSP}$ masses providing the correct relic density was previously unknown. In most of the cases the Sommerfeld effect changes the relic density by a factor of two or even higher relative to the tree-level computation. This underscores the fact that the relic density of TeV scale MSSM dark matter can usually not be predicted correctly without accounting for this effect.

In light of these results a re-investigation of the bounds on the Sommerfeld enhanced scenarios coming from Indirect Detection experiments is imperative. It is likely that so far unexplored regions exist, sufficiently far away from the resonance that they are not excluded, but with Sommerfeld enhanced annihilation cross sections which could be probed by upcoming experiments, e.g.~the Cherenkov Telescope Array (CTA). This will be the subject of a dedicated study in the near future.

\subsubsection*{Acknowledgements}

This work is supported in part by the Gottfried Wilhelm Leibniz programme 
of the Deutsche Forschungsgemeinschaft (DFG) and the Excellence Cluster 
``Origin and Structure of the Universe'' at Technische Universit\"at 
M\"unchen. We further 
gratefully acknowledge that part of this work was performed using the 
facilities of the Computational Center for Particle and Astrophysics (C2PAP) 
of the Excellence Cluster.


\appendix

\section{Thermal effects}
\label{app:thermal}

Freeze-out of dark matter begins when the Universe has cooled to 
the temperature $T_f\simeq m_\chi/20$. For the dark matter masses 
considered in this paper $T_f$ is in the range $50 \ldots 200\,$GeV, 
which includes the temperature $T_c \approx 165\,$GeV 
of the electroweak phase transition. Above the critical temperature $T_c$, 
the Lagrangian mass of the electroweak gauge bosons vanishes and 
so does the neutralino--chargino mass splitting in the wino-like  
region. Furthermore, large thermal masses and mass splittings 
may be generated. While thermal effects on the short-distance 
annihilation process are small \cite{Beneke:2014gla}, the gauge-boson 
mass determines the range of the potential, which is an important 
quantity for the Sommerfeld effect. Furthermore, 
the mass splitting of the lightest 
chargino and neutralino influences the location of the Sommerfeld 
resonance. In the following, we investigate the thermal modification 
of the gauge boson mass and neutralino--chargino 
mass splitting through a combination 
of estimates, analytical calculations and numerical checks.

The relevant temperature range for this investigation is limited 
from above by $T \simeq m_\chi/20$, when freeze-out begins, which 
allows us to treat $T/m_\chi$ as small. The temperature of the Universe 
together with the Boltzmann distribution sets the characteristic 
scale of the three-momentum of the scattering 
dark matter particles to $|\vec{p}\,|
\sim (m_\chi T)^{1/2}$. The Sommerfeld effect 
is caused by ladder diagrams with loop momentum $k$ satisfying 
$k^0 \ll |\vec{k}|\ll m_\chi$ where $k^0$ is determined by the pole of 
the dark matter particle propagator. The characteristic scale of 
$\vec{k}$ is also $(m_\chi T)^{1/2}$ until 
$m_\chi T\sim m_W^2$, where $m_W$ is the mass of the exchanged electroweak 
gauge boson (zero for the photon), at which point the Sommerfeld 
enhancement saturates. When the Universe cools  
below $T_s\simeq m_W^2/m_\chi$, the external momentum $p$ 
continues to decrease while $\vec{k}\sim m_W$ remains constant, and the 
thermal modification of the Sommerfeld effect fades out. Hence the 
temperatures of interest are limited from below by $T_s$ in the range from 
1 to $4\,$GeV. An exception is 
the Sommerfeld enhancement due to photon exchange between the charginos, 
whose effect on the relic density turns off only when the charginos 
decouple from the thermal plasma at a temperature set by the 
neutralino--chargino mass difference.

\subsection{Higgs vacuum expectation value}

We approximate the temperature dependence of the Higgs field 
vacuum expectation value by 
\begin{equation}
\label{vT}
v(T) = v \,\sqrt{1-\frac{T^2}{T_c^2}} \qquad (T<T_c)
\end{equation}
and zero above $T_c$. The critical temperature is taken to be 
$T_c=165\,$GeV, as follows from the effective potential given in 
Ref.~\cite{Quiros:1999jp}.
The expansion of the Universe proceeds adiabatically such that the 
particle masses are given by the standard expressions with 
the instantaneous value $v(T)$.  

\subsubsection{Gauge boson masses}

This implies the temperature-dependent 
mass (squared) $m_W^2(T) = m_W^2 + [\delta m_W^2]_{\rm vev}$ 
for the electroweak gauge bosons, where 
\begin{equation}
\label{MWT}
[\delta m_W^2]_{\rm vev} = -\frac{m_W^2 T^2}{T_c^2},
\end{equation}
and $m_W^2(T) = 0$ above $T_c$. Similarly for the $Z$ boson mass 
$m_Z(T)$.

\subsubsection{Lightest neutralino--chargino mass difference}

The temperature dependence of the neutralino and chargino masses 
is not by itself of interest, since always $m_\chi \gg T$. However, 
the temperature dependence of small mass splittings must be considered, 
since the mass splitting determines, for instance, the location of 
the Sommerfeld resonance, and further appears in the nearly on-shell 
propagator of the two-neutralino/chargino state in the ladder diagrams. 

Close to the pure-wino limit the neutralino--chargino 
mass difference is dominated by the radiatively induced splitting, 
which in the pure-wino limit is given by 
\begin{equation}
[\delta m_{+0}]_{\rm radiative} = 
\frac{1-c_w}{2}\,\alpha_2 m_W \approx 158\,\mbox{MeV}.
\end{equation}
The expression refers to the approximation $M_Z \ll m_\chi$ 
and the numerical value employs the SU(2) coupling $\alpha_2(m_\chi) 
= 0.032810$ at the scale $m_\chi=2.5\,$TeV. Whenever the radiative 
mass splitting dominates over the tree-level splitting, it changes very 
little compared to the pure-wino value. We therefore assume that 
it is proportional to $v$, and implement the temperature-dependence 
by multiplying the zero-temperature radiative contribution 
to the mass difference with $v(T)/v$.

\begin{figure}
\centering
\includegraphics[width=.48\textwidth]{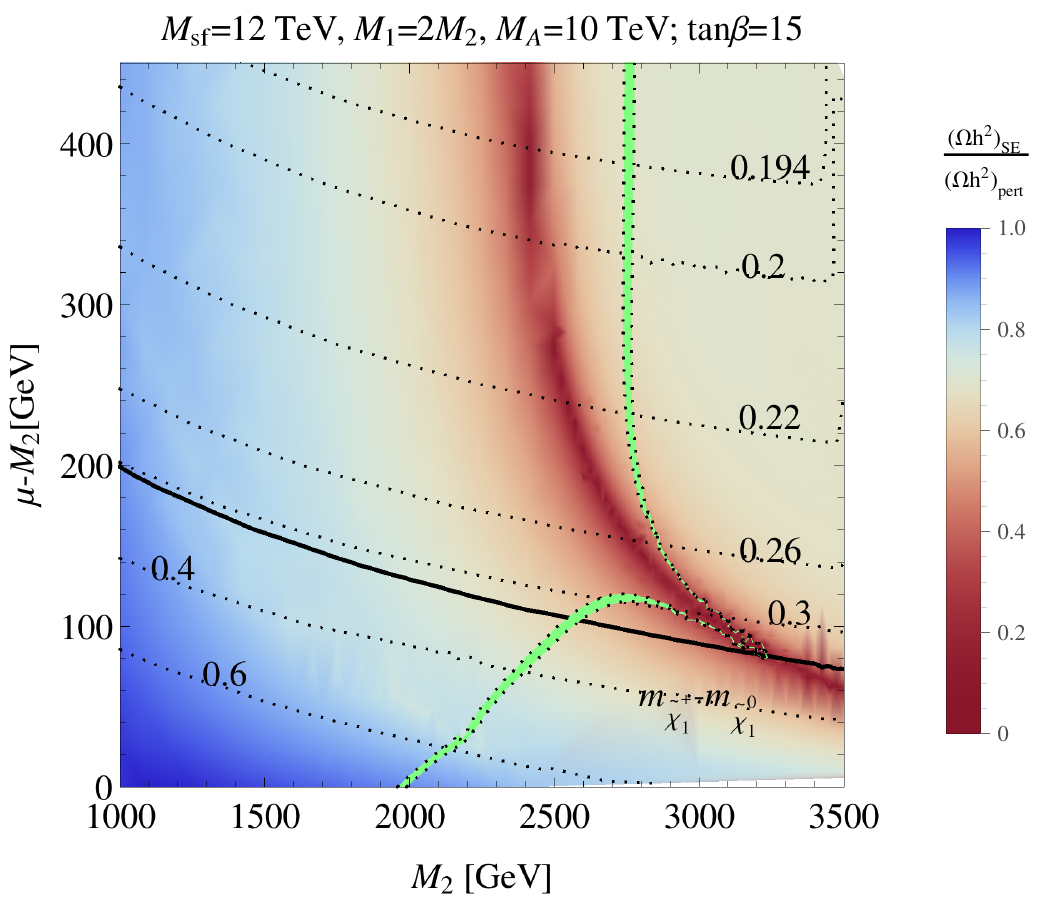}
\includegraphics[width=.50\textwidth]{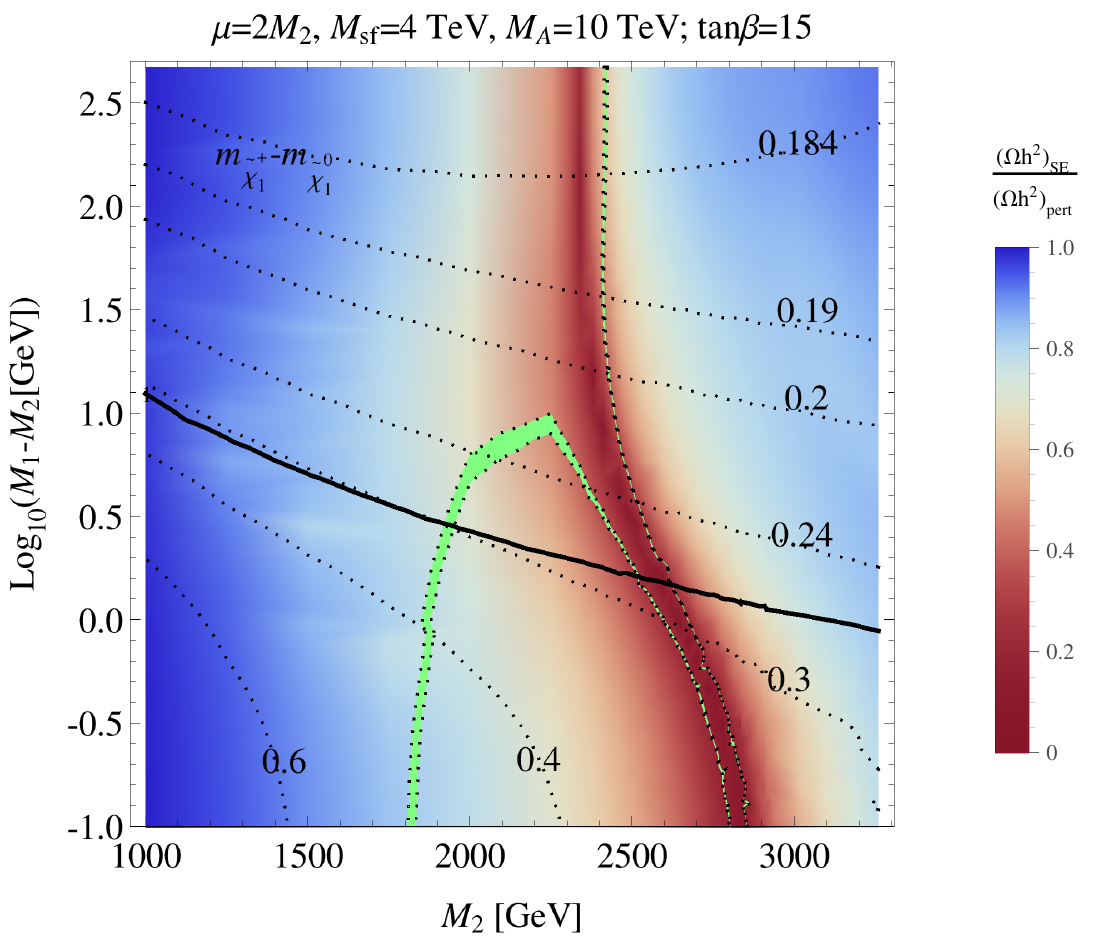}
\caption{Contours of constant zero-temperature 
chargino--neutralino mass splitting in the 
plane of $M_2$ vs. $\mu-M_2$ (left) and $M_1-M_2$ (right) corresponding 
to Figs.~\ref{fig:m2-mu_RD} and~\ref{fig:m2-m1_msf4TeV}. The contours 
include the tree-level and one-loop mass splitting. Above the thick black line 
the mass splitting is radiatively dominated, that is, the one-loop correction 
is larger than the tree-level splitting. The background refers to the 
size of the Sommerfeld effect and the green band to the correct relic 
density (within 2$\sigma$), as in  
Figs.~\ref{fig:m2-mu_RD} and~\ref{fig:m2-m1_msf4TeV}. 
\label{fig:masssplit}}
\end{figure}

In the presence of a Higgsino- or bino- component of the wino-like 
neutralino, there is an additional tree-level mass splitting, which 
can be computed from the mass matrices $X$, $Y$ in 
Sec.~\ref{sec:params}. The dependence on the Higgs vacuum expectation 
value changes from quartic towards the pure-wino limit to quadratic 
when the Higgsino or bino admixture increases. The cross-over to quadratic 
dependence occurs in about the same region in the parameter space shown in 
Figures~\ref{fig:m2-mu_RD} (Higgsino admixture) and~\ref{fig:m2-m1_msf4TeV} 
(bino admixture), respectively, where the tree-level mass splitting 
becomes comparable to and then exceeds the radiative one. In case of 
a mixed wino-Higgsino dark matter particle, the radiative and one-loop 
mass splittings are equal when $\mu-M_2$ (assuming $\mu$ and $M_2$ have 
equal sign) is between approximately 
200$\,$GeV ($M_2=1\,$TeV) and 100$\,$GeV ($M_2=3.5\,$TeV), which includes 
the ``nose'' in Fig.~\ref{fig:m2-mu_RD}, where the ``correct relic density'' 
line is pulled into the Sommerfeld resonance. For mixed bino-wino dark 
matter, equality occurs when $M_1-M_2$ ranges from approximately 
10~GeV to 1~GeV in the range of $M_2$ shown in 
Fig.~\ref{fig:m2-m1_msf4TeV}. This is illustrated quantitatively in 
Fig.~\ref{fig:masssplit}, which shows contours of constant mass splitting 
on the background of Figs.~\ref{fig:m2-mu_RD} and~\ref{fig:m2-m1_msf4TeV}. 

Thus, the neutralino--chargino mass splitting has two components with 
different dependence on the Higgs vacuum expectation value and hence 
different temperature dependence. In the numerical investigation of this 
effect, we separate the tree and one-loop contribution to the mass 
splitting and modify each by its own dependence on $v(T)$.

\subsection{Thermal self-energies}

The second effect on the particle masses arises from their thermal 
self-energies. We discuss the case of electroweak gauge bosons and 
neutralino--chargino mass splittings in turn.

\subsubsection{Electroweak gauge bosons}

The potential generated by electroweak gauge boson exchange is the 
Fourier transform of the gauge boson propagator. Including the full 
gauge boson self-energy, the latter is given by 
\begin{equation}
\label{propW}
\frac{1}{\vec{k}^2+m_W^2(T)+\Pi_{00}(k^0,|\vec{k}|)}\,.
\end{equation}
The 00 component appears, since the spin-independent potential is 
generated by the exchange of the zero component of the gauge field.
The one-loop thermal contribution to the self-energy can be represented 
as 
\begin{equation}
[\Pi_{00}]_{\rm thermal} = \frac{g_2^2}{2\pi^2}\,
\int_{-\infty}^\infty dq^0 \,|q^0|\,n_X(|q^0|)\;
I_X\bigg(\frac{k^0}{|\vec{k}|},\frac{q^0}{|\vec{k}|},
\frac{m_W}{|\vec{k}|}\bigg)\,,
\end{equation}
where $n_B$ ($n_F$) denotes the Bose-Einstein (Fermi-Dirac) distribution, 
and $I_X$ arises from the loop integrand after integration over 
the spatial loop momentum $\vec{q}$. $X=B$ ($F$) must be chosen for 
gauge boson (fermion) loops.

The standard expressions for thermal gauge boson masses, which have 
been employed in previous investigations of thermal modifications of the 
Sommerfeld effect \cite{Cirelli:2007xd, Hryczuk:2010zi} refer to the 
self-energy $\Pi_{00}$ with external momentum $k^0,|\vec{k}|\ll T$, 
while in fact $|\vec{k}|\gg T$. Furthermore $|q^0|\sim T$, since for 
larger $|q^0|$ the integrand is exponentially suppressed by the 
thermal distribution function. The appropriate procedure is therefore 
to expand $I_X$ in $k^0/|\vec{k}|$ and $q^0/|\vec{k}|$.

We are not aware of computations of $[\Pi_{00}]_{\rm thermal}$ for 
general external momentum in the broken electroweak gauge theory. We 
shall therefore estimate the self-energy in the unbroken theory, 
using results from Ref.~\cite{Brambilla:2008cx} for QCD, which is formally 
applicable to our situation when $T \gg m_W$, which may be realised 
at the beginning of freeze-out. Expanding the expressions for 
$I_X$ given in Sec.~III of Ref.~\cite{Brambilla:2008cx}, we find 
\begin{equation}
\label{Pi00}
[\Pi_{00}]_{\rm thermal} = -\frac{g_2^2 T^2}{9}\,,
\end{equation}
up to corrections of order $T^2/|\vec{k}|^2\sim T/m_\chi$. 
This result arises from gauge boson (and ghost) loops only. The fermion 
loop contribution is suppressed. In the broken theory with massive 
gauge bosons in the loop, the expression above will be multiplied 
by some function $f(m_W(T)/|\vec{k}|)$ with $f(0)=1$. It is reasonable 
to assume that a massive propagator will suppress the loop 
integral such that $0<f(x)\leq 1$. The above estimate therefore 
provides an upper limit on the thermal self-energy modification of 
the propagator (\ref{propW}).

The self-energy correction (\ref{Pi00}) is opposite in 
sign and has a smaller coefficient than the usual thermal mass, making 
it less relevant. 
With regard to Eq.~(\ref{propW}) we note that it has the same 
temperature dependence and sign as the Higgs vacuum expectation 
value effect (\ref{MWT}), but the coefficient of 
Eq.~(\ref{Pi00}) is about five times smaller as long as $T<T_c$. Above 
$T_c$, $[\delta m_W^2]_{\rm vev}=-m_W^2$ remains constant and is 
eventually exceeded by the thermal self-energy contribution. This 
is not relevant for the Sommerfeld effect in the dark matter 
mass range considered here, so we may assume that the dominant 
effect on the electroweak Yukawa potential is due to 
Eq.~(\ref{MWT}).\footnote{As an aside we note that in the unbroken gauge 
theory with massless gauge bosons, the negative value of Eq.~(\ref{Pi00}) 
seems to lead to a singularity in $1/(\vec{k}^2+\Pi_{00})$. However, 
Eq.~(\ref{Pi00}) has been derived under the assumption $T\ll |\vec{k}|$, 
implying $|\Pi_{00}|\ll \vec{k}^2$, hence the singularity arises 
for values of $\vec{k}$ outside the approximations made. The thermal 
correction to the propagator is equivalent to 
$-[\Pi_{00}]_{\rm thermal}/\vec{k}^4$ up to corrections beyond the adopted 
approximation, which in coordinate space 
amounts to a correction of the form $\alpha_2^2 T^2 r$ on top of 
the $\alpha_2/r$ Coulomb potential.}

\subsubsection{Neutralino--chargino mass difference}

The thermal one-loop self-energy of a fermion in a vector-like theory with 
massless gauge bosons was considered in Ref.~\cite{Weldon:1982bn}. The 
extension to massive gauge boson exchange and the full electroweak 
theory can be found in Refs.~\cite{Petitgirard:1991mf} 
and~\cite{Quimbay:1995jn}, 
respectively. The latter reference also covers the MSSM, which 
features loop diagrams with gauge boson and Higgs exchange, as well 
as a fermion-sfermion loop.

The thermal correction to the fermion mass is obtained by expanding 
the self-energy in $\vec{k}/m_\chi$, where $k$ is the external momentum, 
and by solving the dispersion relation for $\omega(\vec{k}=0) = 
m_\chi + [\delta m_\chi]_{\rm thermal}$. For the photon radiative 
correction to the chargino mass we find 
\begin{equation}
\label{QEDmasssplit}
[\delta m_{\chi^+}]_{\rm thermal,\gamma} = \frac{\pi \alpha_{\rm em}}{3}\,
\frac{T^2}{m_\chi}\,,
\end{equation}
and zero for the neutralino. This generates a mass difference, which 
is of order 50~MeV at the beginning of freeze-out, smaller than 
but in a similar ballpark as the zero-temperature radiative mass 
splitting. However, the thermal correction decreases rapidly with $T$.

Because of this, we do not evaluate the mass splitting in the full 
MSSM. Rather, we give some estimates. First, to obtain an idea of the 
effect of a non-zero gauge boson mass we generalise the above expression  
to the case of a massive photon using results from 
Refs.~\cite{Petitgirard:1991mf,Quimbay:1995jn}. We find that the 
right-hand side of Eq.~(\ref{QEDmasssplit}) is multiplied by the factor 
\begin{equation}
r(x_T) = \frac{6}{\pi^2} \int_{x_T}^\infty 
dx\,\frac{\sqrt{x^2-x_T^2}}{e^x-1}\left(1+\frac{x_T^2}{2 x^2}\right),
\end{equation}
where $x_T=m_\gamma/T$. The second term never exceeds a few percent of 
the first. The modifying factor equals 1 at $x_T=0$ by definition, 
reaches $r(1)\approx 0.4$ and is exponentially suppressed for $x_T>1$ 
as expected. The electroweak gauge boson contribution to the chargino 
and neutralino self-energy is suppressed by $r(x_T)$ relative to 
photon exchange, but is multiplied by the larger SU(2) coupling. Hence 
it can compete with (\ref{QEDmasssplit}) at the beginning of freeze-out 
if the dark matter mass is above 2.5~TeV, but is exponentially suppressed 
very soon after. Similar conclusions apply to Higgs exchange.

Finally, we consider the fermion-sfermion loop. The relevant case is 
the thermal correction to the fermion propagator, which can be assumed to 
be massless, since the top quark is too heavy to be relevant. We find 
\begin{equation}
\label{sfermionfermionmasssplit}
[\delta m_{\chi}]_{\rm thermal, (s)fermion} = 
\mbox{coupling factors}\times\mbox{const}\times \frac{m_\chi T^4}{
(\msf^2-m_{\chi}^2)^2}\,.
\end{equation}
We recall that to avoid sfermion co-annihilation, we assumed 
$\msf > 1.25 \,m_{\chi}$ in our analysis. 
The last factor is therefore parametrically of order $T^4/m_\chi^3$, 
which makes this contribution negligible compared to the photon 
correction. 
We may therefore assume that the photon correction (\ref{QEDmasssplit}) 
is the only relevant contribution to the neutralino--chargino mass splitting, 
except perhaps at the beginning of freeze-out, where it still provides 
a reasonable estimate up to ${\cal O}(1)$ factors.

\subsection{Effect on the Sommerfeld enhancement and relic density}

We proceed to estimating the thermal modification of the Sommerfeld 
effect and its consequences for the relic density. For wino-like dark matter 
the Sommerfeld effect arises primarily from ladder diagram topologies 
with the exchange of $W$ bosons. The loop momentum is in the potential 
region satisfying $k^0 \ll |\vec{k}|\ll m_\chi$. With $\vec{p}\ll m_\chi$ 
the external momentum of the ladder diagram, we can estimate the 
magnitude of the contribution of each ladder rung from the loop momentum 
region $|\vec{k}| \sim \lambda |\vec{p}\,|$ as 
\begin{equation}
\label{SFI}
I \sim \frac{\pi\alpha_2 m_\chi}{|\vec{p}\,|}\times 
\frac{\lambda^3}{\left(\lambda^2+\frac{m_W^2(T)}{|\vec{p}\,|^2}\right)
\left(\lambda^2+\lambda+\frac{m_\chi \delta m_{+0}(T)}
{|\vec{p}\,|^2}\right)}\,,
\end{equation}
which multiplies the tree annihilation cross section. The Sommerfeld 
enhancement is non-perturbative and large when $I$ becomes of order 1.

Several well-known results follow immediately from this equation. 
(1) When $m_W=0$, $\delta m_{+0}=0$, the largest contribution arises 
from $|\vec{k}| \sim |\vec{p}\,| \sim m_\chi v$ (that is, $\lambda\sim 1$), 
and the Coulomb enhancement $\pi\alpha_2/v$ is recovered. (2) When 
$m_W\not=0$, but still $\delta m_{+0}=0$, the same holds as long as 
$|\vec{p}\,| \gg m_W$. As soon as $|\vec{p}\,| \sim  m_W$ or smaller, 
the largest contribution is obtained from $\lambda \sim m_W/|\vec{p}\,|$, 
and $I\sim \pi\alpha_2 m_\chi/m_W$ independent of the external momentum. 
This is the saturation regime for the Sommerfeld enhancement of the 
Yukawa potential. (3) The neutralino--chargino mass difference provides 
an ${\cal O}(1)$ modification of the Sommerfeld enhancement factor $I$ 
whenever $m_\chi \delta m_{+0} \sim \mbox{max}\,(|\vec{p}|^2, m_W^2)$ 
and reduces or cuts off the enhancement when $m_\chi \delta m_{+0}$ 
is larger than the right-hand side of this relation.

In the following we use the above expression to estimate the impact 
of the thermal modifications of $m_W(T)$ and $\delta m_{+0}(T)$ 
discussed in the previous subsection. In doing so, we correlate the 
external neutralino or chargino momentum with the temperature of the 
Universe according to $\vec{p}^{\;2} \sim m_\chi T = m_\chi^2/x$. 
We further 
support these estimates by implementing the thermal effects into 
our Sommerfeld code as described below.

\subsubsection{Yukawa potential}

We first study the modification of the Yukawa potential generated  by 
$W$ exchange. As discussed above, the leading effect is the temperature 
dependence of the Higgs vacuum expectation value. The thermal self-energy 
correction has the same temperature dependence, but is smaller.
The modification of $I$ is due to 
the $W$ propagator
\begin{equation}
\frac{1}{\mbox{max}\,(m_\chi T, m_W^2)+m_W^2 - T^2/4}\,,
\end{equation}
where we assume $T<T_c$ and approximate $m_W^2/T_c^2\to 1/4$. At the 
beginning of freeze-out when $m_\chi T> m_W^2$, the relative size 
of the thermal correction is $1/(4 x)$ which for $x_f \sim 20$ does 
not exceed $1.2\%$. 
When saturation is reached at $T_s \sim m_W^2/m_\chi$, 
the relative correction is only of order $m_W^2/(4m_\chi^2) 
\sim 0.02\%$ for a reference dark matter mass $m_\chi=2.5\,$TeV. 

An analytic expression 
for the Sommerfeld effect is available in a one-state model, 
when the Yukawa potential is replaced by the so-called Hulth\'{e}n 
potential, which provides a good approximation \cite{Cassel:2009wt}. 
Using this expression we find a maximal change of the Sommerfeld factor 
of $0.2\%$ at the 
beginning of freeze-out and decreasing afterwards, confirming the 
above simple estimate.

The modification of the relic density is expected to be even smaller, 
since the suppression due to the Sommerfeld effect builds up from 
the beginning of freeze-out, where it is least significant, 
until about $x\sim 10^4$, where annihilations terminate, see for instance 
Figure~4 of Ref.~\cite{Beneke:2014hja}. We have implemented the thermal 
modification of the potential in our code to check this explicitly. 
The modification is CPU expensive, since the Sommerfeld-corrected 
cross section, which is thermally averaged for given $T$, must now be 
recomputed for every value of $T$. We created two-dimensional cross 
section tables in velocity and temperature, adopting 59 
temperature points. We then compute the Sommerfeld effect and relic 
density including the temperature-dependent potentials for the  
Higgsino-to-wino trajectory in MSSM parameter space 
considered in Ref.~\cite{Beneke:2014hja}, to which we refer for details on 
these models. For the mostly wino models 8 to 13 
of the trajectory, we find that the relic density change is below one 
permille in all cases, in good agreement with the above estimates, 
and not visible within the numerical accuracy 
of the code.

\subsubsection{Neutralino--chargino mass splitting}

It is evident from the temperature dependence of the two contributions 
(vacuum expectation value and thermal self-energy) that the largest 
relative effect of the thermal correction to the neutralino--chargino mass 
splitting again arises at the beginning of freeze-out. This has two 
immediate consequences. (1) While the Sommerfeld resonance depends 
sensitively on the mass splitting (see, for instance the two-state 
model in Ref.~\cite{Slatyer:2009vg}, which shares the essential features 
regarding the mass splitting with wino-like MSSM models), the resonance 
effect develops sufficiently late after the beginning of freeze-out. 
We therefore conclude that the thermal effect on the resonance region is 
negligible. (2) At the beginning of freeze-out $\lambda\sim 1$ in 
(\ref{SFI}), hence the relative modification of $I$ due to the mass 
splitting is of order 
\begin{equation}
\frac{m_\chi \delta m_{+0}(T)}
{|\vec{p}\,|^2} \sim \frac{[\delta m_{+0}]_{\rm vev+thermal}}
{T} \approx \mbox{few permille,}
\end{equation}
as the thermal correction to the mass splitting is as large as the 
mass splitting itself.
The numerical estimate is based on the assumption that the entire 
zero-temperature mass difference of up to 0.5~GeV vanishes due to the 
vanishing of the Higgs expectation value at $T=T_c$, which 
gives the largest possible effect.

We studied the impact of the temperature-dependent 
neutralino--chargino mass splitting on the relic density with the extended 
numerical code described above for the wino-like trajectory models 
of~Ref.\cite{Beneke:2014hja}. The thermal modification is again in the 
permille range, in agreement with the analytic estimates, reaching 
$0.7\%$ at maximum. Once again we find that the observed thermal effect 
is of the same order at the numerical uncertainties due to sampling and the 
choice of $x_\infty$, hence we can only state that the thermal 
effect is well below 1\% in all cases studied.

\subsubsection{Summary}

We conclude that thermal modifications of the 
Sommerfeld effect change the relic density at most in the upper permille 
range, which is negligible for all practical purposes. We point out 
that our investigation of thermal effects is not complete. For example, 
we did not discuss the direct modification of the neutralino and 
chargino two-particle wave function due to interactions with gauge bosons 
in the thermal plasma, an effect that would be referred to as 
``dissociation'' by soft gauge bosons in the case of bound states. 
Power counting suggests that this effect is of the same order as the 
ones investigated here. However, since all these are 
far smaller than the theoretical uncertainty from 
perturbative higher-order corrections, which is probably a few 
percent, we do not attempt a complete analysis in this work.


\providecommand{\href}[2]{#2}\begingroup\raggedright\endgroup


\end{document}